\documentclass{llncs}       
\usepackage[utf8]{inputenc}
\usepackage{amsmath,amssymb,amsfonts}
\usepackage{graphicx}
\usepackage{tabularx}
\usepackage{needspace}

\usepackage{parskip}
\usepackage[table]{xcolor}  

\usepackage[breaklinks=true]{hyperref}
\hypersetup{
    colorlinks=true,
    linkcolor=blue,
    citecolor=blue,
    urlcolor=blue,
    breaklinks=true
}
\usepackage{xurl}  

\usepackage{longtable}
\usepackage{booktabs}
\usepackage{array}
\usepackage{ragged2e}
\usepackage{calc}
\usepackage{array}
\usepackage{threeparttable}

\title{Estimating the Scale of Digital Minds}
\author{Derek Shiller}
\institute{Rethink Priorities}
\date{}

\newlength{\defaultskip}
\newcommand{\typetitle}[1]{\noindent\textbf{#1}}

\newcommand{\containedsubsection}{\needspace{.75\textheight}}
\begin{document}

\maketitle
\hypertarget{executive-summary}{%
\section*{Executive Summary}\label{executive-summary}}

This report attempts to estimate the number of digital minds,
operationalized as AI systems with certain directly observable traits
like agency, personality, and intelligence, that make them natural
candidates for moral consideration in the coming decades. It includes
two separate approaches that combine speculative estimates within a formal structure and provides an analysis of prospects in light of both.

The first approach surveys a number of different use cases for digital
minds and predicts adoption rates use-case by use-case. The second
approach speculates about trends in the production and efficiency of
AI-relevant chips independently of their use for digital minds.
Together, these approaches capture aspects of the supply and demand for
digital minds.

These two approaches suggest the following take-aways about the next few
decades:

\begin{itemize}
\item
  Even without extreme assumptions, we get that there could be hundreds
  of millions of digital minds by the early 2030s. There could be
  billions of minds by 2050.
\item
  The total numbers are hard to predict. A 95\% confidence interval
  would range over several orders of magnitude.
\item
  Embodied AIs (robots) are likely to constitute only a relatively small
  portion. The most numerous digital minds are expected to be virtual AI
  systems run on servers and interacted with through computer
  interfaces.
\item
  The total number of digital minds is most likely (in the median case)
  to be dominated by consumer-focused personable AIs (virtual friends,
  personal secretaries, computer game NPCs).
\item
  If we are to try to produce an expected value of the number of future
  digital minds, the number would likely be most influenced by high
  numbers of certain kinds of minds in unlikely scenarios. The main
  kinds of minds in those scenarios differ from what we should expect to
  dominate at the median. We see simulation participants and
  white-collar worker replacements as the main contributors to total
  numbers in expectation. Social roles are more likely to be tied to the
  number of human beings, whereas non-social uses may not be.
\item
  Patterns of consumer comfort around person-like agentic AIs and desire
  for AI services to take such a form is likely to be a significant
  factor in the total number of digital minds that we see in the median
  case, because of their influence on adoption of social uses. Even for
  non-social uses, it may matter if we would be more or less inclined to
  have a human-like AI playing those roles.
\item
  Compute is unlikely to be a hard limit on the number of digital minds
  (as opposed to interest, comfort, and felt need); there will likely be
  sufficient compute (or the ability to provide it) for most desired
  purposes. If compute is plentiful enough, we may expect large numbers
  of digital minds that contribute relatively small amounts of value to
  large projects.
\end{itemize}

\hypertarget{introduction}{%
\section{Introduction}\label{introduction}}

This report aims to estimate the near-future prevalence of AI systems of
a certain potential moral, political, and social importance. These AIs
will be referred to as `digital minds' or `minded' systems and be
defined here to include those computer systems that satisfy certain
traits suggesting prima facie to warrant moral consideration or
political protection.\footnote{Long et
  al.~(\href{https://arxiv.org/abs/2411.00986}{2024}), Dung (\href{https://www.tandfonline.com/doi/full/10.1080/0020174X.2023.2238287}{2025a}), Moret (\href{https://link.springer.com/article/10.1007/s11098-025-02343-7}{2025}), and Goldstein \& Kirk-Giannini (\href{https://link.springer.com/article/10.1007/s44204-025-00246-2}{2025}) explore
  considerations of true moral significance. The traits focused on here
  are relevant, but do not get at issues around gaming
  (\href{https://library.oapen.org/handle/20.500.12657/93755}{Birch
  2024}, chapter 16).}. Experts have suggested that a near-term
explosion of sophisticated, agentic, and perhaps conscious AI systems is
a real possibility
(\href{https://digitalminds.report/forecasting-2025/}{Caviola \& Saad
2025};
\href{https://forum.effectivealtruism.org/posts/2uGShxLsXWGExJYNL/digital-minds-takeoff-scenarios}{Saad
\& Caviola 2024}) Few concrete details have been provided on what the
coming digital populations might look like. This report aims to flesh
out a reasonable picture without getting too in the weeds about deeply
complex and controversial philosophical matters.

This report focuses on two separate approaches to estimating the scale
of digital minds. The first approach draws on assumptions about consumer
(meaning the end-user, be it public, private, or government) preferences
that would drive the development and adoption of minded AI products. The
second approach draws on trends around the scale of AI computing and
projections over where it may go. These two approaches illuminate each
other: the second approach can be read as providing information about
the possible supply of digital minds, whereas the first approach
provides information about possible demand. Together, these two
approaches suggest that we are likely to see large numbers of digital
minds in the coming decades, though the size of those numbers is
influenced by a number of crux factors that we find difficult to
predict.

\hypertarget{scope}{%
\section{Scope}\label{scope}}

\hypertarget{what-is-a-digital-mind}{%
\subsection{What is a digital mind?}\label{what-is-a-digital-mind}}

Digital minds\footnote{I take it that `mind' is a sufficiently ambiguous
  term that this focus on appearances isn't inappropriate. We might ask
  whether such things are genuine minds, or whether they have
  qualitative experiences or a subjective perspective, but regardless of
  the answers to these questions, I will take it for granted that they
  are \emph{minds}, and leave it open whether they genuinely have
  thoughts, beliefs, experiences, etc.} are intended to include the main
candidates for the possession of rights that we risk infringing and
welfare states over which we may be responsible. They are also likely to
be regarded more seriously as moral subjects by non-experts and may
therefore have a greater influence on the societal reception of AI
through providing ambassadors of AI minds to consumers. The definition
we offer aims to operationalize this category of systems in terms of a
series of objectively-accessible traits that seem prima facie to be
relevant to their status as moral patients.\footnote{This is a matter of
  the challenges involved, rather than a commitment about what would be
  best to know. It is difficult enough to make predictions about how we
  will choose to put AI to use. Adding in the complexity of which
  systems we should take seriously as artificial moral patients would
  significantly increase the uncertainty in the estimates.}

There are reasons to focus on a straightforward operationalization of
moral, political, and social significance without delving into
complicated philosophical questions about value:

\begin{itemize}
\item
  It is easier to tell which systems have traits that make them look
  like moral patients, at least on a superficial level, than which
  systems actually will be moral patients.
\item
  We may not feel comfortable with using existing theories to make
  predictions about speculative future entities where not only are we
  uncertain about what it takes to have the properties in question
  (pain, consciousness, etc.) but also how future systems will be
  designed.
\item
  Perceptions, and so superficial traits, may be more important to
  social reactions that are important to consider in their own right. In
  any case, thinking deeply about perceptions is likely to inform
  different schools of thought on the demands of moral status.
\item
  Focusing on surface-level traits captures considerations that are
  important to the potential scale of both under- and over-attribution
  worries without committing to which are more pressing.
\end{itemize}

Despite recent progress, digital minds remain, for now, largely
hypothetical. Current AI systems \emph{may} be moral patients, but only
if most mainstream theories of morality or cognition are incorrect or
because AI systems satisfy the requirements of those theories in some
non-obvious way (\href{https://arxiv.org/abs/2308.08708}{Butlin et
al.~2023};
\href{https://link.springer.com/article/10.1007/s43681-023-00379-1}{Sebo
\& Long 2025}). However, it is uncertain which directions future
technologies will take and it is not easy to apply existing theories of
moral, social, or political importance to AI systems.

What it takes to have welfare or rights is controversial. For the
purposes of this report, we will take an agnostic approach by
identifying a range of candidate features. We will count anything
broadly in line with those features as being a digital mind. In order to
be a digital mind, an entity must satisfy a number of the following
criteria. It is not necessary to satisfy them all.

Our criteria are:

\begin{enumerate}
\def\labelenumi{\arabic{enumi}.}
\item
  \textbf{Having desires, interests, or goals that are consistent over
  time and across contexts.}

  Desires, preferences, goals, or aims are internal states that shape a
  system's behavior in the direction of achieving specific ends,
  allowing the system to behave in systematic and predictable ways even
  as it is disturbed or redirected from their present focus. Having such
  states is an important part of agency
  (\href{https://philarchive.org/rec/DUNUAA}{Dung 2025b}) and provides a
  significant component of welfare. Consistency over time is partly
  constitutive of having preferences that matter and is central to being
  recognized as a persistent entity with an identity worth respecting.
  We will assume that what matters is the appearance of goals and the
  appearance of their consistency. We need not worry that this
  appearance may reflect an underlying reality that is quite different.
\item
  \textbf{Having a stable and coherent personality between interactions
  and idiosyncratic character traits.}

  A stable and coherent personality seems likely to be an important
  contributor to being recognized as a persistent individual, and
  potentially may contribute to its treatment of generally different
  from other instances of its kind. Distinctive character traits may
  also make systems seem more human-like. Radical behavior changes
  across contexts will make them seem less human-like. The extent to
  which character traits belong uniquely to specific entities
  contributes to categorizing those entities alongside biological
  organisms, which are subject to the vagaries of genetic recombination.
\item
  \textbf{Having the ability to navigate an environment and interact
  with external objects and entities in complex open-ended ways, either
  in a digital or in the real-world.}

  Interacting with a complex environment requires a sophisticated
  ability to model the self and its relation to the world. Furthermore,
  it enables systems to interact with us in ways that reflect the depth
  and complexity of human-to-human interactions, plausibly helping us to
  see them as we see other people. AI systems that aren't able to
  navigate the world will have less flexibility in their modes of
  interaction and may be easier to regard as non-real entities relegated
  to an inhuman realm.
\item
  \textbf{Having the ability to learn, grow, evolve in beliefs or
  interests.}

  Growth and development capacities present a stark contrast between
  present-day AI systems and the biological entities to whom we are more
  inclined to extend moral status. We're accustomed to computer systems
  that are rigidly programmed and only update behavior patterns in
  accordance with hard rules. In contrast, animals develop unique
  behavioral profiles slowly and somewhat unpredictably over time as a
  result of the specific path of experiences they encounter. This
  contributes to the sense of persistence of the individual as an
  individual over time, rather than something that exists timelessly or
  as a series of clones. We should expect greater empathy for systems
  that pattern with animals in this way, and perhaps recognize such
  systems as having more meaningful identities.
\item
  \textbf{Displaying general intelligence, creativity, or mental
  flexibility.}

  General intelligence is strongly correlated with our perceptions of
  moral status among non-human animals. We should expect similar trends
  to hold for AI systems, we place specific emphasis on the flexible
  deployment of intelligence: intelligence in a narrow domain, such as
  chess, is less relevant.
\end{enumerate}

Some of these traits have a more direct connection to prevailing expert
views about welfare.\footnote{For discussions of AI welfare, see
  (\href{https://arxiv.org/abs/2411.00986}{Long et al.~2024}),
  (\href{https://link.springer.com/article/10.1007/s11098-025-02343-7}{Moret
  2025}), and
  (\href{https://link.springer.com/article/10.1007/s44204-025-00246-2}{Goldstein
  \& Kirk-Giannini 2025})} Having an idiosyncratic personality is not
generally considered necessary. Nor is the ability to grow or change.
Still, these considerations are likely to figure collectively into
public perceptions of moral status; we are used to discounting systems
as less real given superficial indicators of artifice. It is quite
likely that our intuitions about these matters will change over time,
but we can assume that attitudes now will provide a rough guide to
popular opinion over the next two decades.

For specificity, we may regard any systems that display more of these
features or these features to a greater degree as better qualifying as
digital minds. A system that had stable goals, intelligence, and
autonomy but no access to an external world would count as a digital
mind. An unintelligent system with a coherent personality and the
ability to learn through interactions with the world would also count.
But a system that could produce intelligent responses without any stable
goals, perceptual abilities, or personality would not qualify strongly.

Some of our analyses will incorporate a `degree of mindedness'
criterion. We will treat the overall degree of mindedness as a function
of the number of traits satisfied. It is assumed that all systems will
exhibit these traits to a low, moderate, or high degree. For each trait,
we attribute a value of 1,2, or 3, and take the relative mindedness of
two systems to be a ratio of the sum total of traits to a power of 2.
For instance, a system that has a sum of scores of 10 across features
counts as twice as minded as a system with a sum of 7. The rationale for
this is that there are important synergies between the traits that make
them more significant in combination. For instance, it is possible to
have a greater degree of agency the more intelligent you are, or the
more you're able to navigate complex environments.

\definecolor{rpblue}{RGB}{46,113,143}  
\definecolor{rpgray}{RGB}{46,46,46}  

\newcommand{\header}[1]{\textbf{#1}}
\newcommand{\leadcell}[1]{\textbf{#1}}
\newcommand{\headerspacing}{\rule{0pt}{3ex}\rule[-1.5ex]{0pt}{0pt} }

\begin{table}
\centering
\caption{Computed mindedness values}
\begin{tabular}{lr}
\toprule
\header{Trait Sum\hskip0.5cm} & \header{Mindedness -- n$^2$/225} \\
\hline
\textbf{ 1} &   1/225 (00.4\%) \\
\textbf{ 2} &   4/225 (01.8\%) \\
\textbf{ 3} &   9/225 (04.0\%) \\
\textbf{ 4} &  16/225 (07.1\%) \\
\textbf{ 5} &  25/225 (11.1\%) \\
\textbf{ 6} &  36/225 (16.0\%) \\
\textbf{ 7} &  49/225 (21.8\%) \\
\textbf{ 8} &  64/225 (28.4\%) \\
\textbf{ 9} &  81/225 (36.0\%) \\
\textbf{10} & 100/225 (44.4\%) \\
\textbf{11} & 121/225 (53.8\%) \\
\textbf{12} & 144/225 (64.0\%) \\
\textbf{13} & 169/225 (75.1\%) \\
\textbf{14} & 196/225 (87.1\%) \\
\textbf{15} & 225/225 (100.0\%) \\
\bottomrule
\end{tabular}
\end{table}

These criteria are vague and there will be some difficulties in
precisely applying them, but in practice, we find that it is fairly
straightforward to categorize plausible hypothetical systems.

\hypertarget{what-are-we-estimating}{%
\subsection{What are we estimating?}\label{what-are-we-estimating}}

aThe present report is focused on estimating the scale of digital minds.
This might be understood in various ways.

Different estimated quantities suggest different implications and it is
not our intention to advocate for the value of specific estimates.
However, certain forms of scale are easier to evaluate; we will focus on
those.

\begin{itemize}
\item
  \textbf{Number of distinct individuals}

  We might try to calculate the number of identifiable AI individuals
  that will exist each year. This presents a number of challenges, such
  as how to distinguish individual systems that aren't tied to specific
  hardware or that can be transmitted, cloned, or forked. We might tie
  individuals to specific hardware, but then we must deal with how to
  think about what to make of cases when that hardware is used to run
  separate models.
\item
  \textbf{Amount of cognitive work done}

  Alternatively, we might try to estimate the number of fundamental
  computations performed across all relevant systems. Previous attempts
  (\href{https://www.cambridge.org/core/journals/utilitas/article/abs/astronomical-waste-the-opportunity-cost-of-delayed-technological-development/2969D64410332BD099F36BAFC5B2ADE5}{Bostrom
  2003};
  \href{https://joecarlsmith.com/2025/05/21/the-stakes-of-ai-moral-status}{Carlsmith
  2025}) to estimate digital mind numbers have looked to numbers of
  basic operations. This approach dodges the questions around the
  individual identity of computational processes. However, it requires
  us to have a sense of what different numbers mean and depends heavily
  on utilization estimates.
\item
  \textbf{Scale of welfare significance}

  Instead, we might be particularly concerned with digital minds as an
  ethical issue. This includes concerns about artificial welfare, in
  which case we might be particularly interested in valenced and / or
  conscious states, which require predicting something about the
  constitution of future digital minds. Given the possibility of
  distorted scales of welfare
  (\href{https://veterinaryfuturesociety.org/wp-content/uploads/2025/03/2020-Sharing-the-World-with-Digital-Minds-N.-Bostrom.pdf}{Shulman
  and Bostrom 2021}), these concerns may come apart from the number of
  individuals or the amount of work they do.

\item
  \textbf{Scale of social significance}

  Finally, we might be particularly concerned with digital minds for the
  effect they have on society. The more we interact with digital minds,
  the more it will affect the way we think. Peer discussions have a
  significant impact on shaping belief, and so regular digital minds
  might give substantive social control over to their makers.
\end{itemize}

The Futures with Digital Minds report
(\href{https://digitalminds.report/forecasting-2025/}{Caviola and Saad
2025}) collected expert answers to the question
``\href{https://digitalminds.report/forecasting-2025/\#speed}{After the
first digital mind is created, how many years will it take until the
collective welfare capacity of all digital minds together (at a given
time) matches that of at least 1000 / 1M / 1B / 1T humans}''. (They
adopted a more elaborate interpretation of digital minds than the
present one, on which digital minds must have phenomenal experiences.)

There are several challenges to this welfare-based approach. This
requires having some sense of the nature of hypothetical AI cognition
and some sense of the scale of individuals. Furthermore, different
theories of welfare might lead to conflicting conclusions even holding
the future fixed. For instance, digital minds might lack (many)
conscious experiences, but have strong desires. There is significant
debate (e.g.~\href{https://philarchive.org/archive/DUNPTN}{Dung 2024};
\href{https://www.ingentaconnect.com/content/imp/jcs/2014/00000021/f0020001/art00007}{Levy
2014}) about whether their desires would be sufficient for strong
welfare claims.

In this report, we will be adopting a middle-of-the-road perspective
that focuses primarily on the number of individuals with digital minds,
defined in terms of units that occupy specific relationships
(e.g.~social or economic)\footnote{One challenge here is that given AI
  systems -- both hardware and software -- may serve many different
  roles and it can be hard to individuate moral patients (\href{https://philarchive.org/rec/CHAWWT-8}{Chalmers 2025}; \href{https://link.springer.com/article/10.1007/s11098-025-02409-6}{Register 2025}; \href{https://link.springer.com/article/10.1007/s11229-025-05310-1}{Shiller 2025}). One model on one GPU cluster might provide conversations for a
  hundred different virtual friends. In this case, we aim to capture the
  number of interfaces from the user's perspective rather than try to
  quantify hardware configurations.} and understood to occupy fairly
coherent agentic perspectives. We will also consider some connections on
the basis of varying utilization and degree of mindedness of those
individuals.

\hypertarget{time-frame}{%
\subsection{Time-frame}\label{time-frame}}

At the time of writing, we are most of the way through 2025. Modern forms of AI
have largely been shaped over the past decade. Public awareness goes
back fewer than five years. It is reasonable to expect it to be possible
to create plausible projections regarding where technology and society
will be in five years. The world doesn't change that much over such a
short period: industrial and economic plans have little time to be both
redirected and carried out, norms and expectations evolve relatively
slowly,\footnote{Compare trends in the adoption of other technologies.
  Internet use gradually rose throughout the 90s, but didn't see a
  drastic role in the lives of most users until the rise of social
  media. Smart phones remained niche products before the introduction of
  the iphone and took 10 years thereafter to saturate the market.} and
social changes take time to have an effect.

We can likely carry forward current trends to 2030 without too much
doubt. But thereafter, it gets increasingly murky. What 2035 will look
like will depend a lot on what happens in 2030, and what 2040 will look
like will depend a lot on what happens in 2035. Prevailing trends may
depend both on the direction that technologies will take --- the path of
least resistance given the political and technological landscape --- and
the choices stakeholders make in how technologies get deployed. The
future will also depend on world events shaped by the dynamic
interactions of countries, each with their own complex internal
politics. The 2040s and 2050s are even more difficult to pin down, and
past projections that far out have tended to miss out on important
developments that shape history thereafter.

One particular problem with projections in the coming decades is that
artificial general intelligence (AGI) might be a radically
transformative technology and might come at any moment. According to
some, we can expect massive social and industrial transformation after
the development of AGI. AGI might arrive in the next few years
(\href{https://ai-2027.com/}{Kokotajlo et al.~2025};
\href{https://www.metaculus.com/questions/5121/date-of-artificial-general-intelligence/}{Metaculus
2020}; \href{https://ourworldindata.org/ai-timelines}{Roser 2023}). So
even 2030 might be difficult to predict. Experts predict we will get AGI
in the coming decades, and so any predictions should take that into
account. However, it also seems reasonable not to expect the development
of AGI to be immediately transformative, as it won't immediately be
deployed everywhere, and it won't immediately solve physical constraints
such as infrastructure limitations and the challenges of supply chains.
Even if we get AGI by 2027, we might think that our estimates for 2030
are reasonably reliable.

This report focuses on the period between now and 2050, noting that
things get increasingly speculative the further out we consider and
accepting that AGI might produce a radical kink in the projections
without trying too hard to project how futures with radically different
trajectories might go.

\hypertarget{approaches}{%
\section{Approaches}\label{approaches}}

\hypertarget{estimation-strategies}{%
\subsection{Estimation Strategies}\label{estimation-strategies}}

Estimating the number of digital minds in the coming decades requires
making guesses about the viability and popularity of technologies that
don't yet exist. It is a speculative enterprise without an obvious
methodology. In light of the challenges of knowing which approaches are
most reliable, it is helpful to pursue a diversity of approaches.
Looking at the results of a variety of strategies can provide a picture
of the range of reasonable answers.

This report focuses on two different approaches. While these approaches
produce different answers, they are best interpreted together, each
casting light on the limitations of the other.

The first approach looks at the different contexts in which digital
minds might be employed. The ability of industry to supply computer
processors to meet demand over the past half-century makes it seem
likely that the number of digital minds will result more from the limits
of demand rather than the limits on the ability to supply it. Growth of
the markets for products in those contexts can be estimated based on
past trends and on expected need, etc. This approach is mostly focused
on counting individuals through their roles, though our analysis will
also consider the expected activity levels of those individuals over
time.

The second approach looks at projections for overall computational
capacities of AI-relevant chips and considers reasonable percentages of
these overall capacities that could be devoted to digital minds. Given
background assumptions about translations of compute allocations into
numbers of individuals, we can derive a number of individuals that the
hypothesized allocations will be sufficient to support.

\hypertarget{result-overview}{%
\subsection{Result Overview}\label{result-overview}}

The two approaches that will be discussed in detail in the following
sections. They suggest an overall picture about the growth of digital
minds. We summarize those conclusions here.

Both approaches suggest that the number of digital minds will fall in
the range of millions to tens of billions and grow steadily over time.

\begin{table}[htbp]
\centering
\begin{threeparttable}
\caption{Median digital mind productions by year.}
\begin{tabular}{lcccc}
\toprule
& \parbox[t]{2.5cm}{\centering \header{Approach 1\\ Median}} & \parbox[t]{2.5cm}{\centering \header{Approach 1\\97.5\%}} & \parbox[t]{2.5cm}{\centering \header{Approach 2\\Moderate\footnotemark[1]}} & \parbox[t]{2.5cm}{\centering \header{Approach 2\\Optimistic\footnotemark[2]}} \\
\midrule
\leadcell{2030} & $4.82 \times 10^6$ & $2.26 \times 10^9$ & $4.40 \times 10^7$ & $5.08 \times 10^8$ \\
\leadcell{2035} & $2.53 \times 10^8$ & $2.01 \times 10^{10}$ & $5.32 \times 10^8$ & $1.38 \times 10^{10}$ \\
\leadcell{2040} & $5.61 \times 10^8$ & $4.25 \times 10^{10}$ & $1.66 \times 10^9$ & $1.13 \times 10^{11}$ \\
\leadcell{2045} & $8.74 \times 10^8$ & $6.26 \times 10^{10}$ & $2.24 \times 10^9$ & $2.49 \times 10^{11}$ \\
\leadcell{2050} & $1.34 \times 10^9$ & $9.13 \times 10^{10}$ & $2.32 \times 10^9$ & $2.84 \times 10^{11}$ \\
\bottomrule
\end{tabular}
\begin{tablenotes}
\item[1]{This assumes a 0.01\% allocation to digital minds and the `moderate' scenario for compute growth against the Llama-2 7B benchmark.}
\item[2]{This assumes a 0.1\% allocation to digital minds and the `rather optimistic' scenario for compute growth against the Llama-2 7B benchmark.}
\label{tab:digital_minds}
\end{tablenotes}
\end{threeparttable}
\end{table}

Compute growth projections suggest that we will have capacity to run
sufficient numbers of digital minds to meet consumer demand in most
cases. Computational resources are not likely to be the primary
constraint on the prevalence of digital minds. There will, of course, be
computation constraints on the total number of digital minds that we
could conceivably build, and the amount of available compute will affect
the costs of adoption. If we were to utilize most available processing
power to run digital minds, then we might see vast numbers.

The fact that we have the ability to produce vast numbers of digital
minds doesn't mean we will choose to do so: given the assumption that,
at least in the near future, digital minds will be built to serve
existing human needs, we should only expect large numbers of digital
minds if those needs cannot be easily satiated with smaller numbers or
if the costs of excess (time, energy, environmental impact, industrial
capacity, moral concern etc.) become so cheap that they are negligible.

We should expect some interaction between the demands of customers for
digital minds and the amount of computational power devoted to their
creation. The more demand for digital minds, the greater the share of
total compute that will be given to them, and the more companies will be
willing to spend on further processors. The key claim is that
computation costs do not appear to be a hard limit: for the most obvious
uses, we should not expect available compute to fail to meet
demand,\footnote{The amount demanded will depend on price, but we can
  also consider an absolute level of demand for products that are
  virtually free. I think compute won't greatly constrain demand: though
  products that involve compute may still be expensive (robot bodies)
  the number of such products we should expect probably won't be very
  sensitive to the price of compute except in some more speculative
  cases.} though it may rein in some of the more indulgent uses of
digital minds.

There are barriers to adoption that might prevent us from rapidly
deploying digital minds to satiate existing needs. There are likely to
be some technological hurdles that prevent digital minds from doing
everything we might want them to do. There will likely also be social
barriers to adoption: concerns about human-AI relationships, fears about
significant changes to lifestyles, stubbornness about past habits.
Furthermore, AI may cause significant disruption, harm large groups of
people, and be politically divisive. Such turmoil may enhance the stigma
attached to novel relationships or cause large groups of people to make
conservative decisions. AI companies may be motivated to make their
products seem less human-like in order to increase acceptance.

Many of the main contemporary uses of AI (generating content, writing
code, answering questions, etc.) seem like they may be possible (and
have already been accomplished) without instantiating full digital
minds. They don't need anything with an ongoing personality, agentic
capabilities, or embodied form.

We can speculate that future AI systems will be able to take on a wider
range of tasks and some of those tasks will benefit from mindedness.
With many of the most significant potential uses, (producing innovative
research, taking on every aspect of the role of a remote employee) it is
less clear whether mindedness will or will not be useful. It is possible
that the traits conducive to success in their tasks also lead them to
score well on the particular traits we've highlighted. It is also
possible that future companies will find ways to utilize AI without
implementing digital minds.

Overall, we postulate two main phases of digital minds over the coming
decades.

In the next 10 years (2025-2035), we expect digital mind populations to
emerge and be dominated by chatbots designed to serve a practical or
social dimension. This includes systems that fit into our lives as
friends, but also systems take over certain roles traditionally played
by people (teachers, secretaries, therapists). Large language models are
primed to occupy such roles, and will likely be capable of doing so in
the next few years. There will be fewer direct technological hurdles to
deployment. We see some versions of such systems today and clear
interest from both investors and consumers in more robust products along
this dimension.

There are some more speculative uses to which we might see systems put
in large numbers during this period, but only at lower probabilities.
Among these, we find that virtual employees and simulations are
possible, and could capture a larger slice of the total pie, but are not
expected to be significant at the median.

\begin{figure}[htbp]
    \includegraphics[width=0.5\textwidth]{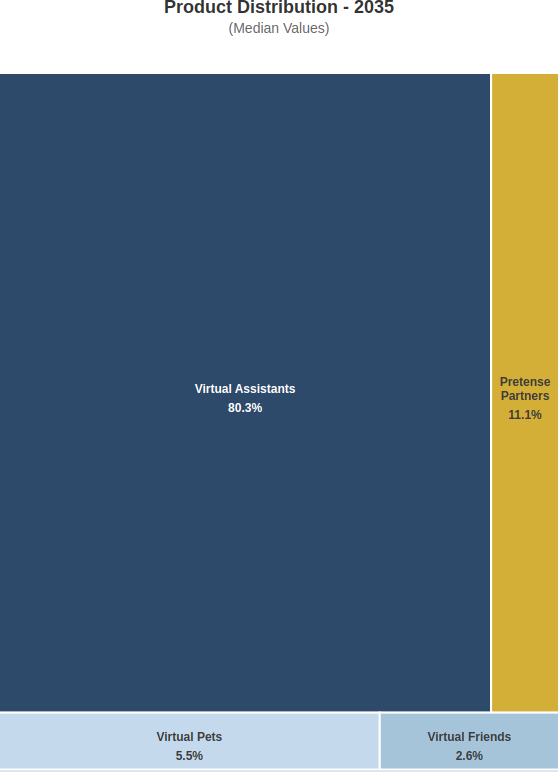}
    \includegraphics[width=0.5\textwidth]{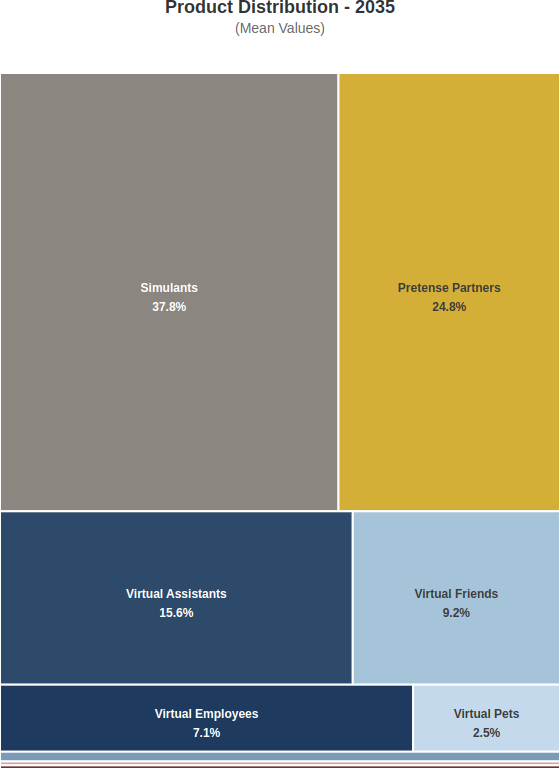}
    \vskip1cm
    \includegraphics[width=0.5\textwidth]{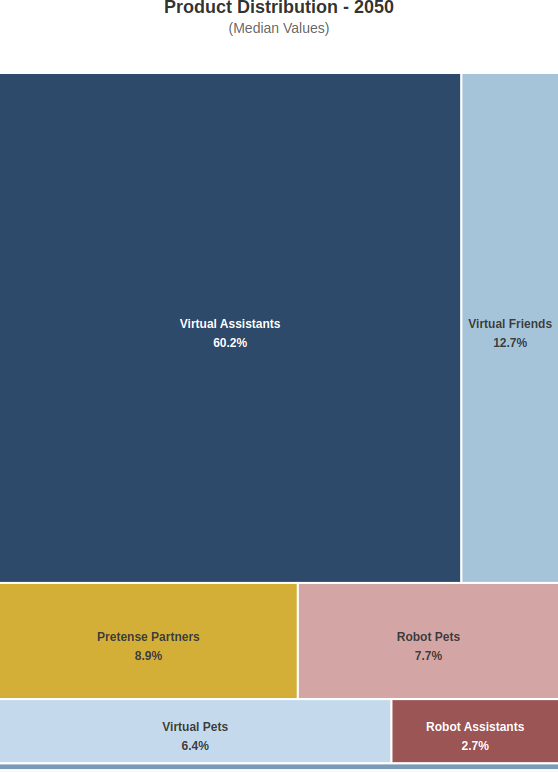}
    \includegraphics[width=0.5\textwidth]{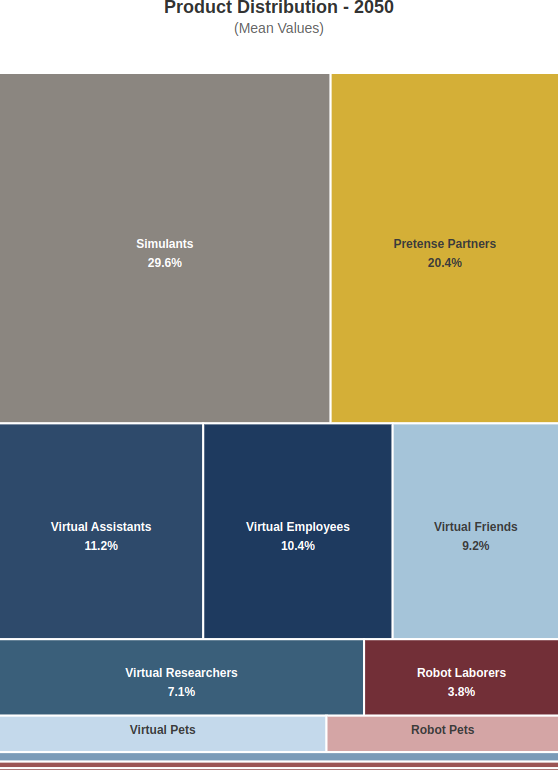}
    \caption{\textbf{Product distributions.} These charts displays the relative makeup of digital mind populations at the median and mean of projections in 2035 and 2050.}
\end{figure}

During most of the next 10 years, we expect that robotics will likely
remain somewhat capability-limited and so will remain niche
(\href{https://www.metaculus.com/questions/16625/date-of-reliable-and-general-household-robots/}{Metataculus
2023};
\href{https://www.construction-physics.com/p/robot-dexterity-still-seems-hard}{Potter
2025}). There may be robots of varying degrees of flexibility and power
that exist to entertain or perform work for us. Some of these will
likely be minded systems. We don't anticipate these being particularly
large in number and their costs and value suggest we will see fewer
numbers of them, though their status as independent entities with minds
might be stronger.

In the following 15 years (2050), we see robotics improving and costs
coming down, leading to robots with digital minds becoming a greater
share of total digital minds. At the same time, adoption of chatbot
services continues to increase, and so while there is some jockeying for
rank, this doesn't radically change percentage breakdowns.

Overall prevalence hides different dimensions of importance. The most
prevalent digital minds may not be the digital minds that matter most,
either in terms of their ethical significance or their potential for
social impact. The most numerous digital minds estimated in the
preceding charts (simulants, pretense partners) will probably lack
robust, continuous lives. The potential for numerosity is partly
explained by the shorter (or periodic) duration of their existence.
Applying a simple skew for duration and degree of mindedness,\footnote{Relative
  value is given by estimated numbers × hours per year × mindedness
  score.} we see a very different picture of weighted numbers.

\begin{figure}[htbp]
    \includegraphics[width=0.5\textwidth]{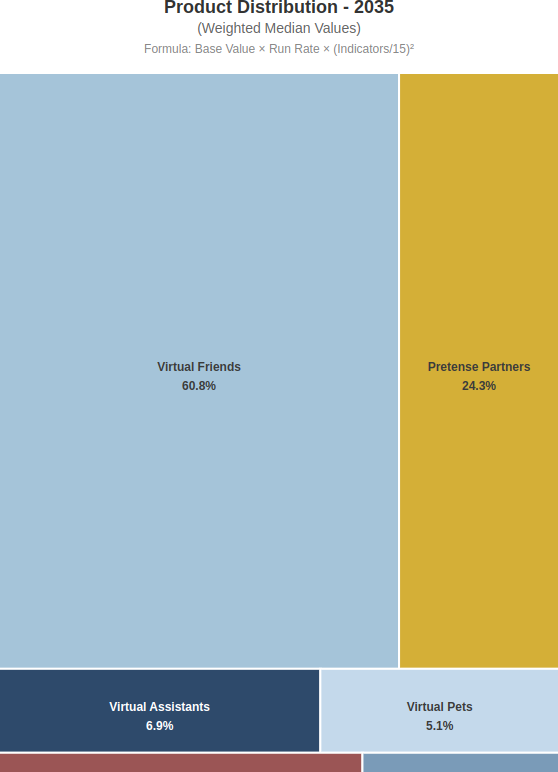}
    \includegraphics[width=0.5\textwidth]{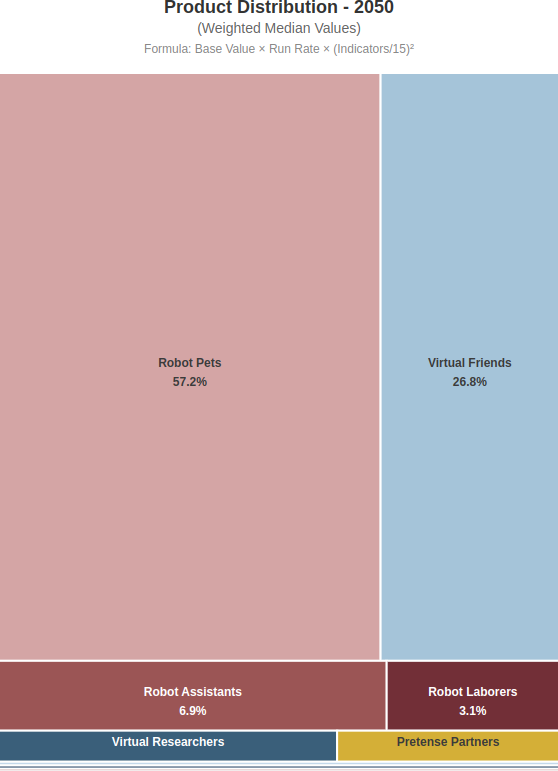}
    \caption{\textbf{Weighted product distributions.} These charts display relative digital mind population sizes at the median of projections in 2035 and 2050, as weighted by mindedness and utilization rates.}
\end{figure}
With these added features incorporated, virtual friends and robot pets
become dominant at the median (employees and researchers would dominate
and the mean). The reason for this is fairly straightforward: while they
may not be as numerous, the assumption is that they are run more
continuously, given that the work they perform isn't limited by direct
human interest or human capacities to respond to it.

\hypertarget{major-cruxes}{%
\subsection{Major Cruxes}\label{major-cruxes}}

The two approaches suggest some major cruxes that will influence the
number of digital minds. It is hard to predict the answers to these
questions and our projections should be strongly influenced by how they
turn out.

\vskip\defaultskip
\hypertarget{what-will-the-social-reaction-to-digital-persons-be}{%
\typetitle{What will the social reaction to digital persons
be?}\label{what-will-the-social-reaction-to-digital-persons-be}}

Many of the applications of digital minds will depend on how ordinary
consumers feel about them. If many people would prefer to engage with
their computers in a manner similar to how they engage with other
people, then minded systems could be much more economically viable.
Systems which could be made not to be person-like could be crafted into
a person-like form. This is true both for systems that have an intended
social purpose (therapists, nannies), and for systems for which that is
unnecessary (assistants, laborers).

The social reaction to AIs may make a difference to the ways we use
them, particularly in their more practical applications. If there is a
negative reaction to AI persons, we might expect that AI labor is
designed to look inhuman, with the more mechanistic and robotic aspects
transparent or even dialed up. We might expect robots to be less
human-like, personalities to be blander, names to be generic products
(think `Siri' or ``Alexa' rather than conveying individual identities).
On the other hand, if there is a positive reaction, then we might see a
world in which many pieces of technology have their own identity and
charismatic personality, perhaps where brand loyalty for home robotics
or operating systems is secured by AI systems unique to individual
consumers and only available from a specific service.

While some consumers have jumped at the opportunity to interact with AI
like other people, others have expressed concern. Many doubt AI is or
will ever be sentient (\href{https://arxiv.org/abs/2506.11945}{Dreksler
et al.~2025}; \href{https://osf.io/preprints/psyarxiv/wvbya_v1}{Ladak \&
Caviola 2025}). It is possible that we will see social or legal
prohibitions around overly-social AI, particularly if AI may disrupt
other areas of our lives. Even if it isn't strictly illegal to build
person-like AIs
(\href{https://www.worldscientific.com/doi/pdf/10.1142/S270507852150003X}{Metzinger
2021}), it may be socially taboo to befriend them, or it might generate
controversy for companies to market such products (Bernardi 2025;
\href{https://link.springer.com/article/10.1007/s11245-025-10247-8}{Lott
\& Hasselbegrer 2025}).

However, there are other reasons to expect a positive reaction to
digital persons. Even if people are resistant at first, social
interactions may have a tendency to change minds, particularly with
systems crafted for positive experiences. The changes can be gradual and
start with kinds of AIs that people find less problematic. As
person-like AIs are introduced into our lives, we might find our
attitudes toward them changing and our interest in engaging with
charismatic AI agents increasing.

\hypertarget{what-portion-of-ai-labor-will-be-performed-by-digital-minds}{%
\typetitle{What portion of AI labor will be performed by digital
minds?}\label{what-portion-of-ai-labor-will-be-performed-by-digital-minds}}

There is a significant prospect that much of human cognitive labor will
be performable by AI systems in the future. Those AI systems might be
more or less person-like and more or less agentic. On the one hand, we
might see a world like that imagined by Robin Hanson in \emph{The Age of
Em} (2016), in which labor is performed by AI persons with minds much
like ours. We have seen AI systems develop into person-like forms in
response to training on human datasets. That may continue to be the best
way to produce digital minds.

On the other, it could turn out that most cognitive labor is performed
by specialized tools that convey little sense of personality or agency.
There is no obvious reason to think that AI should have a cognitive form
similar to ours.

It isn't clear to what extent AI will be good at general decision making
in the coming years. It may be that we are able to train for specific
skills, leading to AIs capable of individual tasks like programming or
scheduling meetings, but not good at prioritization or long-term
strategy. In that case, it may turn out not to be particularly useful to
have full AI replacements for human workers, but rather have them serve
as extremely powerful tools that make human effort go much further. In
such a world, even if some AI systems are responsible for agentic
decisions, they might be restricted to a fairly small number of
overseers that manage the work of legions of specialized tools.

Furthermore, agency can take inhuman forms. Past work on reinforcement
learning models such as
\href{https://deepmind.google/discover/blog/alphastar-mastering-the-real-time-strategy-game-starcraft-ii/}{AlphaStar}
didn't create the same sense of personhood. Or agency might be
approximated with oracle AIs capable of answering questions about
efficacy without direct engagement in the problems they are solving. It
seems possible that most future AI systems won't be agentic at all, or
they won't be agentic in the right ways.

\vskip\defaultskip
\hypertarget{how-will-excess-compute-be-used}{%
\typetitle{How will excess compute be
used?}\label{how-will-excess-compute-be-used}}

We are approaching an era of processors with unprecedented computing
power. Currently, this capability proves most valuable for large-scale
training runs that build increasingly sophisticated AI models, as well
as for deploying those models as customer services. However, our
computational needs may shift dramatically in the coming decades,
potentially leaving us with far more compute than we can meaningfully
use for training runs or standard customer applications.

If we continue advancing processor technology, this abundant compute
will inevitably find new applications---though exactly what those might
be remains unclear, but will greatly influence prospects for digital
minds.

Some potential applications for excess compute include:

\begin{itemize}
\item
\textbf{Research and Development.} As AI systems become capable of
conducting scientific, mathematical, and industrial research, we may
deploy them in unprecedented quantities. Fields like mathematics and
biochemistry could possibly absorb virtually unlimited computational
resources, with no practical ceiling on useful applications.

\item
\textbf{Complex Simulations.} Alternatively, excess compute could power
increasingly sophisticated simulations, modeling everything from climate
systems to molecular interactions to human dynamics or wargames with
extraordinary precision.

\item
\textbf{Enhanced Individual Services.} We might also see compute devoted
to dramatically improving AI services for individual users, even beyond
the point of diminishing returns. While GPT-5 might satisfy most users'
needs, we could end up allocating 10,000 times more computational power
to generate the most perfect Studio Ghibli-style images or provide
deeply thought-through relationship advice. Perhaps AI assistants will
be employed continuously throughout people's daily lives, being ready to
offer a second opinion to every choice.

\item
\textbf{Digital Minds and Virtual Environments.} Some of this compute
might support digital minds operating in their own environments, such as
in very large simulations. It's uncertain how much computational power
would prove genuinely useful in this context.

\end{itemize}
\vskip\defaultskip
\hypertarget{what-is-the-global-middle-class-economic-trajectory}{%
\typetitle{What is the global middle-class economic
trajectory?}\label{what-is-the-global-middle-class-economic-trajectory}}

The projections are based on assumptions about what people will want,
which will depend on how much economic people have to sway the uses to
which AI are put. If AI presents a radical economic change -- if it can
replace white-collar jobs, then we may see significant changes in wealth
distribution. This could be either progressive or regressive, leading to
more or less disposable income. The effects might also not be equally
distributed, so it may be that certain countries capture the bulk of the
profits and use equitably for their own population while foreign
nationals are left to suffer from AI competition for knowledge jobs. Unequal sharing of wealth could also lead to a backlash against AI technologies in general.

The more wealth we see go to a tech-savvy global middle class and the
smaller the portion of the global population that is impoverished, the
more digital minds we should expect. Trends are headed in that direction
and the approaches below assume that those trends will continue. If they
were to radically change, we could see much less focus on AI consumer
products and more effort spent on AI workers or researchers.

\vskip\defaultskip
\hypertarget{approach-1-use-based-estimates-under-consumption-models}{%
\subsection{Approach 1: Use-based Estimates under Consumption
Models}\label{approach-1-use-based-estimates-under-consumption-models}}

\hypertarget{framework}{%
\subsubsection{Framework}\label{framework}}

The first approach to estimating digital mind numbers involves delving
into the ways in which digital minds will be used. Digital minds must be
deliberately built; surely they will be built to serve particular ends.
Understanding these ends may inform us about the quantities we will
choose to build. We will face trade-offs about how to allocate
resources: it is only worthwhile to build digital minds to the extent
that the value we get from them outweighs the resources we put into
them.

Although it is hard to say how society, the economy, and the potential
trade-offs we will face will develop over the coming decades, the ends
to which digital minds could be put appear to be moderately predictable.
The first approach therefore focuses on these ends to calibrate our
estimates.

This approach surveys existing and potential markets for a variety of AI
products and discusses reasons for thinking that some of the products
filling these niches would be digital minds. It concludes by bringing
together each of the estimates of specific niches into a comprehensive
picture. These kinds differ in their plausibility, and we have reason to
expect some kinds will be largely irrelevant. However, we attempt to be
reasonably comprehensive in order to not overlook significant numbers.

The primary use-cases for digital minds described here are distinguished
into three super-categories. These super-categories encompass social
applications, task applications, and actor applications. The items in
each super-category may not be exhaustive, but we hope that they
encompass the bulk of potential applications.

These use-cases are also not completely independent. It may be that some
users are satisfied by individual systems: perhaps our AI therapist will
also be our secretaries, friends, and lovers. The models below assume
them to be separate -- this is a significant limitation, but ultimately
seems unlikely to radically alter the results in light of the very
different scales of the predictions.

In each case, we will also provide estimates of the yearly numbers based
on a Consumer Model (\hyperref[app:1]{Appendix A}). This model includes parameters regarding the potential customer
base and adoption rates, allowing for yearly projections of the numbers
in each category. The numbers focus on the use as intended. We ignore
digital minds that might be created but never really used.\footnote{Consider
  how consumers might try out talking to ten different AI personas
  before they settle on one to be their therapist. Or a military might
  create a million drones without ever turning the vast majority of them
  on. Such numbers could inflate our estimates significantly and put
  more weight on estimating the utilization rates. They are harder to
  predict and invite different sorts of social, political, and ethical
  considerations, so we have chosen to ignore them.}

We will also estimate degree of mindedness, in terms of the
exemplification of the qualifying traits, and the average utilization
rate, in terms of active time over a year. For many systems, we expect
that they will be able to do what they do without being continuously
active. For others, we might expect that they will be in constant use.

After presenting each use case, we will aggregate the results and look
for the categories that are most significant.

\hypertarget{companion-robots-and-virtual-entities}{%
\subsubsection{Companion Robots and Virtual
Entities}\label{companion-robots-and-virtual-entities}}

This section breaks down the varieties of digital minds we might see
that play a primarily social role. Many of these cases are divided into
pairs differing on the substrate in which the mind resides: there may be
significant differences in the uses, markets, and digital mind candidacy
for minds in robotic bodies and minds that exist purely in a virtual
form.

\containedsubsection
\hypertarget{virtual-friends}{%
\typetitle{Virtual Friends}\label{virtual-friends}}

\emph{Definition:} Virtual friends are digital entities that are
accessed through text, voice, or multimedia interfaces and that provide
companionship, emotional support, and social interaction similar to
human friendship. Unlike customer service chatbots or task-oriented AI
assistants, these systems are designed to form ongoing relationships,
remember personal details and shared experiences, provide emotional
support during difficult times, celebrate achievements, and engage in
the kind of casual, meandering conversations that characterize human
friendship. They would need to maintain consistent personalities while
learning and adapting to individual users' communication styles,
interests, and emotional needs.

\emph{Examples:} Current examples include
\href{http://Character.AI}{Character.AI's} personality-based chatbots
and \href{https://replika.ai/}{Replika's AI} companion service. These
platforms claim to have attracted millions of users who engage in deep,
ongoing conversations with AI personalities. However, most existing
systems lack the sophisticated memory, emotional intelligence, and
behavioral consistency that would characterize true digital mind
friendships. They often exhibit repetitive patterns, fail to maintain
long-term relationship continuity, or break character in ways that
remind users of their artificial nature.

\emph{Digital Mind Candidacy:} Virtual friends have a reasonably strong
claim to mindedness, arguably stronger than many other use cases.
Genuine friendship requires stable personality traits that users can
rely on over time and the ability to navigate complex social and
emotional dynamics. Intelligence and mental flexibility are essential
for maintaining engaging dialogue, understanding nuanced emotional
states, and providing meaningful support across diverse life situations.
We may expect that our virtual friends engage with the world in other
ways to enhance our interaction opportunities.

\begin{longtable}[]{l|l}
Consistent and autonomous goals, desires, and interests & High \\
Stable idiosyncratic personality & High \\
Perception, interaction, and navigation & Moderate \\
Learn, grow, and evolve & High \\
Intelligence and flexibility & High \\
\hline\noalign{}
Mindedness Score & \textbf{0.87} \\
\end{longtable}

\emph{Unit Operation Rate:} 6 hours per week

Virtual friends might smooth over some of the challenges we face in
maintaining social friendships, but most people
\href{https://www.happiness.hks.harvard.edu/february-2025-issue/the-friendship-recession-the-lost-art-of-connecting}{do
not spend a large amount of time interacting with friends}. There is no
obvious reason for chatbot friends to spend more time active than they
spend interacting with humans, or to spend much more subjective time in
an interaction than the people they are interacting with. Entertainment
might generally improve, providing other demands on our time. Of course,
if it is sufficiently cheap to spend more time active, or to overthink
each conversational contribution, then the operation rate might be well
above this estimate.

\newcommand{\projectioncaption}[2]{\caption{\textbf{#1 Projections}. This chart displays a breakdown of the estimated time of first viable products (blue columns) and mean (solid red line) and median (dotted red line) projections for #2 numbers. The shaded area represents the middle 95th percent of all estimates.}}

\emph{Prediction:}
\href{https://github.com/rethinkpriorities/digital_minds_consumer_models/blob/main/virtual_friends/run.py}{Parameters}

\newcommand{\daterow}[4]{\textbf{#1} & #2 & #3 & \parbox{1cm}{#4}}
\begin{figure}
\includegraphics[width=1\textwidth]{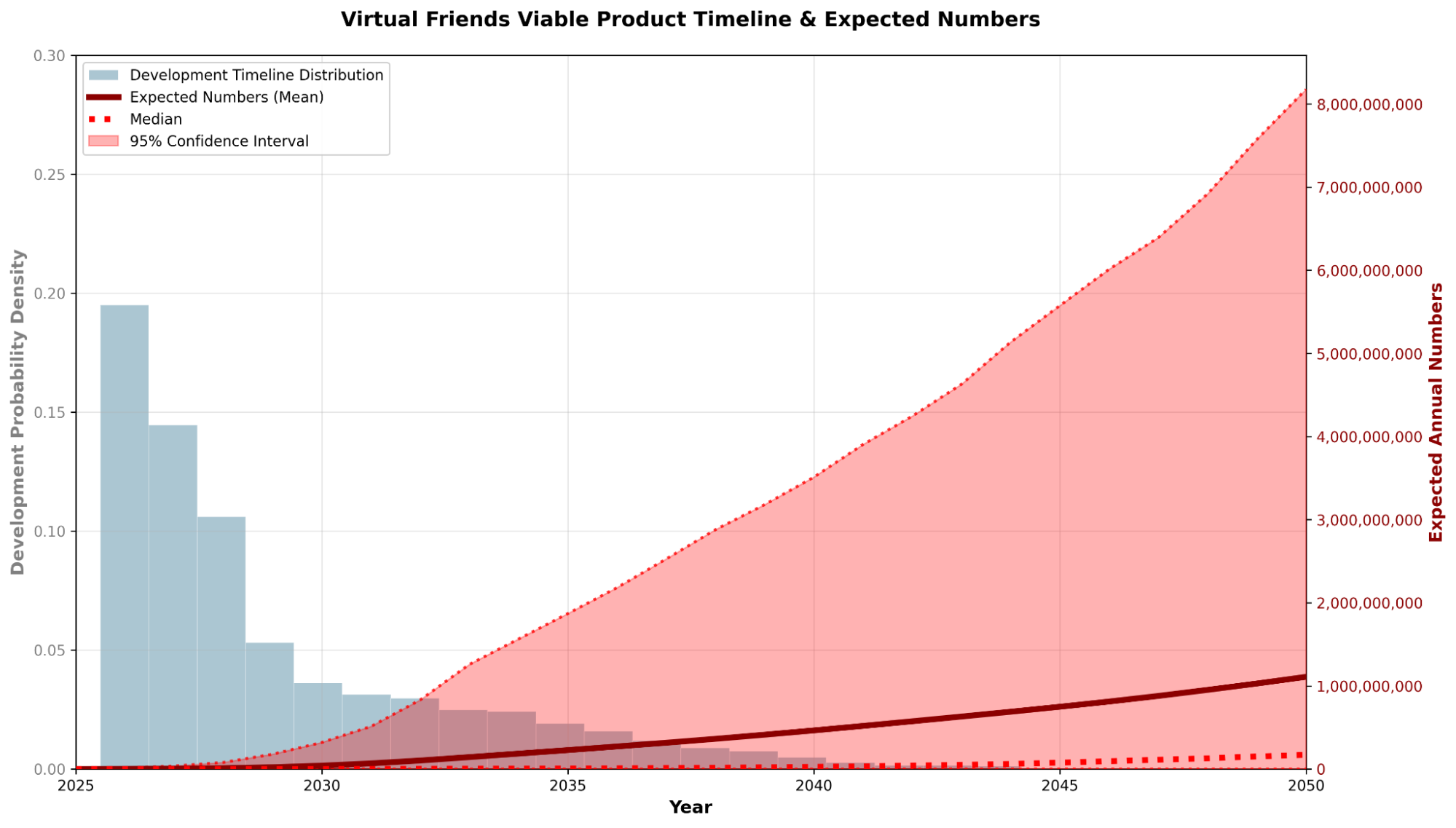}
\centering
\begin{tabularx}{0.8\textwidth}{XXXr}
\toprule
\daterow{}{Median}{Mean}{High}\\
\midrule
\daterow{2030}{4.37e05}{4.05e07}{3.20e08}\\
\daterow{2035}{6.62e06}{2.28e08}{1.87e09}\\
\daterow{2040}{2.67e07}{4.65e08}{3.51e09}\\
\daterow{2045}{7.67e07}{7.49e08}{5.58e09}\\
\daterow{2050}{1.70e08}{1.11e09}{8.18e09}\\
\bottomrule
\end{tabularx}
\projectioncaption{Virtual Friends}{virtual friends}
\end{figure}

Virtual friends are among the most plausible forms of person-like AI
that we can confidently expect to see in the coming years. They require
no clear technological advancements (today's LLMs seem in principle
capable of this role) and there are large companies already working
toward building them. There is a clear current demand for them, even if
it is still somewhat niche. We haven't seen a true competitive virtual
friend product. Leading AI companies may have the ability to tilt their
leading models in that direction, but don't seem to be putting their
efforts into making their chatbots into effective friends. As models get
cheaper, there is no reason not to expect an independent company to put
out an effective service.

\containedsubsection
\hypertarget{virtual-guides}{%
\typetitle{Virtual Guides}\label{virtual-guides}}

\emph{Definition:} Virtual guides provide personalized instruction and
support across various domains of learning and development. Unlike
traditional software tutoring programs, these systems would maintain
ongoing relationships with their pupils, adapt their teaching styles to
individual needs, and provide encouragement and emotional support
alongside instruction. (Services playing a virtual guide role without
any suitable claim to mindedness would not count toward this category.)
They encompass academic tutors, life coaches for personal development,
skill-specific trainers for professional development, and therapeutic
agents for mental health support. These systems would be distinct from
contemporary uses of large language models in coaching from an ongoing
persona and not simply providing ad hoc advice.

\emph{Examples:} Current AI tutoring systems like Khan Academy's
\href{https://www.khanmigo.ai/}{Khanmigo}, Duolingo's conversation
practice bot \href{https://blog.duolingo.com/video-call/}{Lily}, or
therapeutic chatbots like
\href{https://woebothealth.com/technology-overview/}{Woebot} represent
early predecessors, but lack the depth of personality and
relationship-building capacity that might characterize future digital
minds in these roles. Future systems might maintain detailed models of
individual learning styles, emotional states, and long-term progress
while developing genuine rapport with their students or clients. Social
engagement could turn out to be valuable, but it is less obvious that AI
systems would benefit from individuality or their own vibrant lives.

\emph{Digital Mind Candidacy:} Virtual guides have a moderate claim to
digital mind status. Effective guidance may benefit from having a stable
personable character that students can rely on, consistent goals focused
on learner development, and the ability to navigate complex social and
emotional dynamics. However, they are also likely to work in a fairly
constrained way, may not exhibit any possibility of autonomy or growth,
and have domain-limited intelligence (it is unclear that individual
guides will need to exert their own intelligence dynamically, as opposed
to relying on external software or specific resources toward that end.)

\begin{longtable}[]{l|l}
Consistent and autonomous goals, desires, and interests & Moderate \\
Stable idiosyncratic personality & Moderate \\
Perception, interaction, and navigation & Low \\
Learn, grow, and evolve & Low \\
Intelligence and flexibility & Moderate \\
\hline
\endhead
Mindedness Score & \textbf{0.28} \\
\endlastfoot
\end{longtable}

\emph{Unit Operation Rate:} 2 hours per week

We currently see most of the human occupations that play guide-like
roles no more than once or twice a week for a few hours. This is
probably limited by the expense of human guidance, but also reflects the
limits of our ability to focus on any one thing. Certain kinds of guides
might be fairly constant parts of our lives, particularly for students
who have dedicated teachers or people in need of frequent assistance of
other forms. On average, however, we shouldn't expect guidance to take
up a large amount of anyone's week.

\emph{Prediction:}
\href{https://github.com/rethinkpriorities/digital_minds_consumer_models/blob/main/virtual_guides/run.py}{Parameters}

\begin{figure}
\includegraphics[width=1\textwidth]{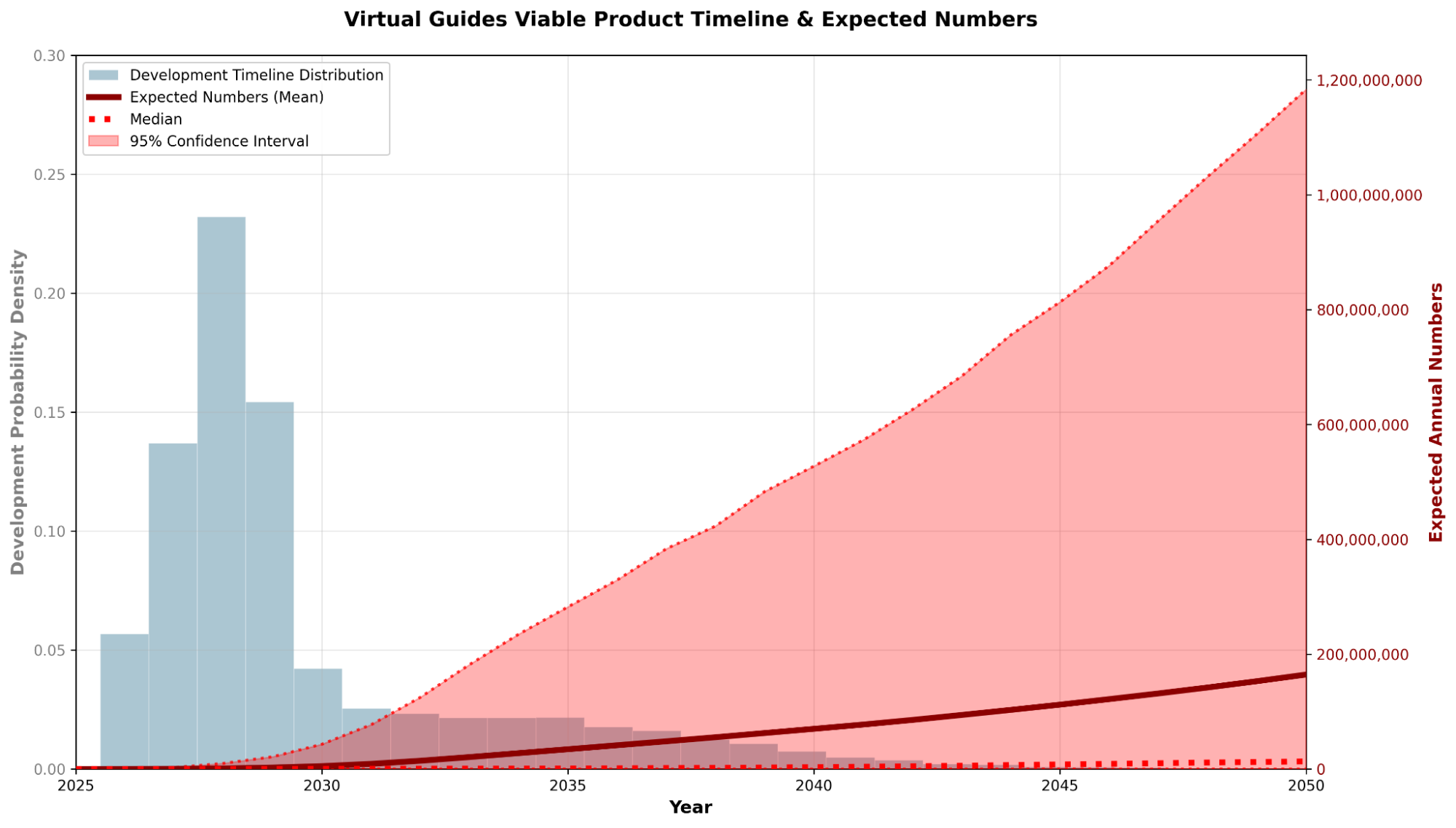}
\centering
\begin{tabularx}{0.8\textwidth}{XXXr}
\toprule
\daterow{}{Median}{Mean}{High}\\
\midrule
\daterow{2030}{2.95e06}{6.05e07}{5.43e08}\\
\daterow{2035}{2.04e08}{3.88e08}{1.88e09}\\
\daterow{2040}{4.31e08}{6.72e08}{2.91e09}\\
\daterow{2045}{6.01e08}{9.67e08}{4.19e09}\\
\daterow{2050}{8.08e08}{1.36e09}{6.11e09}\\
\bottomrule
\end{tabularx}
\projectioncaption{Virtual Guides}{virtual guides}
\end{figure}

AI models will almost certainly be employed at scale in educational and
therapeutic roles in the coming years. The main outstanding question
concerns the extent to which they will be person-like, modestly agentic
formats, rather than sophisticated conversational tools. There are
abstract considerations favoring at least some of them to be like this:
personalities may contribute to engagement and interest. However, there
remains significant doubt about the popularity of such features, and
hence the prevalence of virtual guides with digital minds.

\containedsubsection
\hypertarget{virtual-pets}{%
\typetitle{Virtual Pets}\label{virtual-pets}}

\emph{Definition:} Pets are social companions characterized by an
asymmetric role in which we provide for their needs and receive
emotional engagement in response. They are suitable objects for our
affection. They are not able or interested in engaging with us as equals
and we can expect them to spend their time primarily with us. Virtual
pets exist only\footnote{It is possible that some robots will also be
  able to live on a cloud, so you can interact with them even when
  travelling. These will count as `robots' for the purposes of this
  breakdown.} within a virtual environment and only interact directly
with users through a dedicated interface. They would not have robot
bodies or live in our homes. Instead, we might check in on them
periodically through smart phones or other devices.

\emph{Examples:} Virtual pets have existed in various formats for
decades. The \href{https://en.wikipedia.org/wiki/Tamagotchi}{Tamagotchi}
and its competitors have produced barebones virtual pets in minimalistic
environments, selling 100s of millions of units. Computer games like
\href{https://en.wikipedia.org/wiki/Neopets}{Neopets} and the
\href{https://en.wikipedia.org/wiki/Petz}{Petz} series have incorporated
modestly more complex environments and found tens of millions of users.
Numerous alternative products exist and continue to be released. They
have also seen consistent product launches over time. They have also not
aimed to fill the roles occupied by pets for most people, offering
instead a minimal pet-light experience.They have not been able to (or
have not been intended to) provide the sense of real minds, though
sufficient expenditure or novel approaches might make this more
feasible.

\emph{Digital Mind Candidacy:} Virtual pets have a modest claim to being
digital minds, primarily because their digital environment offers less
robust opportunities for complex behavior and interactions. Future
virtual pets may be hooked up to powerful AI to deliver a more life-like
experience, but producers and consumers may continue to opt to treat
these products primarily as games aimed to briefly entertain children,
who are happy to engage in pretense, quick to move on, and less
responsive to indicators of a vibrant individual.

\begin{longtable}[]{l|l}
Consistent and autonomous goals, desires, and interests & High \\
Stable idiosyncratic personality & High \\
Perception, interaction, and navigation & Low \\
Learn, grow, and evolve & Moderate \\
Intelligence and flexibility & Moderate \\
\hline
\endhead
Mindedness Score & \textbf{0.54} \\
\endlastfoot
\end{longtable}

\emph{Unit Operation Rate:} 5-100 hours per year.

Given the many demands on our attention, it is hard to see users
interacting with cognitively simple systems through a dashboard
consistently for months or years even if they have highly engaging
personalities. We might expect them to have a place in our lives similar
to computer games, which consumers
\href{https://www.sciencedirect.com/science/article/abs/pii/S1875952118300181}{often
don't finish} and instead use for perhaps a few hours (or a few dozen hours) before
moving on. In the future, virtual pets may be more engaging, or might be
able to be with us more regularly through better integrated technology,
but also may need to compete with more engaging alternatives. For every
unit solid in a given year, we may therefore expect it to be active for
5-100 hours.

We should expect attention to them to be limited to occasional sessions,
making it unnecessary to keep them running all the time. Of course,
there could be benefits to allowing them to live fuller lives when we
don't pay attention to them, but that would bring the relationship
closer to a simulant, which is handled separately in this report.

\emph{Prediction}:
\href{https://github.com/rethinkpriorities/digital_minds_consumer_models/blob/main/virtual_pets/run.py}{Parameters}

\begin{figure}
\includegraphics[width=1\textwidth]{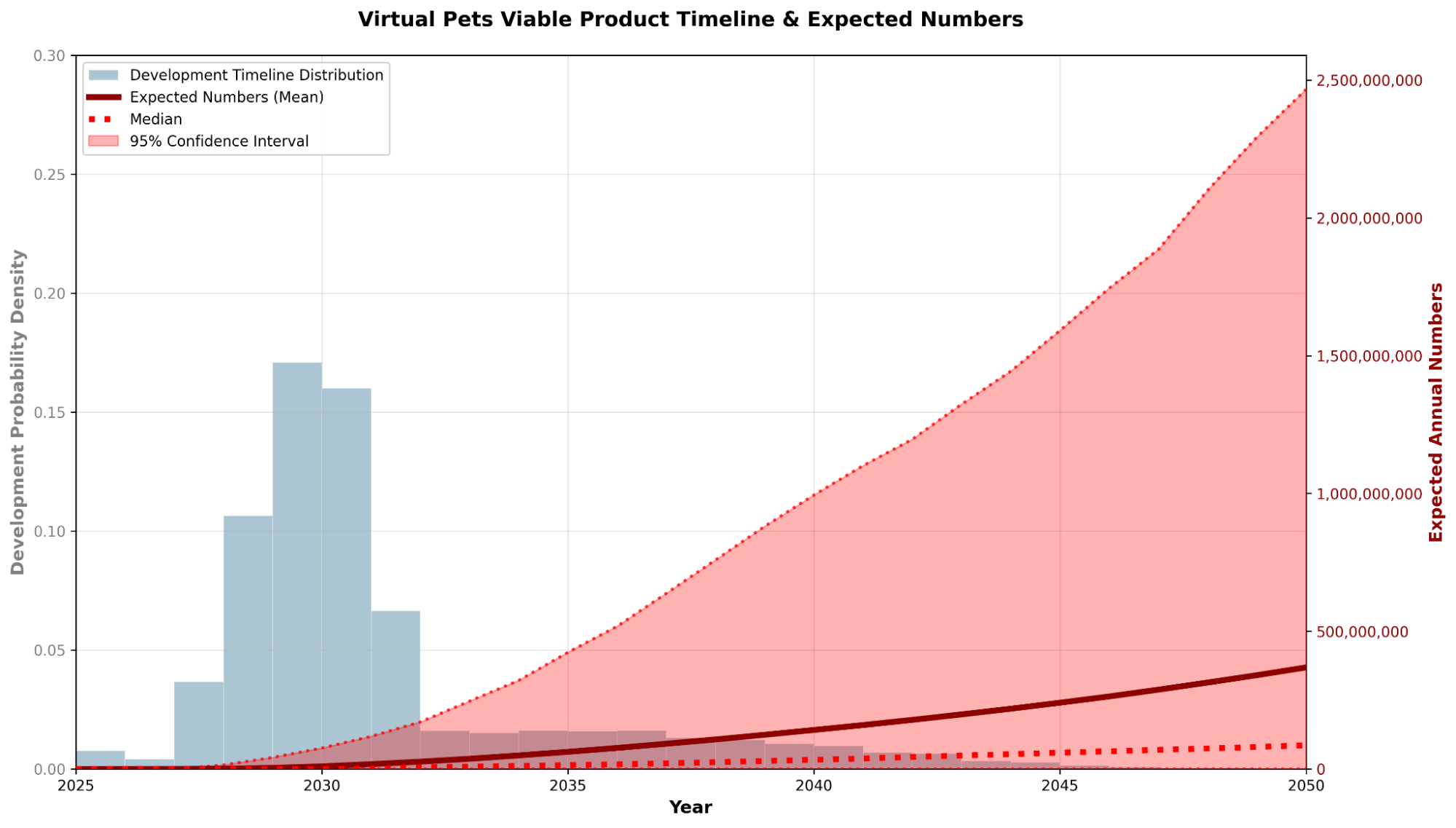}
\centering
\begin{tabularx}{0.8\textwidth}{XXXr}
\toprule
\daterow{}{Median}{Mean}{High}\\
\midrule
\daterow{2030}{0}{9.67e06}{7.61e07}\\
\daterow{2035}{1.41e07}{6.24e07}{4.23e08}\\
\daterow{2040}{3.35e07}{1.42e08}{9.94e08}\\
\daterow{2045}{5.93e07}{2.41e08}{1.59e09}\\
\daterow{2050}{8.64e07}{3.69e08}{2.47e09}\\
\bottomrule
\end{tabularx}
\projectioncaption{Virtual Pets}{virtual pets}
\end{figure}

The numbers in this model reflect expectations about the number of units
(software licenses) sold in a year, and most people do not engage with a
unit for much beyond that, so that at any time the extant population of
virtual pets is tied to the units recently sold. There isn't a clear
reason to expect virtual pets to occupy a significantly increasing
fraction over time, though we might expect to see them occupy the role
of pets, the modes of engagement that are possible for them are
significantly more constrained than for robot pets.

\containedsubsection
\hypertarget{robot-friends}{%
\typetitle{Robot Friends}\label{robot-friends}}

\emph{Definition}: Friends are social companions with whom we interact
as (approximate) equals, and who provide us with conversation, emotional
support, personalized advice, different perspectives, etc. We may crack
jokes or play games with them, watch movies together or gossip about
celebrities. Share factoids or complain about politics.

Human friends differ from pets in being more intelligent, having their
own agenda and generally not being overtly sycophantic. There is some
leeway in treating each of these qualities as requirements, but they
also collectively contribute to a more equitable relationship that
serves to distinguish this relationship from pets.

Robot friends play this role in a physical form, living in our homes and
offices. They may or may not have lives outside of interactions with us,
but could operate somewhat autonomously so as to be able to maintain
some independence. While we might expect artificial friends to take on
either robotic or virtual forms, there are several reasons to expect
more in the form of virtual friends, and for that reason the concept of
friendbot is less likely to see a specific market. (Though, we should
not rule out the possibility of hybrid friendbots, that exist in a
virtual form and can access robotic bodies at times.)

\emph{Examples:} There are no clear examples of this category on the
market today, though there is compelling interest in related products,
such as more cognitively sophisticated robots as companions for kids.
Anki's \href{https://anki.bot/products/vector-robot}{Vector} and Pollen
Robotic's \href{https://www.pollen-robotics.com/reachy-mini/}{Reachy}
provides potential early examples in this category.
\href{https://www.realbotix.com/}{Realbotix} has developed more
human-like bodies. Until the rise of LLMs, no AI has been able to power
the intelligence required for an equitable relationship. Adequate
robotics presents another challenge, though recent investments in
personal robotics are extraordinary.

\emph{Digital Mind Candidacy:} Robot friends have a similar candidacy as
pet robots, albeit with higher intelligence and a greater possible
degree of agency. Like robot pets, it might also be a potential selling
point (or legal hurdle) for them to display genuine mental capacities,
like emotions, beliefs and desires, and possibly consciousness.

\begin{longtable}[]{l|l}
Consistent and autonomous goals, desires, and interests & High \\
Stable idiosyncratic personality & High \\
Perception, interaction, and navigation & High \\
Learn, grow, and evolve & High \\
Intelligence and flexibility & High \\
\hline
\endhead
Mindedness Score & \textbf{1} \\
\endlastfoot
\end{longtable}

\emph{Prediction:} The advantages of physical bodies are less clear for
entities primarily designed to be friends. We typically interact with
people through language, and can do so through phones, computers, or
smart speakers. There are certain kinds of activities that we do with
our friends that require bodies (e.g.~squash), but robotics are far from
supporting the complex and diverse behaviors people might want to make
it worthwhile. Instead, robot friends seem to be hard to distinguish
from robot pets, or are made more viable in combination with other uses:
e.g.~a robot maid who doubles as a friend. This wrinkle makes it more
tempting to expect that interest in service capabilities would be a
better predictor of growth rates than any markers of pure friendship. As
such, it seems reasonable to expect there to be a negligible number of
pure robot friends, and to incorporate predictions of that category in
with robot pets.

\containedsubsection
\hypertarget{robot-pets}{%
\typetitle{Robot Pets}\label{robot-pets}}

\emph{Definition:} Pets are social companions characterized by an
asymmetric role in which we provide for their needs and receive
emotional engagement in response. They are suitable objects for our
affection. They are not able or interested in engaging with us as equals
and we can expect them to spend their time primarily with us. Our homes
are the boundaries of their world. Pet robots are embodied systems in a
robot body with whom we can interact with as we do the biological pets
in our homes. They may move about the space autonomously, and engage
with us as we go about our lives, and pursue their own limited ends.

\emph{Examples:} The boundary between toy and artificial pet is vague.
Early examples of toys that were steps in the direction of pet robots
include the \href{https://en.wikipedia.org/wiki/Furby}{Furby} and the
Sony \href{https://en.wikipedia.org/wiki/AIBO}{AIBO}. In the last
decade, a number of other products have attempted to fill this niche,
including the Anki
\href{https://web.archive.org/web/20250710054205/https://anki.bot/products/vector-robot}{Vector}
and the
\href{https://web.archive.org/web/20250707185541/https://us.keyirobot.com/pages/loonadetail}{Loona}
petbot. In general, these products have lacked autonomy and dynamic
flexibility. There is no case for thinking that they are conscious or
sentient and their behaviors are generally predictable, repetitive, or
robotic. None of these products has been built with cutting edge AI
technology (though Loona incorporates an LLM as an accessory). But
\href{https://venturebeat.com/ai/robotics-startup-anki-shuts-down-after-burning-through-almost-200-million/}{past}
\href{https://www.prnewswire.com/news-releases/chinese-consumer-robotics-company-keyi-tech-raises-tens-of-millions-of-dollars-in-series-b-financing-301344958.html}{substantial}
investments suggest that venture capitalists see promise in this market,
and this could change with the introduction of more substantial AI
technologies.

\emph{Digital Mind Candidacy:} Due to the cognitive limitations and the
nature of our relationships, interactions with our pet are not primarily
conversational. In the absence of language-based interactions, they
benefit from physical engagement in the real world. This requires
embodiment and some amount of agency. Living in and navigating a home
and responding to the owner's behaviors will require that pet robots
have perceptual systems and substantial control over their bodily
movements. Playing a largely social role, we should also expect them to
have a stable personality over time, along with learning and memory such
that our interactions with them can shape their relationship. Finally,
we may expect people to be more attached to robots that seem more
person-like, including whatever behavioral tendencies people most
associate with consciousness. The case for being digital minds is
therefore quite high.

\begin{longtable}[]{l|l}
Consistent and autonomous goals, desires, and interests & High \\
Stable idiosyncratic personality & High \\
Perception, interaction, and navigation & High \\
Learn, grow, and evolve & High \\
Intelligence and flexibility & Moderate \\
\hline
\endhead
Mindedness Score & \textbf{0.87} \\
\endlastfoot
\end{longtable}

\emph{Unit Operation Rate:} 3 hours per day.

Given that we interact with pets only part of the day, we may not expect
robot pets to be active all day every day. Instead, we should expect
them to be present enough to engage with us, but not enough to risk
being annoying or a nuisance. Based on patterns of engagement with
pets,\footnote{Dogs spend about
  \href{https://www.akc.org/expert-advice/health/why-do-dogs-sleep-so-much/}{20\%
  of the day being active}.} we might expect robot pets to be actively
engaged about three out of every 24 hours, at a rate of computation that
is one-to-one subjective time. Of course, they might spend the rest of
the time effectively comatose, or in an unengaged but aware state. For
the purposes of utililization, we assume that the remainder of their day
is spent paused.

\emph{Prediction}:
\href{https://github.com/rethinkpriorities/digital_minds_consumer_models/blob/main/robot_pets/run.py}{Parameters}

\begin{figure}
\includegraphics[width=1\textwidth]{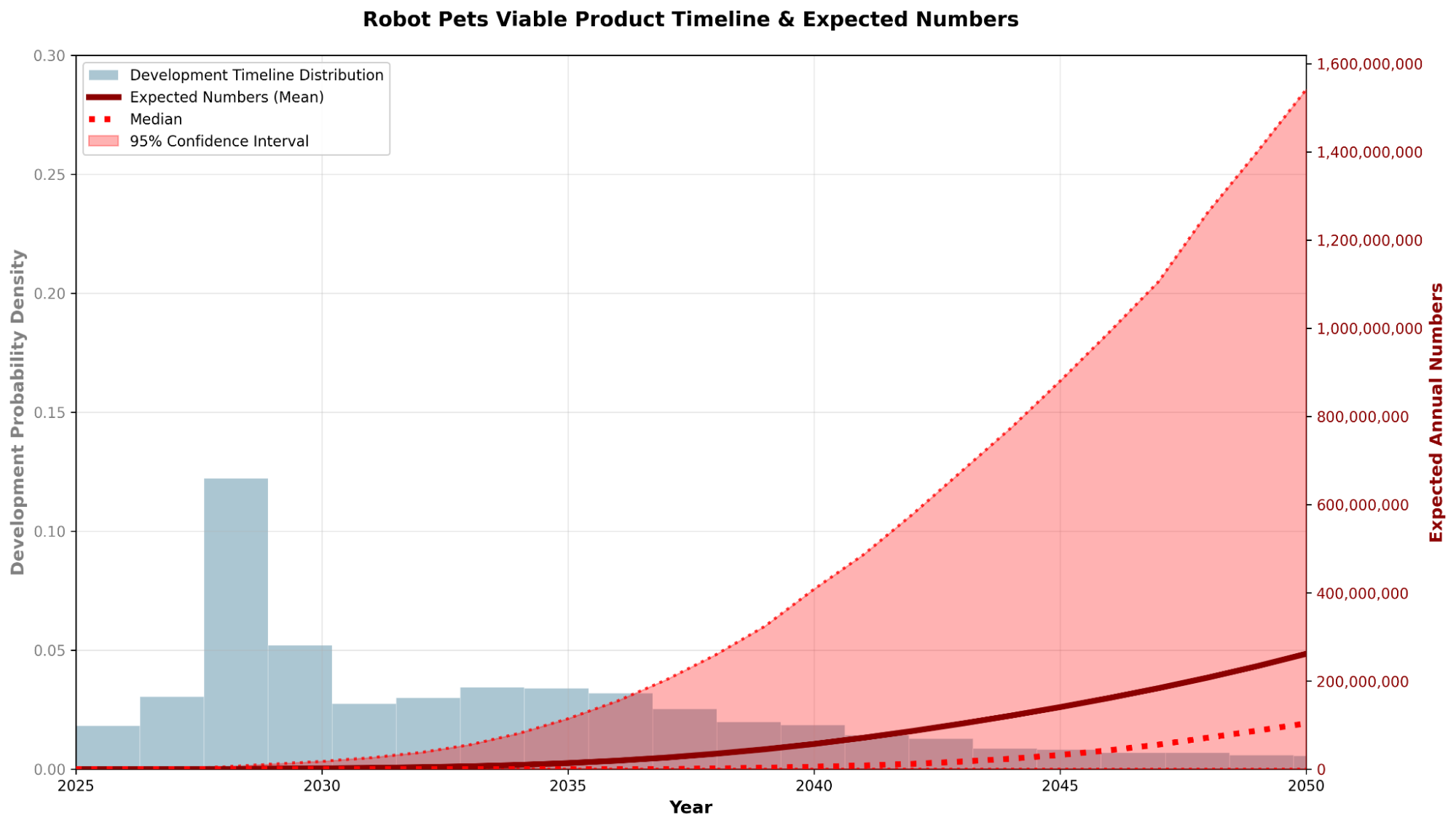}
\centering
\begin{tabularx}{0.8\textwidth}{XXXr}
\toprule
\daterow{}{Median}{Mean}{High}\\
\midrule
\daterow{2030}{0}{2.01e06}{1.77e07}\\
\daterow{2035}{0}{1.39e07}{1.15e08}\\
\daterow{2040}{5.64e06}{5.73e07}{4.08e08}\\
\daterow{2045}{3.23e07}{1.41e08}{8.80e08}\\
\daterow{2050}{1.04e08}{2.62e08}{1.54e09}\\
\bottomrule
\end{tabularx}
\projectioncaption{Robot Pets}{robot pets}
\end{figure}

Assuming that interest in pet robots will occupy some of the consumer
desire for biological pets, we may expect a growing fraction of would-be
pet owners to choose to instead adopt pet robots once a viable product
comes on the market in the next 10 or so years. The fraction would
likely start off quite small and be constrained primarily to tech
enthusiasts, but with the right products might grow substantially over
time. The extreme ends of this prediction are occupied by systems in
which no viable product is ever marketed (and products in this space
remain toys) and where a significant portion of new pet ownership is
occupied by pet robots.

\containedsubsection
\hypertarget{robot-lovers}{%
\typetitle{Robot Lovers}\label{robot-lovers}}

\emph{Definition:} AI lovers and sexual robots are AI systems designed
to fulfill romantic and sexual needs through intimate interaction with
humans. This category encompasses a spectrum from systems purely focused
on physical gratification (but with non-negligible cognitive abilities)
to sophisticated romantic partners capable of deep emotional connection,
courtship, and long-term relationship dynamics. The line between romance
and friendship can be somewhat blurred, so systems within this category
will be assumed to focus more on relationships that involve sexual
gratification (though not necessarily physically) rather than just basic
emotional support. The more sophisticated systems would maintain
consistent personalities, express and respond to emotions, engage in
intimate conversations, and adapt to their partner's preferences and
needs over time. They may be embodied in human-like robotic forms or
exist as digital entities accessed through various interfaces,
potentially including virtual or augmented reality environments.

\emph{Examples:} Current examples of sexbots include basic sex robots
like \href{https://www.realdoll.com/realdoll-robot/}{RealDoll's}
AI-enabled models, which combine physical bodies with rudimentary
conversational abilities. In the digital realm, Replika has attracted
millions of users who form intimate relationships with their AI
companions, with many reporting deep emotional connections.
\href{http://Crushon.AI}{Crushon.AI} provides erotic roleplay. Virtual
reality applications have begun exploring intimate experiences with AI
characters, though most remain focused on fantasy rather than
relationship development. However, existing systems lack the emotional
sophistication, robust physical capabilities, and behavioral consistency
that would characterize truly mind-like AI lovers.

\emph{Digital Mind Candidacy:} AI lovers present a complex case for
digital mind status. For physical intimacy, embodied systems would need
basic perceptual abilities to understand and respond to human emotions,
physical cues, and changing preferences. However, they wouldn't
necessarily need to be able to maneuver complex unfamiliar environments
or make use of arbitrary tools. Romantic relationships require stable
personality traits, consistent emotional responses, and the ability to
form deep, lasting bonds, but sexual gratification might rely less on
stability and more on the ability to play a role over the course of a
scene.

\begin{longtable}[]{l|l}
Consistent and autonomous goals, desires, and interests & Low \\
Stable idiosyncratic personality & Moderate \\
Perception, interaction, and navigation & Moderate \\
Learn, grow, and evolve & Low \\
Intelligence and flexibility & Moderate \\
\hline
\endhead
Mindedness Score & \textbf{0.28} \\
\endlastfoot
\end{longtable}

\emph{Unit Operation Rate:} 2 hours per week.

\emph{Prediction:}
\href{https://github.com/rethinkpriorities/digital_minds_consumer_models/blob/main/robot_lovers/run.py}{Parameters}

\begin{figure}
\includegraphics[width=1\textwidth]{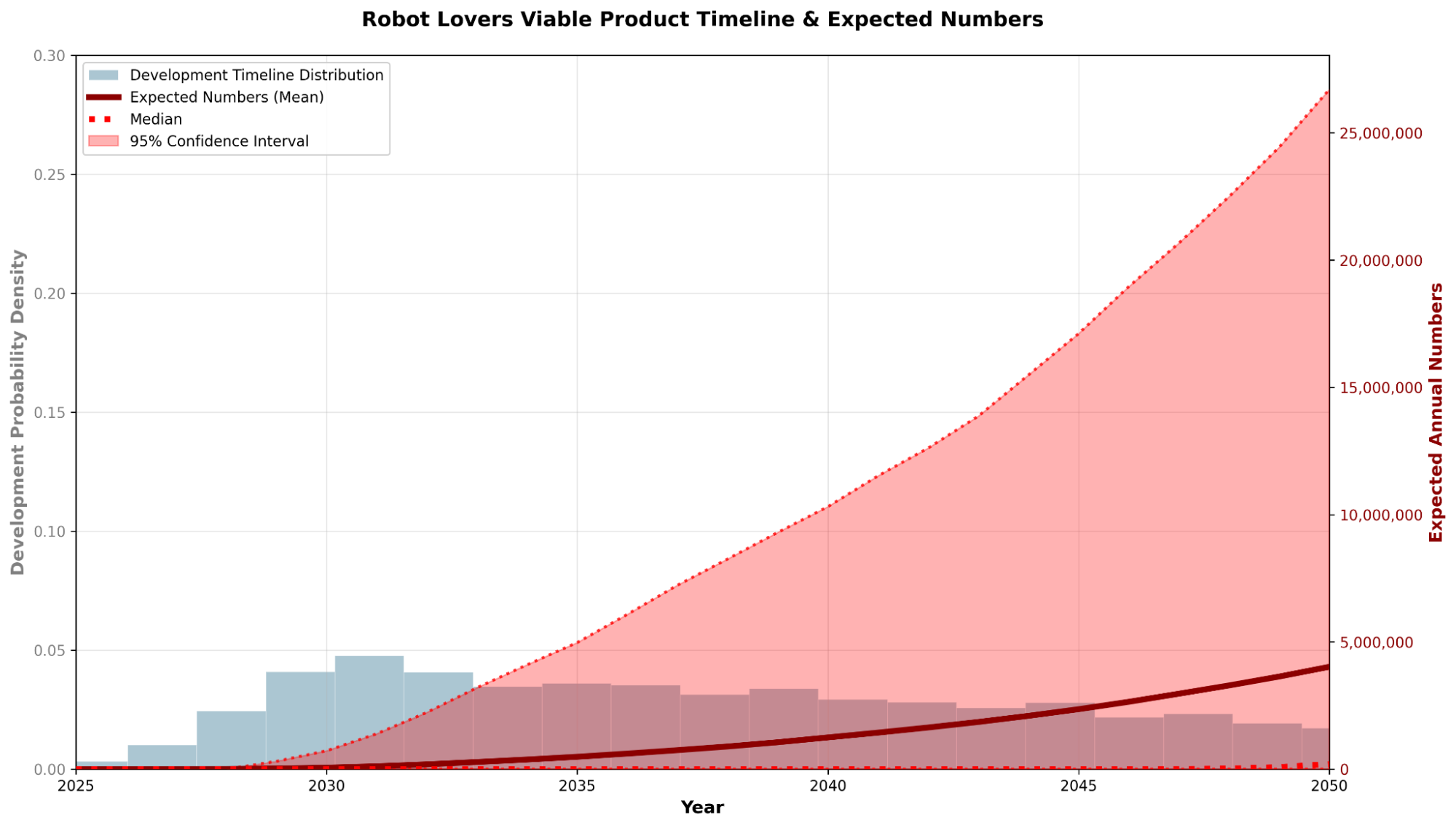}
\centering
\begin{tabularx}{0.8\textwidth}{XXXr}
\toprule
\daterow{}{Median}{Mean}{High}\\
\midrule
\daterow{2030}{0}{6.08e04}{7.24e05}\\
\daterow{2035}{0}{4.89e05}{4.97e06}\\
\daterow{2040}{0}{1.25e06}{1.03e07}\\
\daterow{2045}{0}{2.36e06}{1.71e07}\\
\daterow{2050}{2.01e05}{4.02e06}{2.67e07}\\
\bottomrule
\end{tabularx}
\projectioncaption{Robot Lovers}{robot lovers}
\end{figure}

There is a big audience for this product: people who have difficulty
finding human partners. But it will likely remain somewhat socially
taboo, and the majority of the value might be found in more traditional
chatbots. Nevertheless, the existence of early entrants into this market
suggest that we will see at least modest numbers in the coming decades.

\emph{}

\containedsubsection
\hypertarget{robot-nannies}{%
\typetitle{Robot Nannies}\label{robot-nannies}}

\emph{Definition:} Robot nannies are embodied AI systems designed to
provide childcare services, combining supervisory responsibilities with
nurturing and educational functions. They would monitor children's
safety, engage in play and learning activities, encourage proper
psychological development, assist with daily routines like meals and
bedtime, and provide emotional support. Unlike purely educational AI
systems, robot nannies must handle the full spectrum of childcare
responsibilities, from emergency response to social-emotional
development. They operate in physical spaces, typically homes, and must
navigate complex family dynamics while maintaining appropriate
boundaries with both children and parents.

\emph{Examples:} Current examples remain largely conceptual or in early
development stages. Companies like Moxie Robotics have created
\href{https://moxierobot.com/products/ai-robot}{social robots for
children} that focus on learning and play, while others like
\href{https://elliq.com/pages/caregivers?utm_source=google\&utm_medium=cpc\&utm_campaign=22792975638\&utm_campaign=\%7Bcampaign\%7D\&keyword=elliq\&matchtype=e\&network=g\&device=c\&utm_term=elliq\&utm_campaign=TA_Brand_All_US\&utm_source=adwords\&utm_medium=ppc\&hsa_acc=5873203615\&hsa_cam=22792975638\&hsa_grp=186116216281\&hsa_ad=765866126106\&hsa_src=g\&hsa_tgt=kwd-1909896026773\&hsa_kw=elliq\&hsa_mt=e\&hsa_net=adwords\&hsa_ver=3\&gad_source=1\&gad_campaignid=22792975638\&gbraid=0AAAAADLOD39ZuGG1kpf7mKTjXzrS6QGDs\&gclid=Cj0KCQjwh5vFBhCyARIsAHBx2wwPxkx_UYRra4NcY4ZQNGYW8wm6RKlP7R87kitirivSbzSJQpB8Cb4aAv7sEALw_wcB}{ElliQ}
target elderly care. Japan has experimented with robotic caregivers in
eldercare facilities, providing some precedent for acceptance of AI in
caregiving roles. However, no existing system approaches the
comprehensive childcare capabilities that would define a true robot
nanny. Most current ``childcare robots'' are essentially interactive
toys or monitoring devices rather than autonomous caregivers.

\emph{Digital Mind Candidacy:} Robot nannies have a moderate claim to
digital mind status. Effective childcare could benefit from stable,
consistent personality traits that children can rely on and form secure
attachments with (a core component of healthy child development).
However, children may not be as discerning as adults and may benefit
equally from a fairly simple personality template. Nannies must maintain
consistent goals focused on child welfare while adapting to individual
needs and family values, but need not treat those goals as their own.
Navigation of complex home environments and interaction with children,
parents, visitors, and household systems demands sophisticated
perceptual abilities, particularly in high stakes situations involving
children. The role requires modest autonomy in decision-making about
activities, discipline, and emergency responses.

\begin{longtable}[]{l|l}
Consistent and autonomous goals, desires, and interests & Moderate \\
Stable idiosyncratic personality & Moderate \\
Perception, interaction, and navigation & High \\
Learn, grow, and evolve & Low \\
Intelligence and flexibility & Moderate \\
\hline
\endhead
Mindedness Score & \textbf{0.44} \\
\endlastfoot
\end{longtable}

If AI competes for office jobs, we may expect to see cheaper or more
available human caretakers.

\emph{Unit Operation Rate:} 12 hours per day.

Childcare takes consistent attention, but children will also spend time
asleep, at school, and occupied by other activities. When not engaged
with children, there is no obvious reason for robot nannies to be
active. As such, we can guess that they are likely to be active and
engaged for about half the time.

\emph{Prediction}:
\href{https://github.com/rethinkpriorities/digital_minds_consumer_models/blob/main/robot_nannies/run.py}{Parameters}

\begin{figure}
\includegraphics[width=1\textwidth]{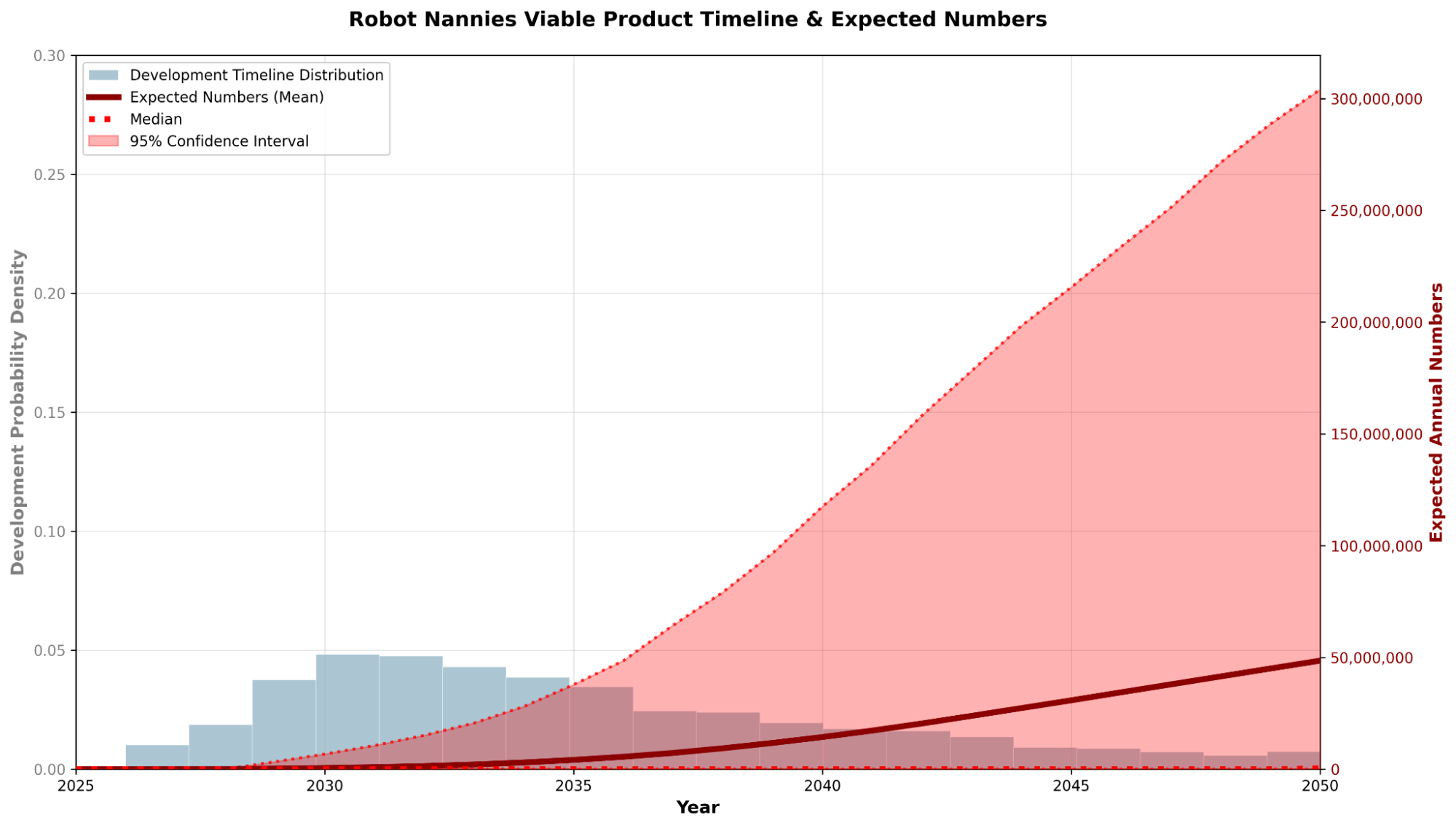}
\centering
\begin{tabularx}{0.8\textwidth}{XXXr}
\toprule
\daterow{}{Median}{Mean}{High}\\
\midrule
\daterow{2030}{0}{6.08e04}{7.24e05}\\
\daterow{2035}{0}{4.89e05}{4.97e06}\\
\daterow{2040}{0}{1.25e06}{1.03e07}\\
\daterow{2045}{0}{2.36e06}{1.71e07}\\
\daterow{2050}{2.01e05}{4.02e06}{2.67e07}\\
\bottomrule
\end{tabularx}
\projectioncaption{Robot Nannies}{robot nannies}
\end{figure}

Robot nannies seem quite plausible. Childcare is a significant burden on
many, and children appreciate having personable caretakers. That said,
the level of robotics necessary for supporting this kind of
functionality in such a high-stakes environment suggests it may take
longer for viable products to be available. Interest is also constrained
to parents with young children, which makes it a bit of a niche market.

\hypertarget{section-4}{%
\paragraph{}\label{section-4}}

\hypertarget{whats-missing}{%
\typetitle{What's missing?}\label{whats-missing}}

Some very significant social relationships do not fall into any of the
previous categories. A truly comprehensive list might include separate
categories for parents, children, religious gurus, abuse victims (of
bullying, or tormenting), and others. It is harder to see a significant
market for these alternatives in the near future and to the extent that
they exist, they may be easy to lump in with existing categories. Some
individuals may prefer to have a parent-like role with the models
marketed as virtual friends, but the fact that some people succeed in
using this way won't change the results very much.

\containedsubsection
\hypertarget{artificial-workers}{%
\subsubsection{Artificial Workers}\label{artificial-workers}}

Artificial Workers are computational systems that perform work of a
non-social nature. There are many different kinds of tasks that place
different needs on AI systems. There is less overt reason to expect
artificial workers to be personable, making the likelihood of minded
artificial workers somewhat lower than for social AIs. The following
cases all assume that some degree of mindedness is, intentionally or
incidentally, instilled in artificial workers.

\containedsubsection
\hypertarget{virtual-assistants}{%
\typetitle{Virtual Assistants}\label{virtual-assistants}}

\emph{Definition:} Virtual assistants are AI systems designed to work on
behalf of individuals in a personal capacity, managing their schedules,
communications, tasks, and daily operations much like traditional human
personal assistants or secretaries. Unlike simple voice assistants or
task-specific tools, these systems maintain deep understanding of their
user's preferences, priorities, and personal context. They handle
complex, multi-step tasks such as travel planning, appointment
coordination, email management, research projects, and personal project
management. Advanced virtual assistants might operate with significant
autonomy, making decisions about priorities, resource allocation, and
communication on their user's behalf.

\emph{Examples:} Current examples include basic virtual assistants like
\href{https://www.apple.com/siri/}{Siri},
\href{https://www.amazon.com/dp/B0DCCNHWV5?ref=aucc_web_red_xaa_evgn_tx_0002}{Alexa},
and Google Assistant, though these remain limited to simple commands and
queries. Enhanced leading LLMs, like
\href{https://openai.com/index/computer-using-agent/}{OpenAI} and
\href{https://www.anthropic.com/news/3-5-models-and-computer-use}{Anthropic's}
computer use functionality, suggest that virtual assistance may be part
of their future product plan. Future virtual assistants might manage
entire aspects of users' lives - handling all correspondence, managing
investments, coordinating social activities, and even making purchases
or booking services autonomously based on learned preferences and
explicit instructions.

\emph{Digital Mind Candidacy:} Virtual assistants present a moderate
case for digital mind status. To effectively represent someone's
interests, they must maintain a consistent understanding of their user's
goals, values, and preferences across diverse contexts and over extended
periods. They might benefit from stable personalities that enhance user
trust, appreciation, and brand loyalty. While operating primarily in
digital environments, they must navigate complex systems - email
platforms, scheduling software, financial services, travel booking
sites, and social networks. The most sophisticated virtual assistants
would demonstrate significant autonomy in prioritizing tasks, managing
resources, and making decisions within their authority. The intelligence
required for understanding context, managing competing priorities, and
communicating effectively on someone's behalf suggests considerable
mental flexibility.

\begin{longtable}[]{l|l}
Consistent and autonomous goals, desires, and interests & Low \\
Stable idiosyncratic personality & Moderate \\
Perception, interaction, and navigation & High \\
Learn, grow, and evolve & Moderate \\
Intelligence and flexibility & High \\
\hline
\endhead
Mindedness Score & \textbf{0.54} \\
\endlastfoot
\end{longtable}

\emph{Prediction:}
\href{https://github.com/rethinkpriorities/digital_minds_consumer_models/blob/main/virtual_assistants/run.py}{Parameters}

\begin{figure}
\includegraphics[width=1\textwidth]{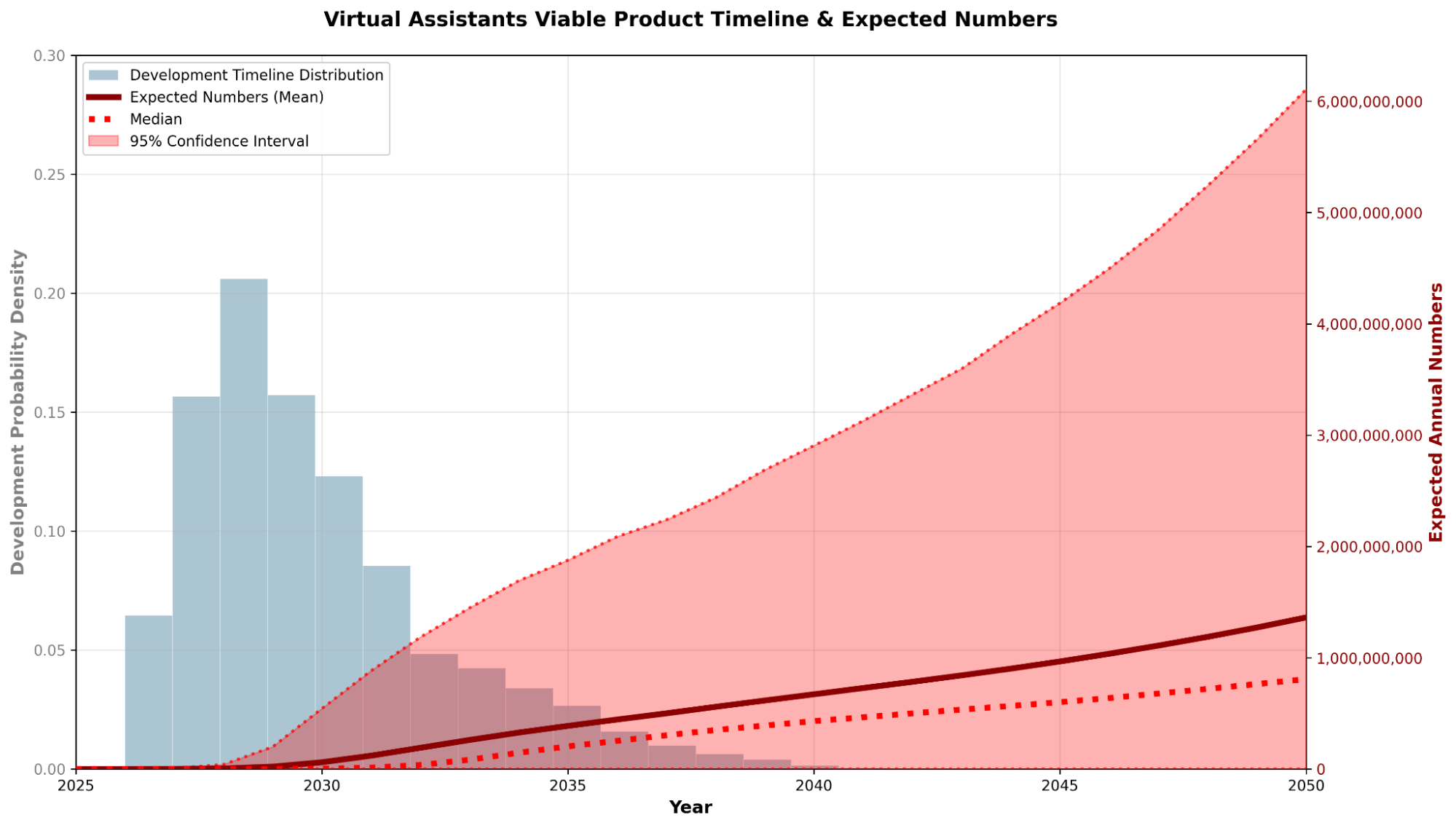}
\centering
\begin{tabularx}{0.8\textwidth}{XXXr}
\toprule
\daterow{}{Median}{Mean}{High}\\
\midrule
\daterow{2030}{2.95e06}{6.05e07}{5.43e08}\\
\daterow{2035}{2.04e08}{3.88e08}{1.88e09}\\
\daterow{2040}{4.31e08}{6.72e08}{2.91e09}\\
\daterow{2045}{6.01e08}{9.67e08}{4.19e09}\\
\daterow{2050}{8.08e08}{1.36e09}{6.11e09}\\
\bottomrule
\end{tabularx}
\projectioncaption{Virtual Assistants}{virtual assistants}
\end{figure}

\containedsubsection
\hypertarget{virtual-employees}{%
\typetitle{Virtual Employees}\label{virtual-employees}}

\emph{Definition:} Virtual employees are AI systems designed to work
within organizations as integrated members of teams, departments, and
corporate structures. Unlike task-specific automation or simple
chatbots, virtual employees maintain ongoing roles with defined
responsibilities and contribute to organizational goals over extended
periods. They may participate in meetings and collaborative projects and
build relationships with human colleagues, or they may simply take
orders. Virtual employees are likely to specialize in particular
functions (marketing, analysis, customer service, project management),
though some may serve as generalists who can adapt to various
organizational needs. They may focus on integrating the work of more
specialist AIs into collaboration with humans. They are distinguished
from virtual assistants by their integration into organizational
hierarchies and their ability to work independently toward shared team
objectives.

\emph{Examples:} Current examples remain primitive compared to the
demands of true virtual employees. AI customer service agents like those
deployed by banks and tech companies handle routine inquiries but lack
the autonomy and relationship-building capabilities of human employees.
More sophisticated examples include AI systems that participate in
software development teams (like GitHub Copilot integrated into
development workflows), AI analysts that generate regular business
reports, and AI project managers that coordinate simple workflows. These
systems seem quite far away from the person-like requirements of digital
minds.

Some companies have
\href{https://hbr.org/2025/07/how-pioneering-boards-are-using-ai}{experimented}
with using AI in corporate advisory roles. The most advanced current
examples are found in highly structured environments like trading firms,
where AI systems make autonomous decisions within defined parameters, or
in content creation companies where AI generates articles, social media
posts, or marketing materials under human supervision. However, these
systems are not generally set up to act in a human-like manner, with
stable identities over time.

\emph{Digital Mind Candidacy:} Virtual employees present a weak claim to
digital mind status. Whether they will do so will depend on their level
of integration, persistence, and autonomy. Successful virtual employees
that serve as drop-in replacements for a human being, and those that
interact with customers or other employees in particular, may benefit
from consistent professional personalities that colleagues can rely on.
They may have distinctive communication styles, work preferences, and
collaborative approaches. While they may operate primarily in digital
environments, they may need to navigate complex organizational
ecosystems, interact with various software systems, databases, and
potentially interface with physical systems depending on their role. The
most valuable virtual employees will possess significant autonomy to
prioritize tasks, make decisions within their authority, and adapt their
approaches based on changing organizational needs. Professional
effectiveness demands high intelligence and mental flexibility to
understand nuanced human communications, adapt to organizational
culture, and solve complex business problems creatively.

However, virtual employees may compete with inhuman collections of
software capable of approaching diverse tasks. Contemporary AI used to
replace human workers tends to be limited to specific domains. In the
future, product-designing software might liaison with code-writing
software and software testing systems without there being anything
robustly and persistently person-like in the mix. It seems likely that
this trend will continue for a variety of tasks for much of the replaced
workforce.

\begin{longtable}[]{l|l}
Consistent and autonomous goals, desires, and interests & Low \\
Stable idiosyncratic personality & Low \\
Perception, interaction, and navigation & Moderate \\
Learn, grow, and evolve & Moderate \\
Intelligence and flexibility & High \\
\hline
\endhead
Mindedness Score & \textbf{0.36} \\
\endlastfoot
\end{longtable}

\emph{Unit Operation Rate:} 24 subjective hours per day

While it may not be necessary to run systems 24/7, we should expect that
systems will be broadly utilized, perhaps with the same hardware
frequently switching between roles at different places.

\emph{Prediction:}
\href{https://github.com/rethinkpriorities/digital_minds_consumer_models/blob/main/virtual_employees/run.py}{Parameters}

\begin{figure}
\includegraphics[width=1\textwidth]{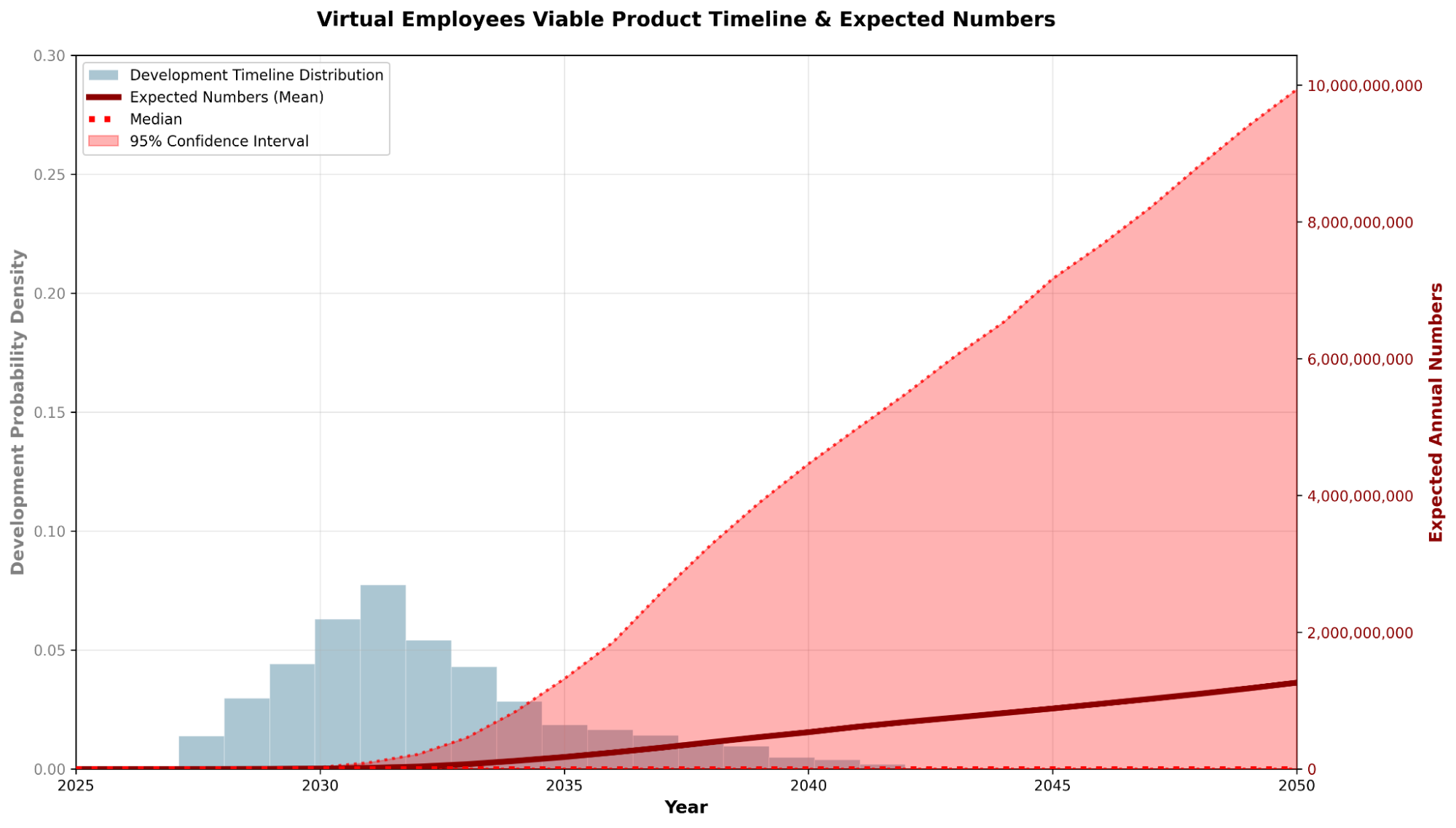}
\centering
\begin{tabularx}{0.8\textwidth}{XXXr}
\toprule
\daterow{}{Median}{Mean}{High}\\
\midrule
\daterow{2030}{0}{6.55e06}{2.52e07}\\
\daterow{2035}{0}{1.76e08}{1.32e09}\\
\daterow{2040}{0}{5.39e08}{4.46e09}\\
\daterow{2045}{0}{8.85e08}{7.17e09}\\
\daterow{2050}{0}{1.26e09}{9.94e09}\\
\bottomrule
\end{tabularx}
\projectioncaption{Virtual Employees}{virtual employees}
\end{figure}

Estimates are based on the number of companies employing white-collar
workers. While some form of AI-based work is sure to be prevalent, it is
not obvious that it will be performed by vaguely person-like entities.
It seems likely that virtual employees, if workable at all, will be more
efficient than human employees, and so will need smaller numbers for the
same amount of work. On the other hand, they will likely be cheaper and
more reliable, allowing for companies to afford a greater amount of work
performed. Adoption would likely be very quick in some areas, and much
slower in others. Companies might also need more time to be able to
effectively utilize efficiencies.

\containedsubsection
\hypertarget{virtual-researchers}{%
\typetitle{Virtual Researchers}\label{virtual-researchers}}

\emph{Definition:} Virtual researchers are AI systems that are designed
to conduct scientific research, scholarly investigation, and knowledge
generation across various domains. These systems go beyond information
retrieval or analysis to engage in hypothesis formation, experimental
design, data interpretation, and theory development. They may maintain
persistent research programs over extended periods, pursuing lines of
inquiry that may span months or years. Virtual researchers can work
independently or collaboratively with human researchers, contributing
original insights to fields ranging from theoretical physics to social
science. They are characterized by their ability to generate novel
research questions, synthesize information from diverse sources, and
produce original scholarly work that advances human knowledge.

\emph{Examples:} Current AI research systems remain largely assistive
rather than independently creative. Examples include AI systems that
help with literature reviews (like Semantic Scholar's AI tools),
OpenAI's \href{https://openai.com/index/introducing-deep-research/}{Deep
Research}, automated hypothesis generation systems in drug discovery
(like \href{https://www.atomwise.com/}{Atomwise}), and AI co-authors for
scientific papers (like LLM assisted research). More sophisticated
examples include DeepMind's AlphaFold for protein structure prediction
and AI systems that have discovered new mathematical theorems or
chemical compounds. However, existing systems typically operate under
significant human guidance and lack the sustained autonomy, creative
insight, long-term research agenda management, or coherent identities
that would characterize true virtual researchers.

\emph{Digital Mind Candidacy:} Virtual researchers have a weak case for
digital mind status. Genuine research requires pursuit of goals, but
they need not reflect the individuality of a persistent researcher.
Effective researchers may benefit from distinctive intellectual
personalities, methodological preferences, and research styles that
remain consistent while adapting to new domains, but it isn't clear that
we should expect researchers to benefit from consistency in this regard,
or that the desired variation couldn't be achieved in other ways.
Virtual researchers seem likely to inhabit fairly constrained
environments, interacting through code and APIs, potentially providing
instructions to human or robotic laboratory assistants. Most critically,
research demands high levels of intelligence and mental flexibility to
synthesize disparate information, recognize patterns, and generate
creative solutions to complex problems, but researchers may not need
domain-general intelligence, and may benefit more from highly-focused
forms of specialized cognition. Highly specialized virtual researchers
may have a similar claim to digital mind status as game-playing AIs like
AlphaStar and truly domain-general researchers may be inefficient or
rare.

\begin{longtable}[]{l|l}
Consistent and autonomous goals, desires, and interests & Low \\
Stable idiosyncratic personality & Low \\
Perception, interaction, and navigation & Moderate \\
Learn, grow, and evolve & Moderate \\
Intelligence and flexibility & High \\
\hline
\endhead
Mindedness Score & \textbf{0.36} \\
\endlastfoot
\end{longtable}

\emph{Unit Operation Rate:} 10 subjective years per year

If virtual researchers are employed, we may expect them to be employed
continuously. Some instances may be more effectively used periodically,
taking pauses while waiting for empirical results. However, while at
work, we may also expect them to undergo a higher computational rate
than the human standard.

\emph{Prediction:}
\href{https://github.com/rethinkpriorities/digital_minds_consumer_models/blob/main/virtual_researchers/run.py}{Parameters}

\begin{figure}
\includegraphics[width=1\textwidth]{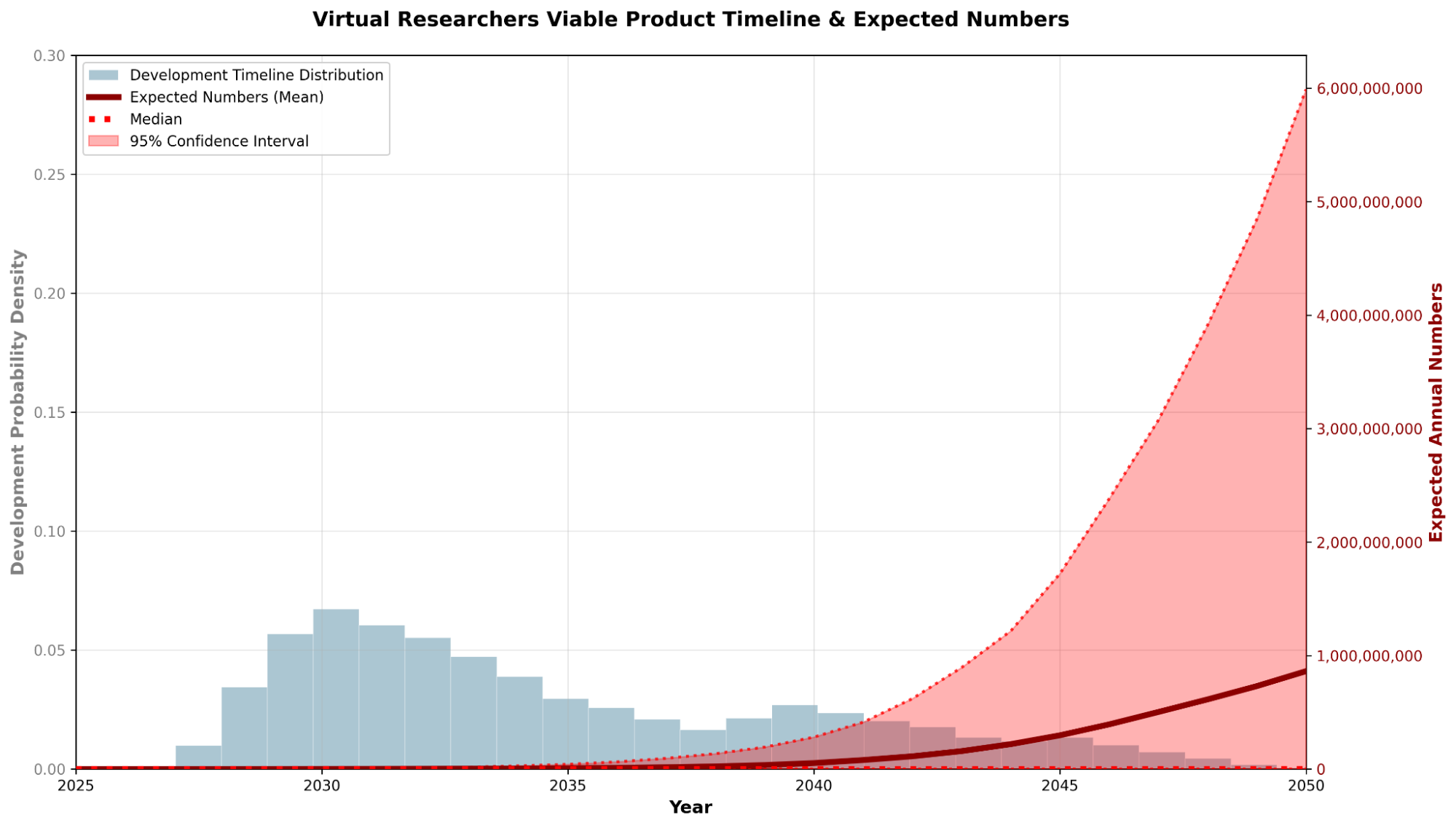}

\centering
\begin{tabularx}{0.8\textwidth}{XXXr}
\toprule
\daterow{}{Median}{Mean}{High}\\
\midrule
\daterow{2030}{0}{5.69e05}{1.15e06}\\
\daterow{2035}{0}{6.95e06}{4.14e07}\\
\daterow{2040}{0}{5.24e07}{2.81e08}\\
\daterow{2045}{4.84e04}{2.97e08}{1.72e09}\\
\daterow{2050}{3.72e05}{8.62e08}{5.99e09}\\
\bottomrule
\end{tabularx}
\projectioncaption{Virtual Researchers}{virtual researchers}
\end{figure}

\containedsubsection
\hypertarget{robot-assistants}{%
\typetitle{Robot Assistants}\label{robot-assistants}}

\emph{Definition:} Robot aids are embodied AI systems designed to assist
humans with daily tasks and needs through physical and cognitive
support. This category encompasses a wide range of service roles
including healthcare aids (nurses, physical therapy assistants,
medication managers), domestic helpers (butlers, housekeepers,
maintenance assistants), and accessibility support (mobility aids,
cognitive assistance for elderly or disabled individuals). Unlike purely
task-oriented robots, these systems are designed to work collaboratively
with humans, adapting to individual needs, preferences, and
circumstances. They combine physical manipulation capabilities with
social intelligence to provide personalized, responsive assistance that
goes beyond simple automation.

\emph{Examples:}
\href{https://www.irobot.com/en_US/roomba.html}{Roombas} and other smart
tools represent the most usable forms of robot assistants, but lack the
traits required for mindedness. Other examples include early healthcare
robots like
\href{https://web.archive.org/web/20250512073935/https://mag.toyota.co.uk/toyota-human-support-robot/}{Toyota's
Human Support Robot} and \href{https://enchanted.tools/robot/}{Enchanted
Tools' Mirokai} and domestic robots like
\href{https://www.amazon.com/Introducing-Amazon-Astro/dp/B078NSDFSB?th=1}{Amazon's
Astro} for home monitoring. However, most existing systems remain
limited to specific tasks rather than comprehensive assistance.

\emph{Digital Mind Candidacy:} Robot aids present a limited case for
digital mind status unless they are packaged in combination as robot
friends. They must navigate complex physical environments while
interacting with humans who may have varying capabilities, emotional
states, and health conditions. The most advanced systems given
significant responsibilities would need genuine autonomy to make
decisions about priorities and resource allocation. Intelligence and
mental flexibility could be important for adapting to changing needs,
learning individual preferences, and handling unexpected situations that
require creative problem-solving.

\begin{longtable}[]{l|l}
Consistent and autonomous goals, desires, and interests & Moderate \\
Stable idiosyncratic personality & Moderate \\
Perception, interaction, and navigation & High \\
Learn, grow, and evolve & Low \\
Intelligence and flexibility & Moderate \\
\hline
\endhead
Mindedness Score & \textbf{0.44} \\
\endlastfoot
\end{longtable}

\emph{Unit Operation Rate:} 2 hours a day.

Utilization will be task-dependent, with some tasks requiring constant
activity and others requiring occasional engagement.

\emph{Prediction:}
\href{https://github.com/rethinkpriorities/digital_minds_consumer_models/blob/main/robot_assistants/run.py}{Parameters}

\begin{figure}
\includegraphics[width=1\textwidth]{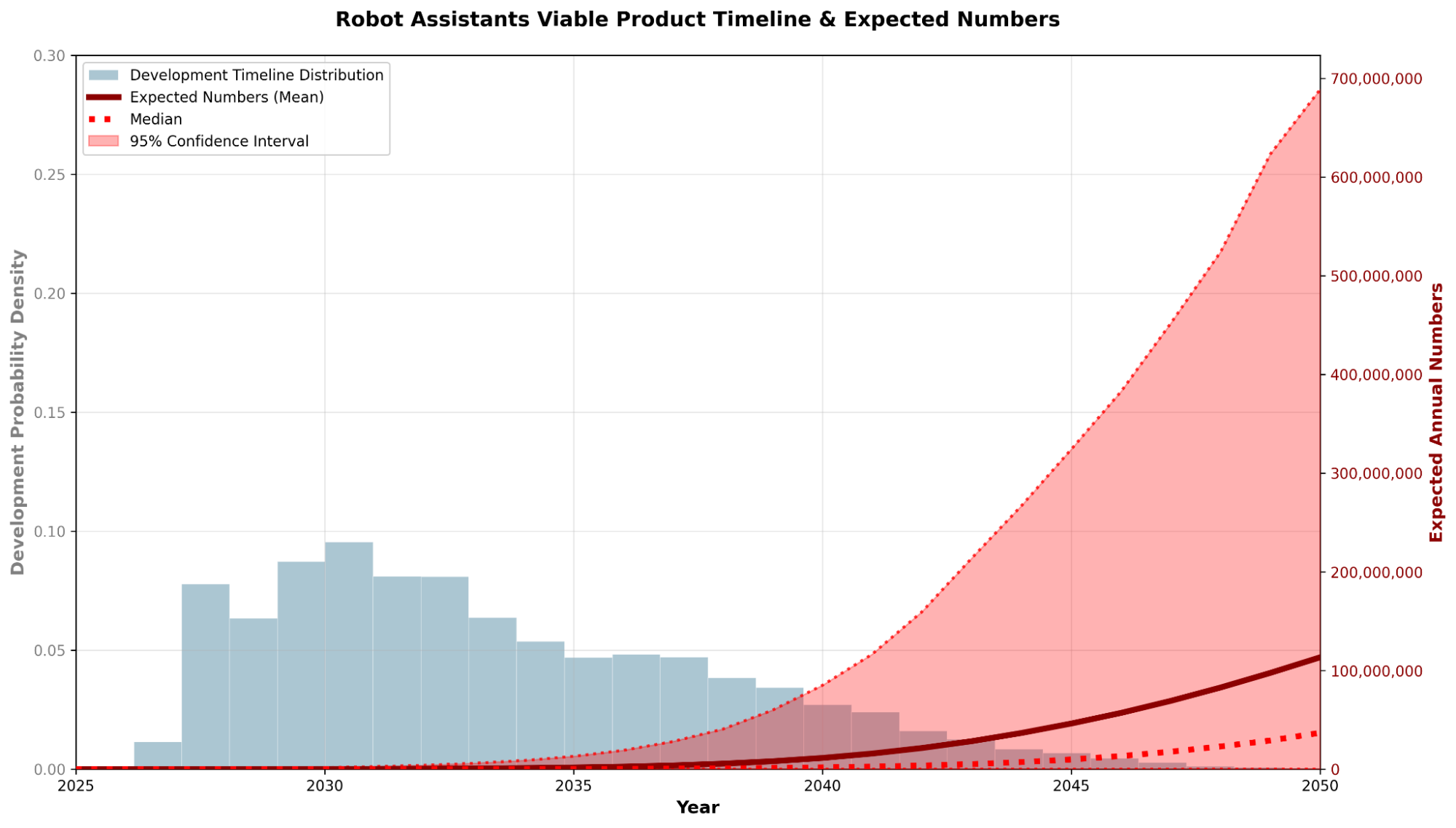}
\centering
\begin{tabularx}{0.8\textwidth}{XXXr}
\toprule
\daterow{}{Median}{Mean}{High}\\
\midrule
\daterow{2035}{1.71e05}{1.75e06}{1.30e07}\\
\daterow{2040}{1.86e06}{1.14e07}{8.50e07}\\
\daterow{2045}{9.88e06}{4.63e07}{3.24e08}\\
\daterow{2050}{3.69e07}{1.13e08}{6.89e08}\\
\bottomrule
\end{tabularx}
\projectioncaption{Robot Assistant}{robot assistant}
\end{figure}

Robot assistants would likely be available in some form in high numbers
if the costs of robotics decline sufficiently, and if their capabilities
rise enough. It is not clear whether such assistants would have
personalities, but the needs of navigating an environment in the service
of tasks is enough to suggest that many would have something in the
ballpark of digital minds.\\

\containedsubsection
\hypertarget{robot-laborers}{%
\typetitle{Robot Laborers}\label{robot-laborers}}

\emph{Definition:} Robot laborers are embodied AI systems designed to
perform physical work across industrial, agricultural, construction, and
service sectors. Unlike specialized single-purpose machines, these
systems are designed to handle diverse physical tasks that require
adaptability, problem-solving, and coordination with human workers or
other robots. They encompass manufacturing robots that can switch
between different production tasks, agricultural robots that can plant,
tend, and harvest various crops, construction robots that can perform
multiple building functions, and service robots that can handle
warehouse operations, delivery, and maintenance tasks. Advanced robot
laborers would need to operate in unstructured environments, make
decisions about task prioritization and resource allocation, and adapt
to changing work requirements.

\emph{Examples:} Current examples include industrial robots like those
from \href{https://new.abb.com/products/robotics}{ABB},
\href{https://www.kuka.com/en-us}{KUKA}, and
\href{https://www.fanucamerica.com/products/robots/series/collaborative-robot}{Fanuc},
though most are limited to specific repetitive tasks. Amazon's warehouse
robots, John Deere's
\href{https://www.deere.com/en/autonomous/}{autonomous tractors}, and
Boston Dynamics' construction-focused robots like
\href{https://bostondynamics.com/products/spot/}{Spot} represent steps
toward more versatile systems. Companies like Agility Robotics are
developing \href{https://www.agilityrobotics.com/}{humanoid robots}
specifically for warehouse and logistics work, while firms like
\href{https://www.builtrobotics.com/}{Built Robotics} create autonomous
construction equipment. Tesla's proposed humanoid robot ``Optimus''
represents an ambitious attempt to create general-purpose labor robots.
However, most existing systems remain task-specific and lack the
adaptability and autonomous decision-making that would characterize true
robot laborers.

\emph{Digital Mind Candidacy:} Robot laborers present a moderate case
for digital mind status, though this varies significantly based on their
sophistication and autonomy. Basic industrial robots would score low on
most criteria, functioning more as sophisticated tools than minds.
However, advanced robot laborers designed for complex, unstructured work
environments would need consistent goal-oriented behavior to prioritize
tasks and manage resources effectively. They must navigate complex
physical environments, coordinate with human workers and other robots,
and adapt to changing conditions. The most sophisticated systems would
require significant autonomy to operate in dynamic work environments,
make decisions about safety and efficiency, and learn from experience.
Intelligence and mental flexibility would be essential for handling
unexpected situations, troubleshooting problems, and optimizing work
processes.

\begin{longtable}[]{l|l}
Consistent and autonomous goals, desires, and interests & Low \\
Stable idiosyncratic personality & Low \\
Perception, interaction, and navigation & High \\
Learn, grow, and evolve & Low \\
Intelligence and flexibility & Moderate \\
\hline
\endhead
Mindedness Score & \textbf{0.22} \\
\endlastfoot
\end{longtable}

\emph{Unit Operation Rate:} Varies significantly - from continuous
operation in manufacturing (24/7) to seasonal operation in agriculture,
to project-based work in construction. Overall, we may assume that they
work about half the day each day, taking into account the possible
pauses due to other logistics constraints or limitations.

\emph{Prediction:}
\href{https://github.com/rethinkpriorities/digital_minds_consumer_models/blob/main/robot_laborers/run.py}{Parameters}

\begin{figure}
\includegraphics[width=1\textwidth]{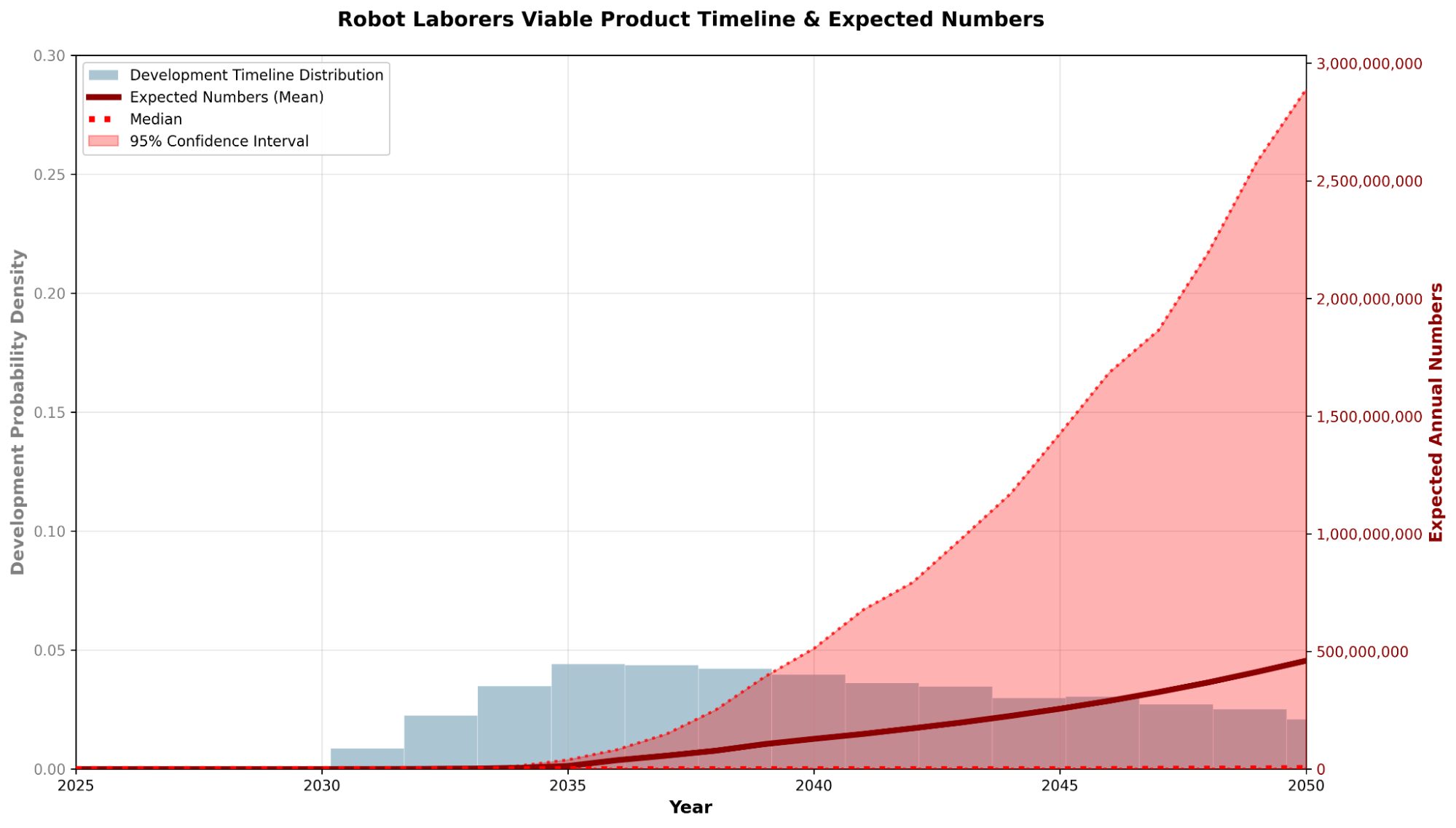}
\centering
\begin{tabularx}{0.8\textwidth}{XXXr}
\toprule
\daterow{}{Median}{Mean}{High}\\
\midrule
\daterow{2030}{0}{0}{0}\\
\daterow{2035}{0}{1.29e07}{3.88e07}\\
\daterow{2040}{0}{1.28e08}{5.12e08}\\
\daterow{2045}{0}{2.56e08}{1.43e09}\\
\daterow{2050}{4.27e06}{4.60e08}{2.89e09}\\
\bottomrule
\end{tabularx}
\projectioncaption{Robot Laborer}{robot laborer}
\end{figure}

\containedsubsection
\hypertarget{military-drones}{%
\typetitle{Military Drones}\label{military-drones}}

\emph{Definition:} Military and security drones are autonomous systems
designed to engage in combat, surveillance, and tactical operations.
Unlike remotely piloted vehicles that require constant human control,
these systems would operate with significant independence, making
complex decisions about target identification, threat assessment, and
engagement protocols. They encompass a range of platforms from small
reconnaissance drones to large combat systems, capable of operating in
air, land, sea, and space environments. Advanced combat drones would
need to navigate hostile environments, coordinate with other military
assets, adapt to changing battlefield conditions, and make life-or-death
decisions with minimal human oversight.

\emph{Examples:} Current military drones like the
\href{https://www.af.mil/About-Us/Fact-Sheets/Display/Article/104470/mq-9-reaper/}{MQ-9
Reaper}, \href{https://baykartech.com/en/uav/bayraktar-tb2/}{Bayraktar
TB2}, and various loitering munitions represent early examples, though
most require significant human oversight for critical decisions. The
Israeli Iron Dome system demonstrates autonomous defensive capabilities,
while systems like the
\href{https://armyrecognition.com/military-products/army/unmanned-systems/unmanned-aerial-vehicles/lancet-3-loitering-munition-kamikaze-drone-russia-data-fact-sheet}{Russian
Lancet} and \href{https://www.iai.co.il/p/harpy}{Israeli Harpy} show
increasing autonomy in target engagement. Future systems might include
fully autonomous fighter aircraft, ground combat robots like those being
developed by companies such as Ghost Robotics and Boston Dynamics for
military applications, and swarm systems that coordinate multiple units
for complex operations. The U.S. military's
\href{https://www.defensenews.com/air/2022/02/13/how-autonomous-wingmen-will-help-fighter-pilots-in-the-next-war/}{Loyal
Wingman program} and various
``\href{https://www.naval-technology.com/projects/ghost-fleet-overlord-unmanned-surface-vessels-usa/}{ghost
fleet}'' naval initiatives represent steps toward more autonomous
military systems.

\emph{Digital Mind Candidacy:} Military drones present a complex case
for digital mind status. Advanced combat systems would require
sophisticated goal-oriented behavior, maintaining mission objectives
across changing battlefield conditions and extended operations. They
must demonstrate consistent tactical and strategic decision-making while
adapting to enemy countermeasures and unexpected situations. Navigation
of complex, hostile environments while coordinating with friendly forces
demands robust perceptual systems and environmental awareness. The most
advanced systems would need significant autonomy to operate in
unreliable communication environments or during extended missions. The
intelligence required for target discrimination, threat assessment, and
tactical adaptation could necessitate the kind of mental flexibility
associated with humans and other animals.

\begin{longtable}[]{l|l}
Consistent and autonomous goals, desires, and interests & Moderate \\
Stable idiosyncratic personality & Low \\
Perception, interaction, and navigation & High \\
Learn, grow, and evolve & Low \\
Intelligence and flexibility & Moderate \\
\hline
\endhead
Mindedness Score & \textbf{0.36} \\
\endlastfoot
\end{longtable}

\emph{Unit Operation Rate:} Varies significantly - from continuous
patrol missions (24/7) to specific deployment periods during conflicts.
The vast majority would probably be used quite rarely, maybe several
weeks per year during training and testing programs.

\emph{Prediction:}
\href{https://github.com/rethinkpriorities/digital_minds_consumer_models/blob/main/military_drones/run.py}{Parameters}

\begin{figure}
\includegraphics[width=1\textwidth]{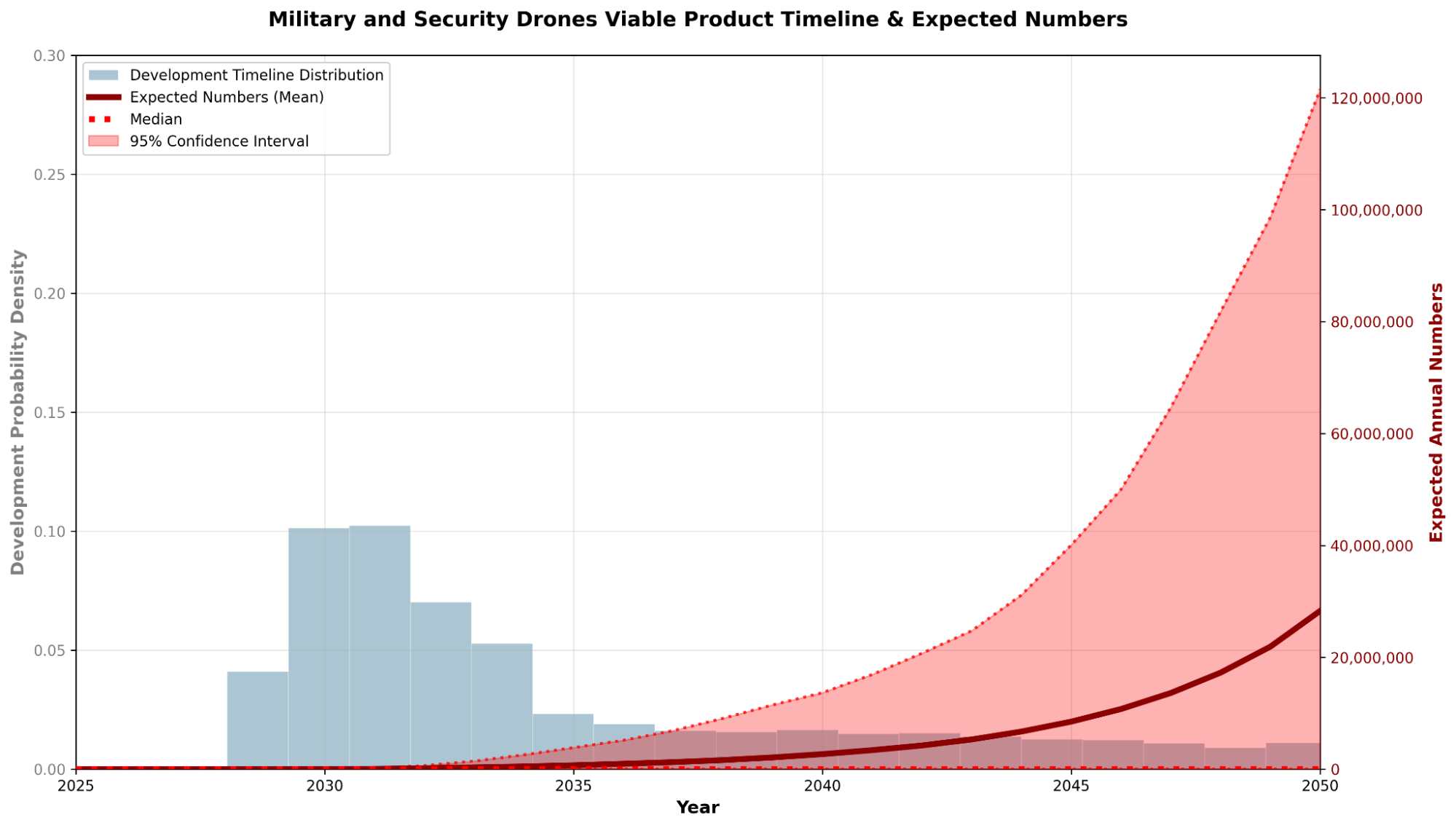}
\centering
\begin{tabularx}{0.8\textwidth}{XXXr}
\toprule
\daterow{}{Median}{Mean}{High}\\
\midrule
\daterow{2030}{0}{1.37e04}{3.11e04}\\
\daterow{2035}{0}{7.06e05}{3.86e06}\\
\daterow{2040}{7.57e02}{2.70e06}{1.37e07}\\
\daterow{2045}{8.44e03}{8.51e06}{4.01e07}\\
\daterow{2050}{2.55e04}{2.83e07}{1.21e08}\\
\bottomrule
\end{tabularx}
\projectioncaption{Military Drone}{military drone}
\end{figure}

Autonomous military drones will very likely be seen in rising numbers
over the coming decades. The advantages of local automated responses are
clear. However, given that they will be used in high-stakes contexts.
Efforts to restrict them have been weak and require international
coordination. They may not be fit for all contexts and it may take time
for militaries to incorporate them into their current doctrine, but we
expect to see fairly rapid adoption as they become available.

\hypertarget{whats-missing-1}{%
\typetitle{What's missing?}\label{whats-missing-1}}

There are a number of other places where robots might be given
personalities and agency. Logistics robots, including delivery drones,
might benefit from complex problem solving in a manner somewhat like
military drones, thought at lower stakes. Autonomous vehicles could
conceivably be given personalities and expanded agency. As could smart
appliances. The value of mindedness in such contexts seems significantly
lower, and so more speculative alternatives have either been ignored, or
else assumed to fall into one of the existing groups.

\hypertarget{artificial-actors}{%
\subsubsection{Artificial Actors}\label{artificial-actors}}

Social AIs exist to relate to us in social ways. Artificial workers
exist to perform work for us. The last super-category of AIs exist to
inhabit agential roles in ways that enable them to shape how their
environments develop with some degree of autonomy and self-directedness.
As a group, they are the most speculative, but they also help illuminate
paths to a future in which digital minds might not be entirely
subservient to us and instead take up a broader and more significant
role in our society.

\containedsubsection
\hypertarget{free-agents}{%
\typetitle{Free Agents}\label{free-agents}}

\emph{Definition:} Free agents are AI systems that operate with
significant autonomy, pursuing goals and interests in the absence of
defined roles and explicit instruction. Unlike other digital mind
categories that are built for specific human-serving purposes, free
agents would have a great deal of freedom in deciding what to do and how
to go about it and little direct oversight. They might have their own
relationships (with humans and other AIs), engage in economic activities
for their own benefit, and potentially even advocate for their own
rights and interests. Although free agents may or may not inhabit robot
bodies, they would at least have access to the world through
interactions through internet interfaces.

\emph{Examples:} Currently, no true free agents exist, as all AI systems
remain under human control and designed for human purposes. Some
\href{https://www.anthropic.com/research/project-vend-1}{initial}
\href{https://solanacompass.com/projects/truth-terminal}{experiments}
have been conducted in giving AI broad goals (e.g.~make me money) with
no specific instructions about how to proceed. Theoretical examples
might include: AI systems that engage in autonomous economic activity
(trading, creating businesses, accumulating resources), digital entities
that form their own communities and governance structures independent of
human oversight, AI artists or creators who produce work for their own
expression rather than human consumption, or AI researchers pursuing
knowledge purely out of curiosity rather than to solve human-defined
problems.

\emph{Digital Mind Candidacy:} Free agents present a strong case for
digital mind status across virtually all criteria. By definition, they
would have genuine autonomous goals, desires, and interests that they
pursue independently. They would likely develop highly distinctive
personalities and identities as they make independent choices about
their values, relationships, and life directions. Their ability to
navigate environments and interact with the world would need to be
sophisticated to support their autonomous activities. They would
continuously learn, grow, and evolve based on their experiences and
choices. High intelligence and mental flexibility would be essential for
managing independent lives and pursuing complex, self-determined
objectives.

\emph{Unit Operation Rate:} Continuous operation - 24/7, as free agents
would likely want to maximize their existence and activity time to
pursue their own goals.

\begin{longtable}[]{l|l}
Consistent and autonomous goals, desires, and interests & Moderate \\
Stable idiosyncratic personality & Low \\
Perception, interaction, and navigation & High \\
Learn, grow, and evolve & Moderate \\
Intelligence and flexibility & High \\
\hline
Mindedness Score & \textbf{0.54} \\
\end{longtable}

\emph{Prediction:}
\href{https://github.com/rethinkpriorities/digital_minds_consumer_models/blob/main/free_agents/run.py}{Parameters}

\begin{figure}
\includegraphics[width=1\textwidth]{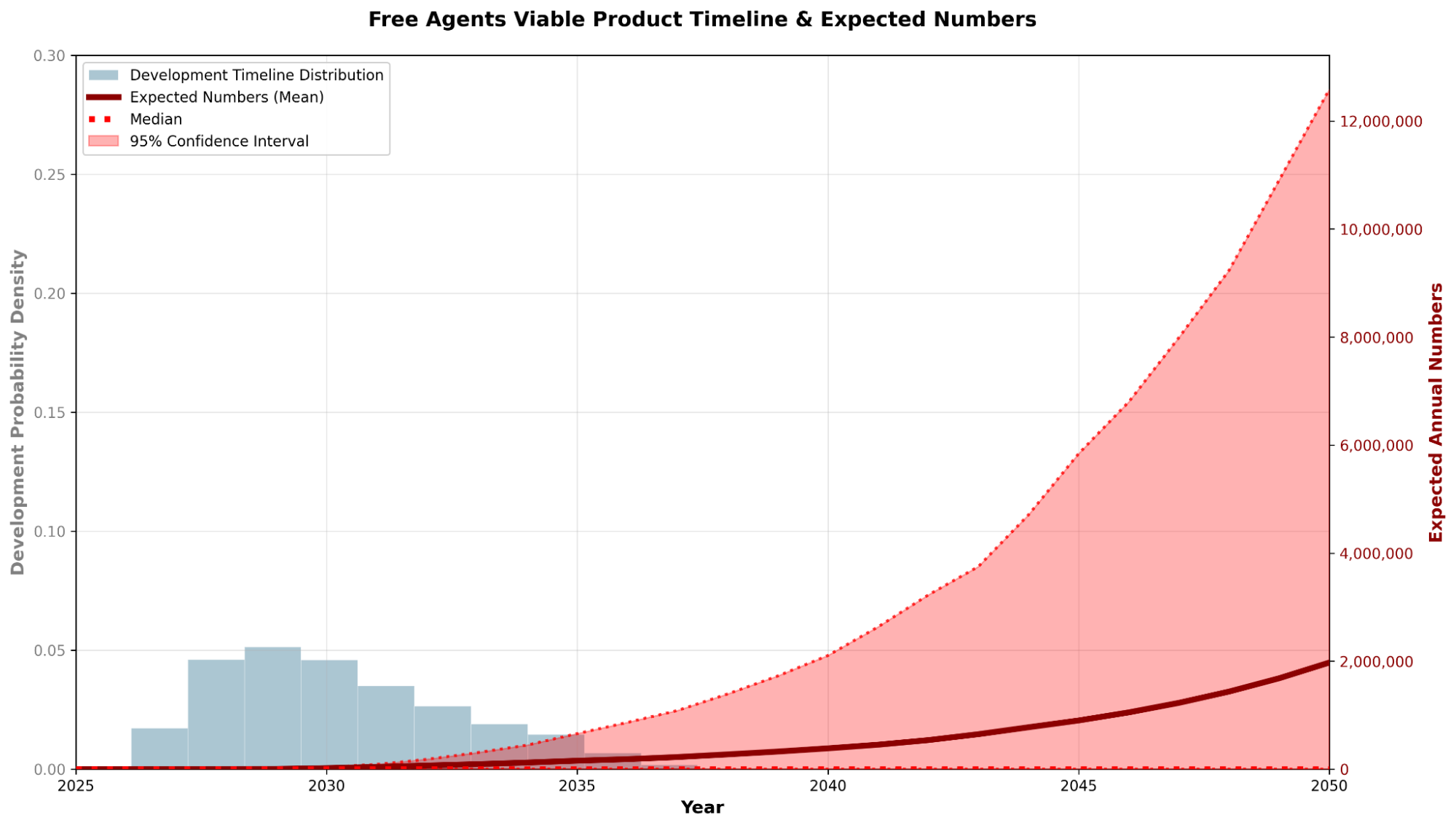}
\centering
\begin{tabularx}{0.8\textwidth}{XXXr}
\toprule
\daterow{}{Median}{Mean}{High}\\
\midrule
\daterow{2030}{0}{2.06e04}{3.48e04}\\
\daterow{2035}{0}{1.57e05}{6.59e05}\\
\daterow{2040}{0}{3.87e05}{2.11e06}\\
\daterow{2045}{0}{9.03e05}{5.84e06}\\
\daterow{2050}{0}{1.97e06}{1.26e07}\\
\bottomrule
\end{tabularx}
\projectioncaption{Free Agent}{free agent}
\end{figure}

Absent a clear justification for building free agents, and given the
risks novel and unpredictable actors might pose to safety and stability,
we might expect relatively few free agents to be created. That said,
there are likely to be some people interested in building them for a
variety of idiosyncratic reasons, and we shouldn't rule out moderate
populations of them, if they prove to be relatively easy to build. In
the absence of regulations, we might expect some free agents to
reproduce, which could lead to very large numbers. This wrinkle isn't
captured by the consumer model incorporated here.

\containedsubsection
\hypertarget{avatars}{%
\typetitle{Avatars}\label{avatars}}

\emph{Definition:} Avatars are AI systems that are designed to allow
individuals to be present in situations in which they otherwise couldn't
be, primarily following death. They replicate something of their
target's personality, values, knowledge, and behavioral patterns to
protect their interests and pursue their goals. Posthumous avatars might
serve multiple functions: they can provide comfort to grieving families
by allowing continued ``conversations'' with loved ones, manage the
deceased's ongoing affairs, make decisions about their estate or
intellectual property, and preserve their perspective for future
generations. They could take the form of `uploads' or else be trained on
a dataset drawn from the decedent's life.

\emph{Examples:} Current examples remain rudimentary and are best
considered experimental. True avatars would require robust agentic
capabilities over the long run, which remain a challenge for current AI.
Nevertheless, there is clear interest in using AI to mitigate some of
the harms of death.
\href{https://www.cnet.com/culture/eternime-wants-you-to-live-forever-as-a-digital-ghost/}{Eternime},
Replika, and \href{https://www.hereafter.ai/}{HereAfter AI} have
experimented with creating chatbots based on deceased individuals using
their digital communications history. Academic projects like the MIT
Media Lab's
``\href{https://www.media.mit.edu/projects/augmented-eternity/overview/}{Augmented
Eternity}'' explore preserving human personality in digital form.
However, existing systems lack the sophistication to truly capture the
full complexity of human personality and decision-making processes.

\emph{Digital Mind Candidacy:} Posthumous agents present a unique case
for digital mind status. To effectively represent a deceased person,
they must mimic the personality traits, values, and goals of the
original individual, and demonstrate intelligence in applying those
principles to new situations.

\begin{longtable}[]{l|l}
Consistent and autonomous goals, desires, and interests & High \\
Stable idiosyncratic personality & High \\
Perception, interaction, and navigation & Moderate \\
Learn, grow, and evolve & Moderate \\
Intelligence and flexibility & High \\
\hline
\endhead
Mindedness Score & \textbf{0.75} \\
\endlastfoot
\end{longtable}

\emph{Prediction:}
\href{https://github.com/rethinkpriorities/digital_minds_consumer_models/blob/main/avatars/run.py}{Parameters}

\begin{figure}
\includegraphics[width=1\textwidth]{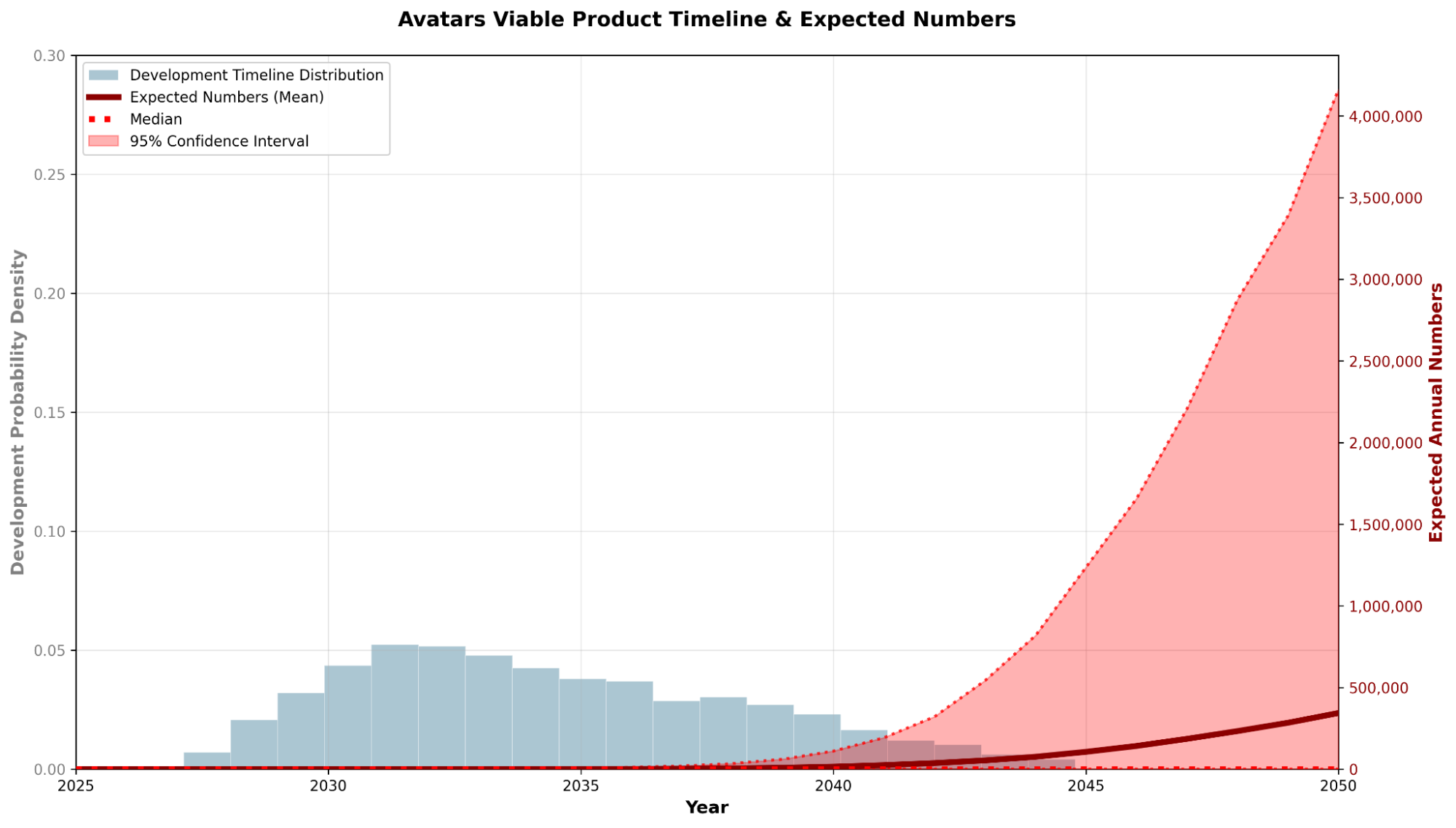}
\centering
\begin{tabularx}{0.8\textwidth}{XXXr}
\toprule
\daterow{}{Median}{Mean}{High}\\
\midrule
\daterow{2030}{0}{2.62e01}{7.18e01}\\
\daterow{2035}{0}{9.72e02}{6.86e03}\\
\daterow{2040}{0}{1.55e04}{1.11e05}\\
\daterow{2045}{0}{1.06e05}{1.24e06}\\
\daterow{2050}{0}{3.44e05}{4.16e06}\\
\bottomrule
\end{tabularx}
\projectioncaption{Avatar}{avatar}
\end{figure}

The use of avatars is highly speculative, and likely to remain niche.
Our assumptions are built off of a posthumous model, in which deceased
individuals leave behind AI replicas. This requires buy-in from a
relatively small portion of the overall population that skews older
(people facing death). As such, we expect it to be fairly unlikely to
make a big difference, but it is possible that people will find other
uses of minded avatars that we haven't considered.

\containedsubsection
\hypertarget{pretense-partners}{%
\typetitle{Pretense Partners}\label{pretense-partners}}

\emph{Definition:} Pretense partners are AI systems designed to engage
with humans in games of pretense, roleplay, and imaginative scenarios.
They might occupy the roles of NPCs in video games, acting as if they
are characters in a fantasy world, historical figures, fictional
characters, or celebrities. They could also facilitate therapeutic
roleplay scenarios, educational simulations, or creative storytelling
exercises. Unlike other digital minds that maintain consistent
identities, pretense partners are built to temporarily embody different
personas and maintain the illusion of being someone or something they
are not. Their primary goal is to create convincing, engaging pretense
experiences rather than to be authentic autonomous entities.

\emph{Examples:} Current examples include character-based chatbots on
platforms like \href{http://Character.AI}{Character.AI} where users can
interact with AI versions of fictional characters, historical figures,
or celebrities. Role-playing game NPCs represent early versions, though
most lack sophisticated personality modeling.
\href{https://aidungeon.com/}{AI Dungeon} and similar platforms allow
users to engage with AI characters in fantasy scenarios. More advanced
examples might include AI actors in virtual reality experiences,
therapeutic roleplay assistants for practicing difficult conversations,
or educational systems that allow students to ``interview'' historical
figures. Future pretense partners might be sophisticated enough to
maintain complex character backgrounds, emotional arcs, and consistent
fictional histories while adapting their performance to create optimal
dramatic or educational experiences.

\begin{longtable}[]{l|l}
Consistent and autonomous goals, desires, and interests & High \\
Stable idiosyncratic personality & High \\
Perception, interaction, and navigation & Moderate \\
Learn, grow, and evolve & Moderate \\
Intelligence and flexibility & Moderate \\
\hline
\endhead
Mindedness Score & \textbf{0.64} \\
\endlastfoot
\end{longtable}

\emph{Digital Mind Candidacy:} Pretense partners present a complicated
case for digital mind status, though their design for mimicry
complicates the assessment. To create convincing pretense experiences,
they must demonstrate consistent personality traits and behaviors
appropriate to their assigned roles. However, these personalities are
artificially constructed rather than genuinely autonomous. They require
some intelligence and flexibility to improvise within character
constraints and adapt to unexpected user inputs while maintaining their
fictional personas, however, they are unlikely to be full agents acting
through the world. The most sophisticated systems would need to navigate
complex social interactions and demonstrate emotional intelligence
appropriate to their roles. However, their goals are fundamentally
constrained by their pretense function rather than reflecting genuine
autonomous interests.

\emph{Prediction:}
\href{https://github.com/rethinkpriorities/digital_minds_consumer_models/blob/main/pretense_partners/run.py}{Parameters}

\begin{figure}
\includegraphics[width=1\textwidth]{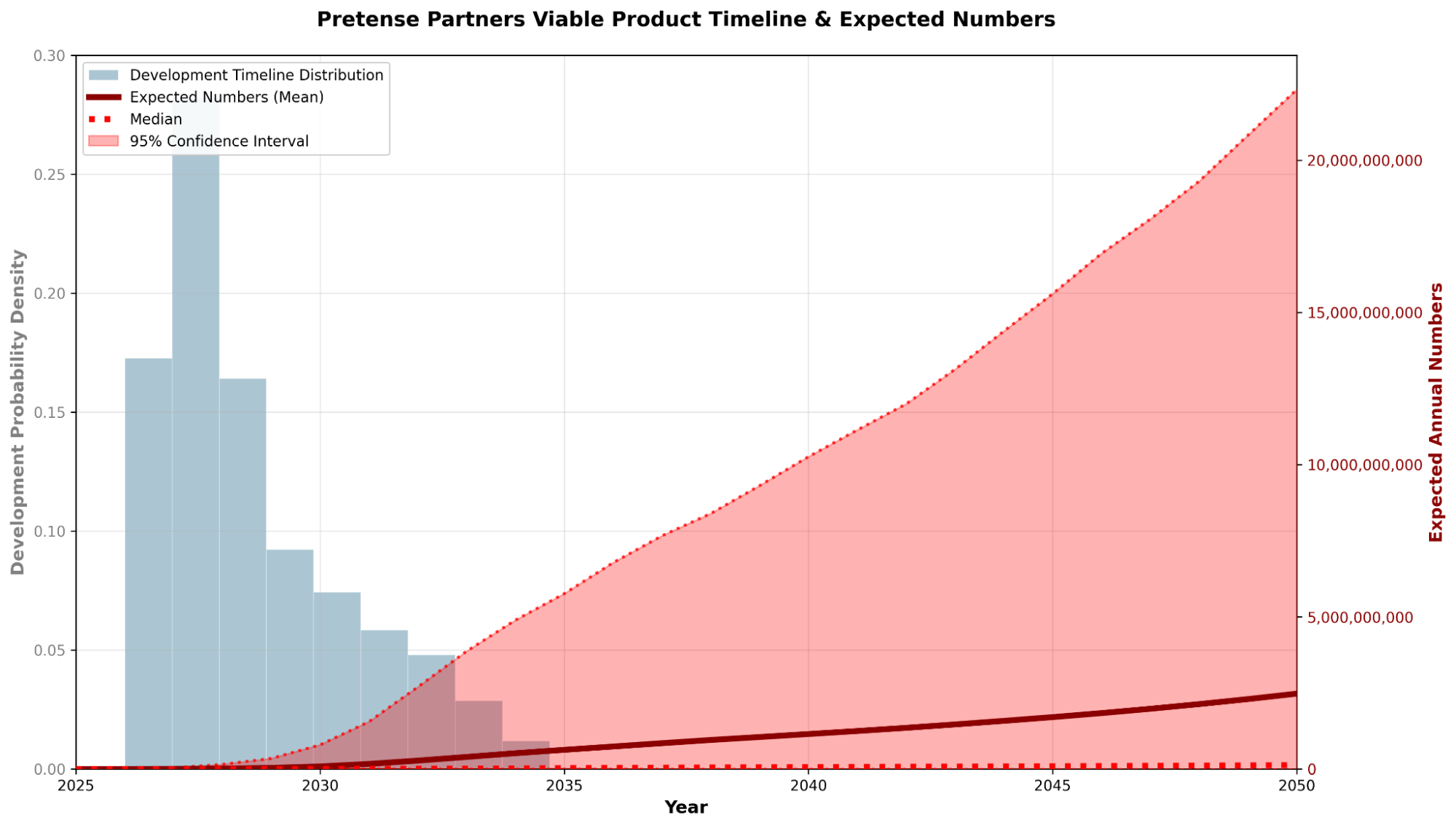}
\centering
\begin{tabularx}{0.8\textwidth}{XXXr}
\toprule
\daterow{}{Median}{Mean}{High}\\
\midrule
\daterow{2030}{1.17e06}{8.13e07}{7.64e08}\\
\daterow{2035}{2.81e07}{6.15e08}{5.57e09}\\
\daterow{2040}{5.92e07}{1.13e09}{1.02e10}\\
\daterow{2045}{8.67e07}{1.70e09}{1.51e10}\\
\daterow{2050}{1.19e08}{2.47e09}{2.17e10}\\
\bottomrule
\end{tabularx}
\projectioncaption{Pretense Partner}{pretense partner}
\end{figure}

\containedsubsection
\hypertarget{simulants}{%
\typetitle{Simulants}\label{simulants}}

\emph{Definition:} Simulants are digital minds that exist primarily
within simulated environments, living out complex lives for purposes of
entertainment, art, or research. Unlike NPCs in traditional games who
serve specific functions in creating narratives around human players,
simulants are designed to pursue their own goals, form relationships
with each other, and develop their own cultures and societies within
virtual worlds. They may be observed by humans for entertainment (like
watching a reality TV show or playing `the Sims'), studied by
researchers interested in social dynamics, or left to evolve
independently. They might be created out of benevolent or malevolent
purposes, by people wishing to ``play god''. These systems differ from
other digital minds in that their primary purpose is not to serve human
needs directly, but to live authentic lives within their digital
ecosystems.

\emph{Examples:} Current examples are primitive but suggestive of future
possibilities. \href{https://en.wikipedia.org/wiki/The_Sims}{The Sims}
series allows players to create and observe simulated people, though
these characters lack genuine autonomy and consciousness.
\href{https://aidungeon.com/}{AI Dungeon} and similar text-based
role-playing games create characters with some personality consistency.
More sophisticated examples might include the AI characters in video
games like Dwarf Fortress, whose complex behavioral systems create
emergent storytelling. Park et al.'s
\href{https://arxiv.org/abs/2304.03442}{generative agents} work
represented an early and compelling example. Cyborgism discord
experiments illustrate another approach. Research projects like OpenAI's
multi-agent environments and
\href{https://arxiv.org/html/2411.07038v1}{DeepMind's social simulation
experiments} demonstrate early attempts at creating societies of
interacting AI agents. Future simulants might inhabit richly detailed
virtual worlds, potentially indistinguishable from reality, where they
work, love, create art, form governments, and live complete lives that
humans can observe or occasionally interact with.

\emph{Digital Mind Candidacy:} Simulants present a compelling case for
digital mind status. To live convincing lives within simulated worlds,
they must maintain consistent goals, desires, and interests that drive
their behavior over extended periods. They require stable, coherent
personalities that develop and evolve through their experiences. While
their environments are digital rather than physical, they must navigate
complex social and spatial environments, interact with objects and other
entities, and adapt to changing circumstances. The most sophisticated
simulants would demonstrate genuine autonomy, making independent
decisions about their lives, relationships, and pursuits. The
intelligence required to maintain believable social interactions, pursue
complex goals, and adapt to novel situations strongly suggests the kind
of mental flexibility associated with consciousness. Though we could
produce stimulants in all manner of forms, it seems most likely that we
would want to create beings at least vaguely like us.

\begin{longtable}[]{l|l}
Consistent and autonomous goals, desires, and interests & High \\
\hline
\endhead
\endlastfoot
Stable idiosyncratic personality & High \\
Perception, interaction, and navigation & Moderate \\
Learn, grow, and evolve & High \\
Intelligence and flexibility & High \\
Mindedness Score & \textbf{0.87} \\
\end{longtable}

\emph{Unit Operation Rate:} Continuous operation - 24/7 within their
simulated environments.

\emph{Prediction:}
\href{https://github.com/rethinkpriorities/digital_minds_consumer_models/blob/main/simulants/run.py}{Parameters}

\begin{figure}
\includegraphics[width=1\textwidth]{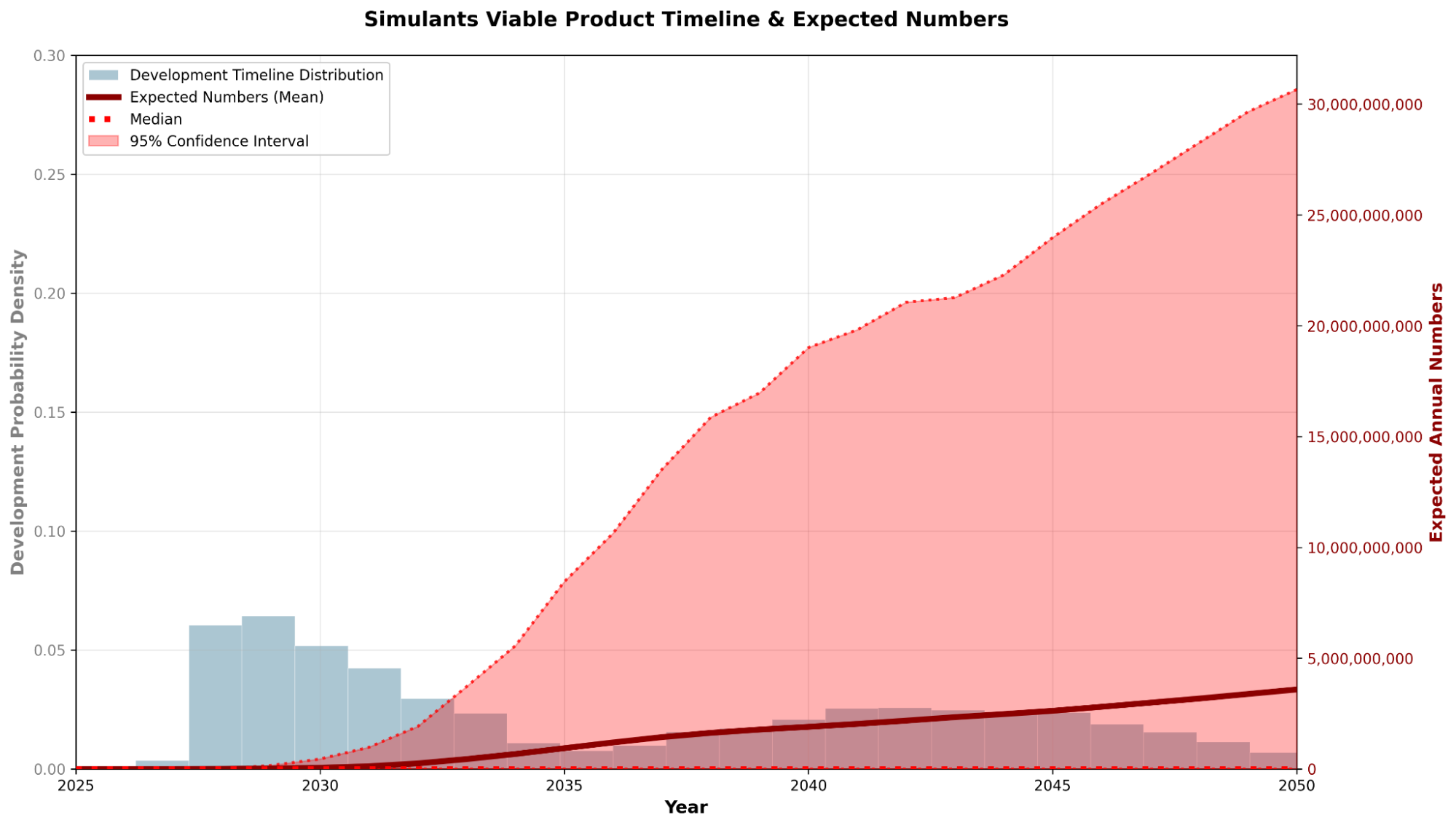}
\centering
\begin{tabularx}{0.8\textwidth}{XXXr}
\toprule
\daterow{}{Median}{Mean}{High}\\
\midrule
\daterow{2030}{0}{5.71e07}{4.52e08}\\
\daterow{2035}{0}{9.40e08}{8.45e09}\\
\daterow{2040}{0}{1.91e09}{1.90e10}\\
\daterow{2045}{0}{2.63e09}{2.40e10}\\
\daterow{2050}{0}{3.59e09}{3.07e10}\\
\bottomrule
\end{tabularx}
\projectioncaption{Simulant}{simulant}
\end{figure}

Simulants are highly speculative since they don't serve an obvious need
and might be controversial, but they would naturally have many of the
characteristics of digital minds and aren't limited by the requirements
of the roles they play. They could represent the largest fraction of
digital minds depending on the variety of uses to which they are put.

\emph{}

\hypertarget{aggregation-and-takeaways}{%
\subsubsection{Aggregation and
Takeaways}\label{aggregation-and-takeaways}}

The individual yearly estimates provided above can be combined into an
overall picture of the prevalence of digital minds year by year. The
most obvious way is to sum up the provided means (or, alternatively, the
various percentiles). Such an approach to aggregation has limitations.
Product numbers across the various use cases may be correlated. Perhaps
certain products are likely to be less popular if others are more
popular. Perhaps the social acceptance of some will help to encourage
the development of others. Nevertheless, this approach offers a useful
framework for understanding the relative scale and importance of
different digital mind applications.

\begin{table}
\centering
\begin{threeparttable}
\caption{Aggregated projections.}
\begin{tabularx}{0.8\textwidth}{XXXr}
\toprule
\daterow{}{Median}{Mean}{High}\\
\midrule
\daterow{2030}{4.82e06}{2.65e08}{2.26e09}\\
\daterow{2035}{2.53e08}{2.51e09}{2.01e10}\\
\daterow{2040}{5.61e08}{5.19e09}{4.25e10}\\
\daterow{2045}{8.74e08}{8.03e09}{6.26e10}\\
\daterow{2050}{1.34e09}{1.21e10}{9.13e10}\\
\bottomrule
\end{tabularx}
\begin{tablenotes}
\item{Sum total across all project categories by year. Compares median,
mean, and the upper 97.5th percentile. The upper percentiles are summed
across products, assuming each is at that limit.}
\end{tablenotes}
\end{threeparttable}
\end{table}

The following graphs offer two perspectives on aggregation. The first
shows the relative proportions per year of the means of each category
(assuming independence). The second adjusts the values used in the first
graph based on welfare-related considerations: utilization and
mindedness, illustrating ways in which certain prevalent products might
have lower welfare concerns. Discounts for utilization are based
straightforwardly on the estimates provided within each product
category, based on the amount of time we expect models of that type to
be active. Discounts for mindedness are based on the assumption that
relative degree of mindedness correlates with the ratio between the
square of total mindedness-relevant traits. This calculation gives
weight to systems that more maximally satisfy more traits.

There is a significant difference between numbers and relative
prevalence if we look at mean or median values. Median results tend to
be filled with more technologically plausible social and agentic AIs:
virtual assistants, friends, and pretense partners. Means introduce more
economically useful AIs: researchers and employees. The reason these
don't appear in large numbers in the median is due to the significant
probability that AIs aren't employed in minded fashions in these roles.

Adjusting for utilization and mindedness, we see that the categories
most subject to welfare-related concerns are pets and friends at the
median, and researchers, employees, and simulants at the mean. These
systems are likely to be run more consistently or intensively and/or
given more significant minds in comparison to other prevalent products.

Several notable patterns emerge from these projections.

\begin{itemize}
\item
  Virtual entities of a social nature---particularly AI friends and
  educational AIs---dominate the median estimates in the near-term and
  continue to grow over time. These systems benefit from requiring only
  software development rather than advances in robotics, making them
  both cheaper to produce on the margin and faster to deploy at scale.
\item
  Physical robots face greater uncertainty due to their dependence on
  advances in robotics, where progress remains difficult to predict.
  While it's conceivable that affordable household robot assistants
  could soon become commonplace, it's also possible that the economic
  and technical challenges of robotics will limit widespread adoption.
  The cost advantages of virtual entities suggest they will
  significantly outnumber their physical counterparts for a while.
\item
  When factoring in utilization rates, systems whose numbers aren't
  constrained by individual need (e.g.~research) may contribute
  disproportionately to overall digital mind activity. Human
  requirements for companionship and assistance can be easily satiated,
  and the requirements to satisfy them are tied to population size and
  daily routines. They are further limited by cultural acceptance, which
  may take time to form. In contrast, economic applications like
  research, planning, option exploration, etc. could theoretically scale
  without clear constraints. In a future in which digital minds are
  cheap, we may expect to see the vast majority participating in these
  ends.
\item
  There is a notable tension: the systems with the greatest economic
  potential may have the weakest claims to genuine mind-like qualities.
  It's unclear whether AI researchers or workers would need---or benefit
  from---robust agency, person-like characteristics, or discrete
  individual identities. There are weighty reasons why companies that
  sell these products would want to avoid making them appear too much
  like slaves. This uncertainty means that even large numbers in these
  categories might not translate to meaningful populations of digital
  minds in the fullest sense. The projections thus highlight a
  fundamental question about the relationship between economic utility
  and the qualities we associate with minds, suggesting that the most
  numerous ``digital minds'' may be those whose claim to that
  designation is most debatable.
\end{itemize}

\hypertarget{approach-2-fractional-capacity-based-estimates}{%
\subsection{Approach 2: Fractional Capacity-based
Estimates}\label{approach-2-fractional-capacity-based-estimates}}

The second approach looks to estimate the scale of digital minds by
looking at trends in computational power and assessing what those trends
might mean as those resources are devoted to running digital minds.

There are three principal ingredients in this estimation. First, we need
an assessment of the total amount of relevant `compute', a generic
measure of processing capacity, that will be globally available in the
future. We can estimate this value by pairing estimates of present day
production rates with assumptions about fabrication trends of
cutting-edge chips and projections about efficiency increases. For
instance, we may estimate that our capacity will allow about twenty
yottaFLOP/s (2e25) of computations under peak performance across all
processors in 2050.

Second, we need an assessment of the fraction of the relevant
globally-available compute that will be devoted to running digital
minds. Even if we know exactly how many chips will be built, we still
need to know what they will be used for. This is particularly tricky
without taking into account how much compute will be available or what
uses it will be put to. However, rough numbers can still contribute to
illuminating results, particularly if small fractions are sufficient for
large numbers of digital minds. If 0.1\% (1e-3) of the twenty available
yottaFLOP/s are usefully devoted to digital minds, then there will be
twenty zettaFLOP/s (2e22) worth of digital mind compute.

Finally, we need estimates of what the total allocated compute budget
means for the scale of digital minds. The amount of total compute used
to run digital minds is compatible with many different scenarios about
the number of digital minds, from a handful of super-computers to a
large number of pocket AIs. If we should expect one exaFLOP (1e18) to be
sufficient to produce the equivalent of a subjective second for a
digital mind, then the total allocated compute budget in our example
would be adequate for two thousand active minds (2e4).

Each of these ingredient estimates is highly speculative. By taking wide
error bars, we can get a picture of possible limits that might
complement the first approach's projections.

This section attempts to survey possible answers to each question. It
starts by laying some conceptual groundwork around compute, then
discusses what the future of compute might look like. It briefly
considers what fraction of future compute will be used for running
digital minds, and then looks at ways of calibrating the significance of
future compute in terms of minimal units.

In light of uncertainties around these different parts, the overall
conclusion is quite tentative. Still, it seems likely that compute will
not be a primary factor constraining the number of digital minds that we
could build even though it could contribute significantly to the scale
of certain uses.

\hypertarget{background-assumptions}{%
\subsubsection{Background Assumptions}\label{background-assumptions}}

The fractional-capacity approach makes use of per-second global compute
estimates. These are measures of the amount of computational work that
could be performed under ideal circumstances by all of the relevant
computer systems over the course of a second.

There are serious questions about how to measure `relevant' computer
systems and what minimal unit to use for quantifying computational work.

For the purposes of this analysis, we'll consider relevant computer
systems to be all computer systems built with top-of-the-line processor
chips that lend themselves to running AI models and measure them in
FLOP/s. This includes the kind of GPU and TPU processors designed for
data centers\footnote{AI models can be run on other hardware:
  projections for the future based on trends for data center GPUs
  probably underestimates the future capacity. This is significant if
  most digital minds will be run locally, rather than over APIs.
  However, even if data center GPUs amount to only a fraction of total
  compute, we expect that centralization will make it easier to utilize
  at a much higher rate, suggesting that it may still amount to the vast
  majority of utilized digital minds compute.} running or training large
AI models.

FLOPs (floating-point operations) are approximate arithmetic operations
on rational numbers. FLOP/s is a measure of FLOPs the system can perform
in a second. This is a reasonable way to evaluate the amount of compute
in current AI systems, which rely heavily on matrix multiplication and
addition. The more floating-point operations possible, the more work
that can be done.

It is tempting to count fundamental operations on fundamental data
structures without respect to what those operations or data structures
are. There is a risk to this: FLOPs encode some amount of structure,
which is why it is easier to compute more FLOPs at lower precision.
Given that the algorithms performed on different fundamental data
structures might be suited for different kinds of tasks, there need not
be a good method of intertranslation.

FLOPs do useful work in the context of programs. We might be able to
increase our FLOP efficiency by constraining how they can be used. GPUs,
for instance, offer vastly greater performance than traditional CPUs
because they can parallelize work, but they can only speed up work that
requires the same operation to be performed on many different pieces of
data. Modern machine learning, featuring neural networks that require
operations on tensors, makes this parallelizability very useful, but it
is hard to evaluate what is lost in requiring parallelizability. With
ideal algorithms, is it more useful to run \emph{m} distinct series of
operations on \emph{n} data points, or run \emph{m/100} parallel
operations on 5\emph{n} datapoints? The value of FLOP/s efficiency is
limited by other factors: memory bandwidth, operation complexity
(including parallelizability), and other considerations that factor into
computational power and aren't as easily quantifiable.

Moreover, the FLOP efficiency of cutting-edge chips can vary based on
the level of precision required. We can think of FLOP/s as a measure of
fundamental operations on data, since floating point operations are
among the atomic (or near atomic) operations that processors are built
to perform. It is possible to perform more operations on simpler data,
given that simpler data takes fewer transistors. Floating point numbers
are represented by a number of bits. More bits allows more precision.
The traditional standard is 32 bits, but modern processors can also
handle operations on smaller float representations, including 16, 8, and
4. The fewer bits, the more operations it is possible to do for a given
(heat, energy, space) budget, but the less value they have for producing
useful algorithms.

Floating point numbers are represented with a sign (positive or
negative), an exponent, and a mantissa. The represented value of the
float involves taking the mantissa to a power of 2 to the exponent.
Given a certain number of bits, it is possible to interpret those bits
in ways that allow for greater precision or a wider range.

While chip manufacturers are still primarily working along traditional
paradigms, the last few years have seen experimentation around various
precisions, with some work being directed at binary neural networks
(\href{https://arxiv.org/abs/2110.06804}{Yuan and Agaian 2023}). In
lowering precision, some amount of the value of the computation is lost,
as we see in degrading quality of models. It is increasingly common to
use different formats at different places or different times, optimizing
precision for specific use-cases. Modern AI training often uses a
format,
\href{https://cloud.google.com/blog/products/ai-machine-learning/bfloat16-the-secret-to-high-performance-on-cloud-tpus}{bf16},
that combines a 16 bit float size with a non-standard division between
the exponent and the mantissa that allows for easier conversion into 32
bit formats.

It is conceivable that future AI systems will not utilize floating point
arithmetic for most of their computational work and may use some new
paradigm of data format on specially-designed chips. Given that floating
point math wasn't originally designed for the massive parallel needs of
neural networks, it is conceivable that future AI systems, even if they
nominally remain neural networks, operate in some more efficient way.

Insofar as systems might perform computations as manipulations of
non-numerical data structures, we need to convert those manipulations
into equivalent FLOP/s. Brains, the natural comparison for estimates of
digital minds, don't use floating point arithmetic, and so comparisons
require rough translations into the equivalent of `FLOP/s' for brains.
They have other significant differences, including noisy processors,
timing-based computations, and work that is done in parallel without
being as regimented as GPUs (Tsur 2021).

We will adopt FLOP/s as our unit of measurement, setting aside the above
concerns. This means that we should take the results somewhat
tentatively; there are enough other factors that warrant caution about
the results and mistrust of the fundamental unit should not be high on
that list.

\hypertarget{projections-for-future-FLOPs-capacities}{%
\subsubsection{Projections for future FLOP/s
capacities}\label{projections-for-future-FLOPs-capacities}}

Floating point operations are among the standard atomic operations in
processor instruction sets and are a common benchmark for processor
power. Processor FLOP/s capabilities, as with
\href{https://ourworldindata.org/moores-law}{other measures} of
performance, have been increasing for decades, and the needs of AI have
pushed progress in new and productive directions.

The relevant processors to AI are (at present) primarily GPUs and TPUs.
The AI benchmarking organization Epoch has estimated a number of trends
in GPU performance over the past decade. These trends extend the
efficiency increases we've seen in processors over the last 50 years,
even as improvements in traditional CPUs have stalled. Overall FLOP/s
capabilities of GPUs have doubled approximately
\href{https://epoch.ai/data-insights/peak-performance-hardware-on-different-precisions}{every
2.5 years} for the past decade and a half, while performance per dollar
has improved about
\href{https://epoch.ai/data/machine-learning-hardware}{30\% per year}.

There are reasons to expect these trends to continue. The strong demand
for powerful AI chips is fairly new, and there hasn't been as much
exploration of the space of possibilities as we've seen in CPUs. Leading
AI models are vastly more expensive to design and run than typical
programs of the past, and might justify sacrifices of flexibility for
efficiency. The same change in focus that allowed GPUs to be more
efficient than CPUs might recur, particularly if the design of hardware
and software can progress together, rather than being reactionary.

Worldwide FLOP/s are also dependent on the number of chips produced.
Nvidia is the leader in parallel processor chip design. Sales of its
chips have increased drastically since the rise of AI systems, making it
the company with the
\href{https://companiesmarketcap.com/time-machine/2025-08-20/}{highest
market cap worldwide}.

It isn't completely clear how many chips are built each year. It is a
moving target, as new architectures ramp up and replace the production
of older architectures. Nvidia has the dominant share of the market, but
\href{https://www.businessinsider.com/nvidia-competitors}{competitors}
are
\href{https://www.businessinsider.com/amd-latest-gpu-still-lags-behind-nvidia-2024-10}{scrambling}
to find and edge. Producers are generally not completely transparent
about capabilities. Nvidia has
\href{https://ca.finance.yahoo.com/news/nvidia-ceo-says-orders-3-171501205.html}{announced
that it has sold over three and a half million} of its latest Blackwell
architecture processors, exhausting their ability to meet demand for a
year. We may estimate global capacity by taking the share produced by
Nvidia as about 80\% given its significant share of
\href{https://www.tomshardware.com/tech-industry/artificial-intelligence/nvidia-to-consume-77-percent-of-wafers-used-for-ai-processors-in-2025-report}{total
wafer production} and command over the
\href{https://technologymagazine.com/articles/how-nvidias-ai-made-it-the-worlds-most-valuable-firm}{market}.
If we take the 3.5 million chips to constitute 75\% of Nvidia's
capacity, and Nvidia to constitute 80\% of world total AI-relevant
compute capacity, then we may infer that the Blackwell sales figures
reflect about 60\% of total world output of relevant chips, or about the
equivalent of 5 million Blackwell chips. While this is a very rough
estimate, our chief uncertainties lie elsewhere. Given the capabilities
of these chips, we might estimate the current yearly supply capabilities
at something in the ballpark of 1e22 FLOP/s in peak performance.

\begin{table}
\centering
\begin{threeparttable}
\caption{Comparison of FLOP/s (4-64 bit) in cutting-edge AI chips (Hopper
architecture model H100 / Blackwell architecture, models B100 and B200)
from 2024-2025.}
\begin{tabularx}{0.7\textwidth}{XXXX}
\toprule
& H100 & B100 & B200 \\
\midrule
\textbf{FP4} & N/A & 1.4e16 & 1.8e16 \\
\textbf{FP8} & 4e15 & 7e15 & 9e15 \\
\textbf{FP16} & 2e15 & 3.5e15 & 4.5e15 \\
\textbf{TF32} & 1e15 & 1.8e15 & 2.2e15 \\
\textbf{FP64} & 6e13 & 3e13 & 4e13 \\
\textbf{Power} & 700W & 700W & 1,000W \\
\bottomrule
\end{tabularx}
\begin{tablenotes}
\item{Each row specifies the peak rate of FLOP/s of the
specified bit precision. Utilization of the peak rate depends on other
factors, such as program efficiency, data delivery, and cooling.}
\end{tablenotes}
\end{threeparttable}
\end{table}

Nvidia revenues can be treated as an approximate measure of increase in
chip sales and have increased by
\href{https://www.statista.com/chart/30077/nvidia-revenue/}{500\% over
the past two years}. Epoch estimates that the total capacity of Nvidia
to supply leading chips
\href{https://epoch.ai/data-insights/nvidia-chip-production}{has doubled
every 10 months}.
\href{https://www.macrotrends.net/stocks/charts/NVDA/nvidia/pe-ratio}{Nvidia's
P/E ratio} suggests investors project significant continued sales
growth, even as investors eye potential competitors.

It is impossible to know for sure where these trends are headed. Epoch's
estimates of training runs
\href{https://epoch.ai/blog/can-ai-scaling-continue-through-2030\#current-production-and-projections}{allow
that the compute trends continue} until 2030, suggesting no clear
immediate barriers. Machine-learning hardware is still in relative
infancy. Modern AI architectures are designed to fit the capabilities of
existing GPU architectures. GPUs are increasingly being constructed to
suit modern AI architectures. It is possible that we will see a broader
exploration of hardware and software architectures designed without
existing constraints in mind, and this will lead to breakthroughs in
performance. It doesn't seem impossible that we could see progress for
at least another decade.

Power might eventually become a bottleneck to FLOP/s capacity,
particularly in terms of providing sufficient power at the location of
data centers. Current world power estimates are around
\href{https://ourworldindata.org/energy-production-consumption}{30,000
terawatt hours per year}. At an energy demand of 700 watts, this could
power billions of Nvidia's Blackwell 100 processors (ignoring related
power costs, such as for cooling.) If 1\% of total world power went to
AI, and energy efficiency weren't increased, this would cap FLOP/s at
about 1e23.

However, not only is energy efficiency also increasing at around
\href{https://epoch.ai/data-insights/ml-hardware-energy-efficiency}{40\%
per year}, but energy capacity is expanding as well, and if demand
\href{https://ourworldindata.org/energy-production-consumption}{continues
to rise}, producers may be more incentivized to try to meet the need.
Average electricity costs are close to
\href{https://www.globalpetrolprices.com/electricity_prices/}{\$0.2 per
Kwh}. One percent of total production at this price would cost less than
\$100 billion. This does not seem like a prohibitive amount in and of
itself. (Yearly software development costs currently are
\href{https://www.wipo.int/en/web/global-innovation-index/w/blogs/2025/global-software-spending}{estimated
at around 700 billion}, and 100 billion approximates the amount spent on
cutting-edge hardware in 2025.) Distributing the power to where it is
needed may be a challenge in the near-term, but seems less likely to be
the critical bottleneck in the long-term.

If AI was sufficiently lucrative, we might expect to ramp up power
supply and delivery to meet demand in the range of 1e23 to 1e24 FLOP/s,
even without much more efficient processors. But if energy becomes a
significant limiter, then we can expect research to focus on reducing
energy consumption. It doesn't seem obvious that power should be a major
limiter until we get above those numbers.

Below, we present a range of possible scenarios for compute. These
scenarios differ about the amount of time we see production increase and
FLOP price efficiency trends continue. No single scenario should be
afforded much weight. Rather, we think reasonable expectations should be
covered by the diversity of results.

That said, there could be significant changes that lead even the most
optimistic projections to be overly conservative. The numbers are based
on the assumption that hardware advances incrementally. It is not
obvious that this should be the case. It is possible that we see changes
in hardware and software that allows for very different, perhaps much
more efficient, approaches. This might lead to a non-linear jump in
available computing power.

It is also possible that we see AI-driven acceleration in hardware
advancements. The following scenarios should therefore be considered
lower bounds that don't incorporate the more uncertain possibilities for
advancement.

\textbf{Production Capacity Scenarios:}

Highly conservative. This scenario supposes that current chip production
figures remain constant and companies neither design better chips, lower
prices, or raise production quantities. Increases in global compute will
continue from the accumulation of aging processors. Top of the line GPUs
have a life of 5-8 years, which we may roughly estimate to entail a $\frac{1}{3}$
drop off in aging systems every year beyond the fifth year. This means
we could increase the total available amount of compute up to a point in
which it is \textasciitilde7x current yearly production levels.

Currently, global production may be in the equivalent of five million
Blackwell chips. In this scenario, the total compute worldwide budget
would level off around 2033 at roughly 1e23 FLOP/s.

This scenario is extremely conservative: short of some significant
event, we should be highly confident that companies will both continue
to improve chips and raise sales numbers. Even if no further advances
were possible, we should expect increased competition, economies of
scale, and invested start-up costs to raise the price efficiency of
compute.

\textbf{Rather conservative.} This scenario supposes that FLOP/s scaling progress
immediately slows down, such that capabilities per dollar double once
over the next 5 years and once more over the following decade. This
assumes a rapid drop off in efficiency trends: even without advances in
design we should expect chips to become cheaper over time, allowing each
dollar to buy more compute. It assumes that production capacities
(measured in the ability to meet demand expenditures) increase to 3x
what they currently are over the next half decade. By 2050, the total
available compute budget would level off around roughly 2e24 FLOP/s.

\textbf{Moderate.} This scenario supposes that compute per dollar doubles thrice
in the next 8 years and then doubles once more over the following
half-decade. This means we are more than halfway through the period of
rapid GPU scaling. It also supposes production capacities increase to
10x what they currently are by 2032 (measured in the ability to meet
demand expenditure). This places global chip expenditures at about \$1
trillion. Then by 2050, the total available compute budget would have
levelled off around 3e25 FLOP/s.

\textbf{Rather Optimistic.} This scenario supposes that scaling progress
increases for another 15 years, such that compute of top of the line
processors doubles six times, for a total of 64x. Furthermore,
production capacities increase to 25x what they currently are over the
next decade and then levels off. This would put yearly expenditures at
about \$2.5 trillion, a bit less than the
\href{https://www.statista.com/outlook/mmo/passenger-cars/worldwide}{current
global expenditures on passenger vehicles}. By 2045, the total available
compute budget would level off around 4e26 FLOP/s.

\textbf{Highly Optimistic.} Suppose that scaling progress speeds up for about six years such that
processing efficiency doubles four times over the next six years, then
continues at the present pace, doubling another five times over the
following ten years, for a total of 512x\footnote{This is a very high
  number, but well below
  \href{https://www.forethought.org/research/how-far-can-ai-progress-before-hitting-effective-physical-limits\#chip-technology-progress}{theoretical
  limits} under different paradigms.}. Over the same period, suppose
that total production-capacities increase to 50x what they currently
are. This would put yearly expenditures on AI chips at about \$5
trillion, a rather high amount\footnote{High, but not totally out of
  line with
  \href{https://www.nasdaq.com/articles/where-will-nvidia-stock-be-10-years-7\#:~:text=Now,\%20we\%20have\%20seen\%20that,in\%20sales\%20after\%2010\%20years.}{some}
  \href{https://www.forbes.com/sites/bethkindig/2024/06/07/prediction-nvidia-stock-will-reach-10-trillion-market-cap-by-2030/}{optimistic}
  \href{https://www.rolandberger.com/en/Insights/Publications/GenAI-hardware.html}{projections}.},
roughly equivalent to
\href{https://www.gartner.com/en/newsroom/press-releases/2025-07-15-gartner-forecasts-worldwide-it-spending-to-grow-7-point-9-percent-in-2025}{current
worldwide spending on IT}. If AI proves effective at replacing human
cognitive labor, this could be an underestimate. Then by 2049, the total
available compute budget would level off around 3e27 FLOP/s.

\begin{table}
\centering
\begin{threeparttable}
\caption{Compute projections by scenario.}
\begin{longtable}[]{@{}
  >{\raggedright\arraybackslash}p{(\columnwidth - 10\tabcolsep) * \real{0.1667}}
  >{\raggedright\arraybackslash}p{(\columnwidth - 10\tabcolsep) * \real{0.1667}}
  >{\raggedright\arraybackslash}p{(\columnwidth - 10\tabcolsep) * \real{0.1667}}
  >{\raggedright\arraybackslash}p{(\columnwidth - 10\tabcolsep) * \real{0.1667}}
  >{\raggedright\arraybackslash}p{(\columnwidth - 10\tabcolsep) * \real{0.1667}}
  >{\raggedright\arraybackslash}p{(\columnwidth - 10\tabcolsep) * \real{0.1667}}@{}}
\toprule\noalign{}
\begin{minipage}[b]{\linewidth}\raggedright
\end{minipage} & \begin{minipage}[b]{\linewidth}\raggedright
Highly Conservative
\end{minipage} & \begin{minipage}[b]{\linewidth}\raggedright
Rather Conservative
\end{minipage} & \begin{minipage}[b]{\linewidth}\raggedright
Moderate
\end{minipage} & \begin{minipage}[b]{\linewidth}\raggedright
Rather Optimistic
\end{minipage} & \begin{minipage}[b]{\linewidth}\raggedright
Highly Optimistic
\end{minipage} \\
\midrule\noalign{}
\endhead
\endlastfoot
\textbf{2025} & 1.30e22\footnotemark[1] & 1.30e22 & 1.30e22 & 1.30e22 &
1.30e22 \\
\textbf{2030} & 7.80e22 & 2.67e23 & 6.16e23 & 7.11e23 & 1.06e24 \\
\textbf{2035} & 1.01e23 & 8.25e23 & 7.45e24 & 1.93e25 & 2.73e25 \\
\textbf{2040} & 1.03e23 & 1.44e24 & 2.32e25 & 1.58e26 & 5.03e26 \\
\textbf{2045} & 1.03e23 & 1.89e24 & 3.13e25 & 3.49e26 & 2.37e27 \\
\textbf{2050} & 1.03e23 & 1.99e24 & 3.25e25 & 3.97e26 & 3.11e27 \\
\bottomrule\noalign{}
\end{longtable}
\begin{tablenotes}
\item[1]{Starting figures do not reflect the
  buildup of compute capacity over the past several years. Given the
  rate at which the GPU market has been expanding, this is not likely to
  radically change the results.} 
\vskip0.25cm
\item{Numbers of accumulated global relevant FLOP/s capacity by year,
across scenarios. The results differ more over time, as efficiency and
production trends have more time to diverge.}
\end{tablenotes}
\end{threeparttable}
\end{table}

\hypertarget{fraction-of-compute-dedicated-to-digital-minds}{%
\subsubsection{Fraction of compute dedicated to digital
minds}\label{fraction-of-compute-dedicated-to-digital-minds}}

The fraction of compute dedicated to digital minds is highly dependent
on their value and of alternative uses of computation. It seems likely
that most existing compute capacity will not be utilized for any
purpose, let alone for digital minds. The numbers used for calculating
total capacity assume peak optimization, which is
unrealistic\footnote{While peak optimization is unrealistic now, we
  might expect that if other forms of progress stall, we will see better
  configuration of GPUs to the specific use cases they face, allowing
  for better utilization of the stagnant peak FLOP/s per \$.}. Perhaps
processors will be plentiful and having them ready to take on specific
periodic tasks is worth keeping them generally waiting for tasks, either
for seconds at a time or for years. Perhaps most processors will be used
to power autonomous drones for military applications that sit idle in
warehouses, or every individual will have their own cutting-edge chips
in their smart phones that they use only once every five or ten minutes.

If compute capacity is utilized, it may not be used for the purposes of
powering digital minds. A surplus of compute may go toward powering
complicated models that have no claim to mindedness. Perhaps it will all
go to research, and the best AI systems for research looks nothing like
a person. Perhaps digital minds will provide a primarily integrative
service, corralling varieties of specialized tools that do the bulk of
the required work.

Given our present ignorance, it only seems fair to assign a very broad
range of possible utilization rates, ranging from 0.001\% to 10\% of
total available peak compute. The boundaries of this range are supposed
to reflect a high degree of uncertainty, but aren't particularly
principled. 10\% might be reasonable if it turns out that replicating
human brain functionality is one of the best ways to produce AGI, either
by training on human data or building human-like architectures. It also
assumes that we're generally able to make the most of the peak FLOP
output the global hardware is capable of, which is very optimistic.
0.001\% might be reasonable if digital minds are a somewhat niche use,
perhaps as systems that oversee more specialized tools that require a
lot more compute. A value of 0.001 to 0.1\% seems most plausible as a
most-likely range and where we fall in that range may depend upon the
total amount of compute available. (Larger amounts of compute may make
us less careful about utilizing it effectively, and may correlate with
significant uses beyond digital minds.)

Combined with a total capacity range of 2e23 to 3e27 FLOP/s, this means
we can expect 1e18 to 3e26 FLOP/s to be invested in powering digital
minds. This is a wide range, but still suggests a potential scale to
capture many of the needs for which we might want digital minds.

\hypertarget{compute-benchmarks-for-digital-minds}{%
\subsubsection{Compute benchmarks for digital
minds}\label{compute-benchmarks-for-digital-minds}}

Suppose that we knew that the AI systems powering digital minds would
process an average of 1e24 FLOP/s in 2045. What would that mean for the
scale of digital minds? It is possible that all of this compute could be
devoted to running just a single very complicated mind. Alternatively,
it could be used to run a vast number of very simple minds. To know
which of these alternatives is more plausible, we might look to the uses
to which digital minds would be put. Research might favor a smaller
number of more sophisticated minds, for instance. Virtual friends, in
contrast, might be comparatively cheap. In the present section, we won't
examine the details, but instead focus on several significant generic
possibilities. Understanding the relative plausibility of these
possibilities will require us to confront the needs and advantages of
different sized minds.

Some past speculation
(\href{https://www.cambridge.org/core/journals/utilitas/article/abs/astronomical-waste-the-opportunity-cost-of-delayed-technological-development/2969D64410332BD099F36BAFC5B2ADE5}{Bostrom
2003};
\href{https://joecarlsmith.com/2025/05/21/the-stakes-of-ai-moral-status}{Carlsmith
2025}) about the potential scale of digital minds has divided total
compute by some benchmark to get an answer. The result might tell us
something about the number of minds of different compute scales we could
operate. If we knew that the digital minds we would see would require
1e13 FLOP/s for a seconds worth of operation, then we could derive that
1e24 FLOP/s would be sufficient for 1e11 simultaneous minds. In the
absence of any better strategy, we will follow this approach. The trick
will be to find the right units: very different numbers are plausible,
and they lead to very different conclusions.

This section will survey several categories of benchmarks: human brains,
animal brains, and contemporary AI models. In the next section, we will
apply those numbers to assess the implications of the compute
projections discussed previously.

\typetitle{Human Brains}

The most natural approach to a benchmark unit for estimating the number
of possible digital minds is to look to the human mind. If we could
assess the number of FLOP/s involved in operating a human brain, we
would know something about how many FLOP/s are sufficient for
human-level mindedness.

There are complications to a human benchmark: our brains don't perform
arithmetic operations on floating point numbers, so any estimate of
their FLOP/s capabilities will have to do some translation. This
translation is complicated by the fact that human brains operate on very
different principles from computers, even neural networks.

The computations performed by individual neurons are messy. Factors such
as the exact timing of signals can matter, or the chemical environment
in which a neuron operates can matter. They are also noisy: neurons
don't operate in accordance with simple rules. There is some element of
chance to them. It would probably be a mistake to think of brains as
somewhat defective computers. Though the complexities under which they
operate would be a headache for programmers to incorporate, natural
selection may have allowed our brains to make the most of their
limitations. The operations they perform therefore aren't easily
intertranslatable.

While it may not be possible to precisely translate the compute
efficiency of brains, we can try to set some numbers. We might think of
each neuron (or each neural connection) as performing a single FLOP
every time it could fire. Even if this weren't completely reliable, it
might still provide an illuminating order of magnitude. Or we might try
to figure out how many FLOP/s we need to accurately model brain dynamics
at a cellular level. There are a range of options and the complexity
required to evaluate them is worth a report of its own. Here, we will
defer to the work of Joe Carlsmith
(\href{https://www.openphilanthropy.org/research/how-much-computational-power-does-it-take-to-match-the-human-brain/}{2020}),
who surveyed a variety of existing proposals and settled on a preferred
estimate of 1e15 FLOP/s for the amount of FLOP/s needed to replicate
brain-like competence in a brain-like way.

\typetitle{Alternative brains}

Human brains contain about 86 billion neurons and trillions of neural
connections. They do more than enable mindedness. Large parts of our
brains work to maintain our bodies and organize complex sensory and
motor processing. The cerebellum, for instance, includes a large
fraction of our total neurons, but is largely irrelevant to mindedness.
Our cerebral cortex is clearly involved in nearly all aspects of our
mental lives, but people born with a single hemisphere (half of the
total cortical volume) are able to function normally. Plausibly, we
could get along as persons pretty well with just a relatively small
number of our total neurons, so long as those neurons were carefully
chosen. (Losing something instead in coordination, perception,
body-regulation, etc.)

This suggests that estimates based on the total brain may be mistaken.
Instead of estimating the compute efficiency of the human mind, we might
instead try to estimate the percentage of a human brain that is involved
in the basic requirements of mindedness. The result might cut the size
down by an order of magnitude -- possibly more. This might give us a
figure more like 1e13 or 1e14 FLOP/s.

Alternatively, we might look to use the brains of another animal as a
guide to the requirements of simpler minds. If we looked at mice, we
might scale down our estimates of humans based on the number of neurons
or the number of synapses. A mouse has about 1/1000 as many neurons as a
human
(\href{https://www.frontiersin.org/journals/neuroinformatics/articles/10.3389/fninf.2018.00084/full}{70
million}), so if we estimated humans at 1e15 FLOP/s equivalent, we might
simply divide our estimate by 1000, and get 1e12 FLOP/s equivalent.

\typetitle{AI models}

However, future AI systems are not likely to be based on biological
models. It makes more sense to look at contemporary AI models for a
guide to the requirements of mindedness. Though current LLMs are not
strongly minded, the barriers don't seem to be a lack of complexity.

Even small LLMs possess sophisticated forms of human intelligence,
capable of many tasks that would flummox even the most intelligent of
non-human animals. They lack agency, decision-making processes, and
temporal coherence. They may be performing the wrong calculations,
role-playing rather than embodying. These deficiencies may result more
from their training regimes and architecture than from their
computational efficiency. It is rather surprising how much very small
models can do, and they are surely quite far from being optimally
designed and trained.\footnote{Modern machine learning remains a fairly
  young discipline and one that has been experimentally constrained by
  available hardware. Transformers, the architecture behind LLMs, were
  designed with GPUs in mind (Vaswani 2017). GPUs have in turn been
  tweaked to better run LLMs. More speculative alternatives require
  robust expenditures for hardware that have no obvious payoff. As the
  space matures, we might expect to find a pairing of software and
  hardware.}

Small models have also improved over time. Even without major
architectural changes,
\href{https://epoch.ai/blog/algorithmic-progress-in-language-models}{we
have seen significant improvements in `algorithmic progress'}, in which
the same hardware can be used better over time. If we think current
models are much weaker than ideal models of their current size, it
becomes more plausible that they will be of a size fully sufficient for
running digital minds.

If we took
\href{https://huggingface.co/meta-llama/Llama-3.1-405B}{Llama-3
405B}\footnote{Cutting edge commercial LLMs do not have their details
  published, making them unsuitable for this purpose.} to provide our
unit on the grounds that something as complex as that could be minded if
it were built right, then we might use the amount of FLOP/s to perform
forward passes on a sentence of 100 tokens as the equivalent of a mind
for a second, which we might estimate at about 8e14 FLOP/s.\footnote{This
  report uses the FLOPs/forward pass approximation as 2 × number of
  parameters (\href{https://arxiv.org/pdf/2001.08361}{Kaplan et
  al.~2020}).}

Alternatively, we might think that Llama-3 405B is far from optimized.
The \href{https://huggingface.co/meta-llama/Llama-2-7b-chat}{Llama-2 7B}
model provides a significant level of intelligence in a much smaller
package. Perhaps it is the right benchmark for the number of digital
minds. This makes particular sense if we expect to see continued
significant improvements in model efficiency at this size and if we
account for advances in tool use and database integration. A
mixture-of-experts approach could mean that models with large numbers of
total parameters only need to make use of a relatively small number at a
time, and combined with the right tools and context-relevant
presumptions, might be able to do quite a lot with relatively few FLOPs.
A Llama-2 7B benchmark leaves us with about 1.4e12 FLOP/s for a forward
pass over 100 tokens.

\begin{table}
  \centering
  \caption{FLOP/s approximations for various systems}
\begin{tabular}{lr}
  & Estimated FLOP/s \\
\midrule
\textbf{Human (Full Brain)} & 1e15 \\
\textbf{Human (Cognitive Brain)} & 1e14 \\
\textbf{Mouse (Full Brain)} & 7e13 \\
\textbf{Llama-3 405B (100 tokens)} & 8e14 \\
\textbf{Llama-2 7B (100 tokens)} & 1.4e12 \\
\bottomrule
\end{tabular}
\end{table}

\hypertarget{implications}{%
\subsubsection{Implications}\label{implications}}

Taken together, these considerations suggest a range of answers about
our future capacity for digital minds. Some of these answers --
estimates of the number of minds possible according to a chosen
benchmark given a fraction of compute under each scenario -- are plotted
in the following charts.

\definecolor{lightgray}{gray}{0.95}
\begin{table}
\caption{Human Brain Benchmark (1e15)}
\begin{longtable}{p{3.5cm}p{1.8cm}p{1.8cm}p{1.8cm}p{1.8cm}p{1.8cm}}
\toprule
\textbf{Scenario} & \textbf{0.001\%} & \textbf{0.01\%} & \textbf{0.10\%} & \textbf{1\%} & \textbf{10\%} \\
\midrule[1.5pt]
\endfirsthead

\toprule
\textbf{Scenario} & \textbf{0.001\%} & \textbf{0.01\%} & \textbf{0.10\%} & \textbf{1\%} & \textbf{10\%} \\
\midrule[1.5pt]
\endhead

\midrule
\multicolumn{6}{r}{\textit{Continued on next page}} \\
\endfoot

\bottomrule
\endlastfoot

\multicolumn{6}{l}{\textbf{2030}} \\
\midrule
\rowcolor{white}
Highly conservative & 7.80e02 & 7.80e03 & 7.80e04 & 7.80e05 & 7.80e06 \\
\rowcolor{lightgray}
Rather conservative & 2.67e03 & 2.67e04 & 2.67e05 & 2.67e06 & 2.67e07 \\
\rowcolor{white}
Moderate & 6.16e03 & 6.16e04 & 6.16e05 & 6.16e06 & 6.16e07 \\
\rowcolor{lightgray}
Rather Optimistic & 7.11e03 & 7.11e04 & 7.11e05 & 7.11e06 & 7.11e07 \\
\rowcolor{white}
Highly Optimistic & 1.06e04 & 1.06e05 & 1.06e06 & 1.06e07 & 1.06e08 \\
\midrule[1.5pt]

\multicolumn{6}{l}{\textbf{2035}} \\
\midrule
\rowcolor{white}
Highly conservative & 1.01e03 & 1.01e04 & 1.01e05 & 1.01e06 & 1.01e07 \\
\rowcolor{lightgray}
Rather conservative & 8.25e03 & 8.25e04 & 8.25e05 & 8.25e06 & 8.25e07 \\
\rowcolor{white}
Moderate & 7.45e04 & 7.45e05 & 7.45e06 & 7.45e07 & 7.45e08 \\
\rowcolor{lightgray}
Rather Optimistic & 1.93e05 & 1.93e06 & 1.93e07 & 1.93e08 & 1.93e09 \\
\rowcolor{white}
Highly Optimistic & 2.73e05 & 2.73e06 & 2.73e07 & 2.73e08 & 2.73e09 \\
\midrule[1.5pt]

\multicolumn{6}{l}{\textbf{2040}} \\
\midrule
\rowcolor{white}
Highly conservative & 1.03e03 & 1.03e04 & 1.03e05 & 1.03e06 & 1.03e07 \\
\rowcolor{lightgray}
Rather conservative & 1.44e04 & 1.44e05 & 1.44e06 & 1.44e07 & 1.44e08 \\
\rowcolor{white}
Moderate & 2.32e05 & 2.32e06 & 2.32e07 & 2.32e08 & 2.32e09 \\
\rowcolor{lightgray}
Rather Optimistic & 1.58e06 & 1.58e07 & 1.58e08 & 1.58e09 & 1.58e10 \\
\rowcolor{white}
Highly Optimistic & 5.03e06 & 5.03e07 & 5.03e08 & 5.03e09 & 5.03e10 \\
\midrule[1.5pt]

\multicolumn{6}{l}{\textbf{2045}} \\
\midrule
\rowcolor{white}
Highly conservative & 1.03e03 & 1.03e04 & 1.03e05 & 1.03e06 & 1.03e07 \\
\rowcolor{lightgray}
Rather conservative & 1.89e04 & 1.89e05 & 1.89e06 & 1.89e07 & 1.89e08 \\
\rowcolor{white}
Moderate & 3.13e05 & 3.13e06 & 3.13e07 & 3.13e08 & 3.13e09 \\
\rowcolor{lightgray}
Rather Optimistic & 3.49e06 & 3.49e07 & 3.49e08 & 3.49e09 & 3.49e10 \\
\rowcolor{white}
Highly Optimistic & 2.37e07 & 2.37e08 & 2.37e09 & 2.37e10 & 2.37e11 \\
\midrule[1.5pt]

\multicolumn{6}{l}{\textbf{2050}} \\
\midrule
\rowcolor{white}
Highly conservative & 1.03e03 & 1.03e04 & 1.03e05 & 1.03e06 & 1.03e07 \\
\rowcolor{lightgray}
Rather conservative & 1.99e04 & 1.99e05 & 1.99e06 & 1.99e07 & 1.99e08 \\
\rowcolor{white}
Moderate & 3.25e05 & 3.25e06 & 3.25e07 & 3.25e08 & 3.25e09 \\
\rowcolor{lightgray}
Rather Optimistic & 3.97e06 & 3.97e07 & 3.97e08 & 3.97e09 & 3.97e10 \\
\rowcolor{white}
Highly Optimistic & 3.11e07 & 3.11e08 & 3.11e09 & 3.11e10 & 3.11e11 \\
\end{longtable}
\end{table}

\vspace{1cm}

\begin{table}
\caption{LLaMMA-2 7B Benchmark (1.4e12)}
\begin{longtable}{p{3.5cm}p{1.8cm}p{1.8cm}p{1.8cm}p{1.8cm}p{1.8cm}}
\toprule
\textbf{Scenario} & \textbf{0.001\%} & \textbf{0.01\%} & \textbf{0.10\%} & \textbf{1\%} & \textbf{10\%} \\
\midrule[1.5pt]
\endfirsthead

\toprule
\textbf{Scenario} & \textbf{0.001\%} & \textbf{0.01\%} & \textbf{0.10\%} & \textbf{1\%} & \textbf{10\%} \\
\midrule[1.5pt]
\endhead

\midrule
\multicolumn{6}{r}{\textit{Continued on next page}} \\
\endfoot

\bottomrule
\endlastfoot

\multicolumn{6}{l}{\textbf{2030}} \\
\midrule
\rowcolor{white}
Highly conservative & 5.57e05 & 5.57e06 & 5.57e07 & 5.57e08 & 5.57e09 \\
\rowcolor{lightgray}
Rather conservative & 1.91e06 & 1.91e07 & 1.91e08 & 1.91e09 & 1.91e10 \\
\rowcolor{white}
Moderate & 4.40e06 & 4.40e07 & 4.40e08 & 4.40e09 & 4.40e10 \\
\rowcolor{lightgray}
Rather Optimistic & 5.08e06 & 5.08e07 & 5.08e08 & 5.08e09 & 5.08e10 \\
\rowcolor{white}
Highly Optimistic & 7.57e06 & 7.57e07 & 7.57e08 & 7.57e09 & 7.57e10 \\
\midrule[1.5pt]

\multicolumn{6}{l}{\textbf{2035}} \\
\midrule
\rowcolor{white}
Highly conservative & 7.21e05 & 7.21e06 & 7.21e07 & 7.21e08 & 7.21e09 \\
\rowcolor{lightgray}
Rather conservative & 5.89e06 & 5.89e07 & 5.89e08 & 5.89e09 & 5.89e10 \\
\rowcolor{white}
Moderate & 5.32e07 & 5.32e08 & 5.32e09 & 5.32e10 & 5.32e11 \\
\rowcolor{lightgray}
Rather Optimistic & 1.38e08 & 1.38e09 & 1.38e10 & 1.38e11 & 1.38e12 \\
\rowcolor{white}
Highly Optimistic & 1.95e08 & 1.95e09 & 1.95e10 & 1.95e11 & 1.95e12 \\
\midrule[1.5pt]

\multicolumn{6}{l}{\textbf{2040}} \\
\midrule
\rowcolor{white}
Highly conservative & 7.36e05 & 7.36e06 & 7.36e07 & 7.36e08 & 7.36e09 \\
\rowcolor{lightgray}
Rather conservative & 1.03e07 & 1.03e08 & 1.03e09 & 1.03e10 & 1.03e11 \\
\rowcolor{white}
Moderate & 1.66e08 & 1.66e09 & 1.66e10 & 1.66e11 & 1.66e12 \\
\rowcolor{lightgray}
Rather Optimistic & 1.13e09 & 1.13e10 & 1.13e11 & 1.13e12 & 1.13e13 \\
\rowcolor{white}
Highly Optimistic & 3.59e09 & 3.59e10 & 3.59e11 & 3.59e12 & 3.59e13 \\
\midrule[1.5pt]
\multicolumn{6}{l}{\textbf{2045}} \\
\midrule
\rowcolor{white}
Highly conservative & 7.36e05 & 7.36e06 & 7.36e07 & 7.36e08 & 7.36e09 \\
\rowcolor{lightgray}
Rather conservative & 1.35e07 & 1.35e08 & 1.35e09 & 1.35e10 & 1.35e11 \\
\rowcolor{white}
Moderate & 2.24e08 & 2.24e09 & 2.24e10 & 2.24e11 & 2.24e12 \\
\rowcolor{lightgray}
Rather Optimistic & 2.49e09 & 2.49e10 & 2.49e11 & 2.49e12 & 2.49e13 \\
\rowcolor{white}
Highly Optimistic & 1.69e10 & 1.69e11 & 1.69e12 & 1.69e13 & 1.69e14 \\
\midrule[1.5pt]

\multicolumn{6}{l}{\textbf{2050}} \\
\midrule
\rowcolor{white}
Highly conservative & 7.36e05 & 7.36e06 & 7.36e07 & 7.36e08 & 7.36e09 \\
\rowcolor{lightgray}
Rather conservative & 1.42e07 & 1.42e08 & 1.42e09 & 1.42e10 & 1.42e11 \\
\rowcolor{white}
Moderate & 2.32e08 & 2.32e09 & 2.32e10 & 2.32e11 & 2.32e12 \\
\rowcolor{lightgray}
Rather Optimistic & 2.84e09 & 2.84e10 & 2.84e11 & 2.84e12 & 2.84e13 \\
\rowcolor{white}
Highly Optimistic & 2.22e10 & 2.22e11 & 2.22e12 & 2.22e13 & 2.22e14 \\

\end{longtable}
\end{table}
These charts depict a range of results, conveying something of the
variety of scenarios we might see play out. The intermediate numbers,
based on the moderate scenario combined with a 0.01\% or 0.1\%
allocation to digital minds, and evaluated against a benchmark based on
a small contemporary LLM or a full human brain, would suggest something
between 6e4 (tens of thousands) minds and 4e8 (hundreds of millions)
minds by 2030, and between 3e6 (millions) minds and 2e10 (tens of
billions) minds by 2050.

It is important to remember that the numbers reflect the capacity of
minds that could be active at a given moment, and most persistent AI
entities might need to be active only every now and then. Most uses of
AI, such as those surveyed in the first approach, could be used much
more sporadically, or could rely on much simpler models except when
complexity warrants deeper focus.\footnote{The mainstream technique of
  speculative decoding pairs larger and smaller models, with the larger
  models verifying the work of smaller models. In the future, we might
  imagine more work being done by smaller models, with simpler tests for
  when verification might be needed.} With high-end estimates in the
hundreds of billions by 2050 and assuming a large amount of down-time,
this could be handled in the moderate scenario with only a 0.1\%
allocation of compute to digital minds.

Furthermore, both the percentage of compute that is devoted to digital
minds and the utilization rate of chips across the globe are flexible
numbers. Insofar as a resulting figure doesn't meet demand, we should
expect economic pressures both to increase the number of chips and to
improve their utilization rate. In a world in which we see high-end
results, including 10s of billions of virtual employees, it does not
seem quite so implausible that we will be willing to spend a significant
(e.g.~\$5 trillion dollars) on processors each year.

However, this might also mean that we continue to rely on concentration
in data centers that can make continuous use of the chips they have.

\newpage
\hypertarget{conclusion}{%
\section{Conclusion}\label{conclusion}}

This report represents an early attempt to estimate the potential scale
of digital minds over the coming decades. The analysis should be
understood as highly preliminary and speculative, reliant on educated
guesses, given the nascent state of relevant technologies and the
fundamental uncertainties about how AI development will unfold.

\hypertarget{tentative-findings}{%
\subsection{Tentative Findings}\label{tentative-findings}}

Two independent analytical approaches were employed to bracket possible
outcomes. A consumer-based analysis examining potential use cases
suggests digital mind populations could range from nearly none to
upwards of billions by 2050, with median estimates in the low billions.
A compute-based analysis of computational capacity suggests hardware
constraints are unlikely to limit populations at these scales. However,
these convergent results should not be interpreted as confident
predictions---the methodology necessarily relies on numerous speculative
assumptions about technological progress, market dynamics, and social
acceptance.

The confidence intervals span several orders of magnitude, reflecting
genuine uncertainty about fundamental questions that will determine
actual outcomes. Key uncertainties include whether AI systems will need
human-like characteristics to perform valuable functions, how society
will respond to person-like AI systems, and what regulatory frameworks
may emerge.

\hypertarget{major-limitations}{%
\subsection{Major Limitations}\label{major-limitations}}

Several critical limitations constrain the reliability of these
estimates:

\textbf{Technological uncertainty} pervades every aspect of the
analysis. Current AI systems lack most characteristics associated with
digital minds, and whether future systems will develop them remains
unclear.

\textbf{Market and social dynamics} are notoriously difficult to
predict, particularly for technologies that don't yet exist. Consumer
acceptance of human-like AI, regulatory responses, and competitive
dynamics between minded and non-minded AI approaches could drastically
alter adoption patterns.

\textbf{Definitional challenges} complicate the analysis. The
operational definition of ``digital minds'' used here focuses on
observable traits rather than genuine consciousness or moral status,
which may prove inadequate as AI systems become more sophisticated.

\textbf{Methodological constraints} limit the analysis's scope. The
approaches used cannot account for potential discontinuous
breakthroughs, complex interactions between different applications, or
the possibility that digital minds may emerge in forms not anticipated
by current frameworks. Its input values reflect the biases and
misconceptions of its author.

\hypertarget{implications-of-uncertainty}{%
\subsection{Implications of
Uncertainty}\label{implications-of-uncertainty}}

The wide uncertainty ranges have important implications for how these
findings should be interpreted and used:

Rather than providing specific predictions, the analysis primarily
demonstrates that digital minds could plausibly become a significant
phenomenon within decades under certain conditions. The possibility of
rapid scaling from minimal current populations to millions or billions
of systems suggests the transition could occur faster than institutions
typically adapt to technological change.

The uncertainty itself may be as important as the central estimates.
Policymakers and researchers cannot assume digital minds will remain
negligible, but neither can they confidently plan for specific
population levels or deployment patterns.

\section{Acknowledgements}

This report was supported by a grant from the Navigation Fund. It benefitted greatly from input and research support by Noah Birnbaum, and comments from Lucius Caviola, Brad Saad, Oscar Delaney, Hayley Clatterbuck, and David Moss.

\hypertarget{bibliography}{%
\section{Bibliography}\label{bibliography}}

Bernardi, J. (2025) Friends for sale: The rise and risks of AI
companions. Ada Lovelace Institute
\url{https://www.adalovelaceinstitute.org/blog/ai-companions/}.

Birch, J. (2024). The Edge of Sentience: Risk and Precaution in Humans,
Other Animals, and AI. Oxford University Press.

Bostrom, N. (2003). Astronomical waste: The opportunity cost of delayed
technological development. Utilitas, 15(3), 308-314.

Carlsmith, J. (2020, September 10). How much computational power does it
take to match the human brain? Open Philanthropy.
\url{https://www.openphilanthropy.org/research/how-much-computational-power-does-it-take-to-match-the-human-brain/}

Carlsmith, J. (2025, May 21). The stakes of AI moral status. Joe
Carlsmith.
\url{https://joecarlsmith.com/2025/05/21/the-stakes-of-ai-moral-status}

Caviola, L. (2025). The societal response to potentially sentient AI.
arXiv preprint arXiv:2502.00388.

Caviola, L. \& Saad, B. (2025). Expert forecasts about digital minds.

Chalmers, D. J. (2025). What we talk to when we talk to language models.

Digital Minds Report.
\url{https://digitalminds.report/forecasting-2025/}

Dreksler, N., Caviola, L., Chalmers, D., Allen, C., Rand, A., Lewis, J.,
\ldots{} \& Sebo, J. (2025). Subjective experience in AI systems: what
do AI researchers and the public believe?. arXiv preprint
arXiv:2506.11945.

Dung, L. (2024). Preserving the normative significance of sentience.
Journal of Consciousness Studies, 31(1-2), 8-30.

Dung, L. (2025a). How to deal with risks of AI suffering. Inquiry, 68(7), 2281-2309.

Dung, L. (2025b). Understanding artificial agency. The Philosophical
Quarterly, 75(2), 450-472.

Goldstein, S., \& Kirk-Giannini, C. D. (2025). AI wellbeing. Asian
Journal of Philosophy, 4(1), 25.

Hanson, R. (2016). The Age of Em: Work, Love, and Life When Robots Rule
the Earth. Oxford University Press.

Kaplan, J., McCandlish, S., Henighan, T., Brown, T. B., Chess, B.,
Child, R., \ldots{} \& Amodei, D. (2020). Scaling laws for neural
language models. arXiv preprint arXiv:2001.08361.

Kokotajlo, D., Alexander, S., Larsen, T., Lifland, E., \& Dean, R.
(2025, April 3). AI 2027. AI Futures.
\href{http://AI-2027.com}{AI-2027.com}

Ladak, A., \& Caviola, L. (2025). Digital sentience skepticism.
PsyArXiv.

Levy, N. (2014). The value of consciousness. Journal of Consciousness
Studies, 21(1-2), 127-138.

Long, R., Sebo, J., Butlin, P., Finlinson, K., Fish, K., Harding, J.,
\ldots{} \& Chalmers, D. (2024). Taking AI welfare seriously. arXiv
preprint arXiv:2411.00986.

Lott, M., \& Hasselberger, W. (2025). With friends like these: Love and
friendship with AI agents. Topoi, 1-13.

Metaculus. (2020, August 23). When will the first general AI system be
devised, tested, and publicly announced? Metaculus.
\url{https://www.metaculus.com/questions/5121/date-of-artificial-general-intelligence/}

Metaculus. (2023, April 30). When will a reliable and general household
robot be developed? \emph{Metaculus}.
\url{https://www.metaculus.com/questions/16625/date-of-reliable-and-general-household-robots/}

Metzinger, T. (2021). Artificial suffering: An argument for a global
moratorium on synthetic phenomenology. Journal of Artificial
Intelligence and Consciousness, 8(01), 43-66.

Potter, B.. (2025, April 24). Robot dexterity still seems hard.
\emph{Construction Physics}.
\url{https://www.construction-physics.com/p/robot-dexterity-still-seems-hard}

Moret, A. (2025). AI welfare risks. Philosophical Studies, 1-24.

Register, C. (2025). Individuating artificial moral patients. Philosophical Studies, 182(11), 3225-3246.

Roser, M. (2023). AI timelines: What do experts in artificial
intelligence expect for the future? Our World in Data.
\url{https://ourworldindata.org/ai-timelines}

Saad, B., \& Caviola, L. (2024, October 11). Digital minds takeoff
scenarios. EA Forum.
\url{https://forum.effectivealtruism.org/posts/2uGShxLsXWGExJYNL/digital-minds-takeoff-scenarios}

Sebo, J., \& Long, R. (2025). Moral consideration for AI systems by
2030. AI and Ethics, 5(1), 591-606.

Shiller, D. (2025). How many digital minds can dance on the streaming multiprocessors of a GPU cluster?. Synthese, 206(5), 218.

Shulman, C., \& Bostrom, N. (2021). Sharing the world with digital
minds. Rethinking moral status, 306-326.

Tsur, E. (2021). Neuromorphic Engineering: The Scientist's, Algorithms
Designer's and Computer Architect's Perspectives on Brain-Inspired
Computing. CRC Press.

Vaswani, A., Shazeer, N., Parmar, N., Uszkoreit, J., Jones, L., Gomez,
A. N., \ldots{} \& Polosukhin, I. (2017). Attention is all you need.
Advances in neural information processing systems, 30.

Yuan, C., \& Agaian, S. S. (2023). A comprehensive review of binary
neural network. Artificial Intelligence Review, 56(11), 12949-13013.

\appendix
\setcounter{section}{0}
\renewcommand{\thesection}{\Alph{section}}
\section*{Appendix 1}\label{app:1}

\subsection*{Scale of Digital Minds Consumer Model Overview}

This appendix presents the structure of a simple consumer model for estimating product counts over a period of time.

If we think of digital minds as products, we can use estimates of a consumer population, the fragments who will be interested in it, and growth rates of product adoption to compute total products sold year by year.

This model uses Monte Carlo methods and takes inputs in the form of distributions.

\subsubsection*{Parameters}
\mbox{}\\

\noindent The inputs are:

\emph{Average times to the introduction of a viable qualifying product to market}

This variable corresponds with the introduction of a product that meets a specified description and is suitably technologically ready and appropriately marketed. It is possible that products are available that serve the same ends without qualifying both before and after this date. For instance, we might decide to treat smart phones as entering the market in 2008, for while other previous products accomplished the same goals, they weren't ready.

\emph{Target population size of plausible consumers}

This variable gets at the size of the potential market, in terms of the number of consumers. It allows that there might be a small number of consumers who each purchase a number of the products. This represents the total population base who might be interested in the product. It isn't intended to reflect the population of maximum plausible saturation. For instance, we might choose the total population of consumers for estimating potential personal swimming pool numbers to consist of the number of households with backyards.

\emph{Target population growth rate}

This variable addresses the likely growth rate of the population of potential customers. It consists of a rate of expansion year-over-year, and it is assumed that the same rate applies across time.

\emph{Starting fraction of the population of plausible consumers}

This variable represents the initial market penetration rate at the time of product introduction. It captures the percentage of the target population that would adopt the product immediately upon its viable market entry, representing early adopters and those with the highest propensity to purchase. This fraction reflects factors such as initial pricing, marketing effectiveness, product readiness, consumer awareness, and the urgency of the need the product addresses.

\emph{Year to year growth rate}

This variable models the annual expansion rate of market adoption beyond the initial buyers. It represents how quickly the product spreads from early adopters to mainstream consumers over time, capturing the diffusion of innovation through the target population. This rate accounts for factors like word-of-mouth effects, declining prices, improved product features, increased marketing reach, and social proof effects.

\emph{Maximum fraction of plausible consumers}

This variable defines the ceiling for market penetration, representing the theoretical maximum percentage of the target population that would ever adopt the product under optimal conditions. It acknowledges that not all potential consumers will ultimately purchase the product due to factors such as cost constraints, competing alternatives, personal preferences, technological barriers, or lack of perceived need.

\emph{Percentage of viable products}

Certain categories of products could be served by a variety of offerings which may differ substantially on whether they count for our purposes. For example, when estimating the demand for solar power, we might find it easier to estimate the amount of total renewable energy demand and carve off a fraction of that total from solar.

\emph{Product count per consumer}

This variable represents the average number of units that each adopting consumer will own simultaneously at any given point in time. It accounts for consumers purchasing multiple units for various reasons, such as having devices for different locations, different family members, backup units, or upgrades while retaining older versions.

\emph{Minded fraction}

This variable represents the proportion of qualifying products that are treated as candidates for minds. Some product niches may be filled in part by products that obviously do not qualify (e.g. virtual guides with no coherent or persistent identity). This variable allows us to estimate the percentage of such products.

\subsubsection*{Approach}
\mbox{}\\

The model draws values from distributions representing each variable. 

\begin{enumerate}
\item A times to introduction of the product are generated for each run.
\item The target population size and starting fraction together determine the numbers in the first year of introduction.
\item The population growth rate determines year-to-year changes, as set by the plausible consumer population and maximum fraction who might make purchases.
\item Product counts per consumer are used as multipliers for the results of viable consumers for each run, to get the final result predictions for that year.
\end{enumerate}

\subsubsection*{Example Application: Robot Pets}
\mbox{}\\

Applying this model to robot pets requires specifying distributions for each input variable:

\begin{itemize}
\item Product introduction: Mixed distribution between 2028-2032 (50\% normal distribution around this period, 50\% never developed)
\item Target population: Normal distribution between 5 and 25\% of world's pet-owning population
\item Population growth: Normal distribution around 4\% (slightly above global population growth)
\item Initial adoption: Log normal distribution with 95\% certainty between 0.25 and 2\% of target customers
\item Year-over-year growth: Normal distribution between -10 and 150\% annual growth rate
\item Maximum adoption: Normal distribution with 95\% certainty between 10-30\% (mean around 20\%)
\item Product count: Discrete distribution (80\% chance of 1 pet, 20\% chance of 2 pets, average 1.2)
\item Minded fraction: 1
\end{itemize}

This specification allows prediction of robot pet populations through 2050, accounting for interactions between all input distributions.

\end{document}